%% file: main.tex
\documentclass[acmsmall,screen,nonacm]{acmart}

\input{macros}
\setcopyright{acmcopyright}
\begin{document}
\renewcommand{\lstlistingname}{}
\renewcommand{\thelstlisting}{}
\title[]{Translating Large-Scale C Repositories to Idiomatic Rust}

\author{Saman Dehghan*}
\authornote{* equally contributing lead authors}
\affiliation{
  \institution{University of Illinois at Urbana-Champaign}
  \country{USA}
}
\email{samand2@illinois.edu}

\author{Tianran Sun*}
\affiliation{%
  \institution{Shanghai Jiao Tong University}
  \country{China}
}
\email{seriousss@sjtu.edu.cn}

\author{Tianxiang Wu}
\affiliation{
  \institution{University of Illinois at Urbana-Champaign}
  \country{USA}
}
\email{tw50@illinois.edu}

\author{Zihan Li}
\affiliation{
  \institution{University of Illinois at Urbana-Champaign}
  \country{USA}
}
\email{zihanl19@illinois.edu}

\author{Reyhaneh Jabbarvand}
\affiliation{
  \institution{University of Illinois at Urbana-Champaign}
  \country{USA}
}
\email{reyhaneh@illinois.edu}

\begin{abstract}
Existing C to Rust translation techniques fail to balance quality and scalability: transpilation-based approaches scale to large projects but produce code with poor safety, idiomaticity, and readability. In contrast, LLM-based techniques are prohibitively expensive due to their reliance on frontier models (without which they cannot reliably generate compilable translations), thus limiting scalability. This paper proposes \approach, a fully automated pipeline for effective and efficient repository-level C to idiomatic safe Rust translation. Evaluating on a diverse set of $23$ C programs, ranging from \num{27} to \num{13200} lines of code, \approach can generate fully compilable Rust code for all and achieve $87\%$ functional equivalence (passing \num{1063099} assertions out of \num{1221192} in test suites with average function and line coverage of $74.7\%$ and $72.2\%$). Compared to six prior repository-level C to Rust translation techniques, the translations by \approach are overall safer (fewer raw pointers, pointer arithmetic, and unsafe constructs), more idiomatic (fewer Rust linter violations), and more readable. When the translations cannot pass all tests to fulfill functional equivalence, human developers were able to complete the task in $4.5$ hours, on average, using \approach as debugging support. 

\end{abstract}

\maketitle

\input{Sections/Introduction}
\input{Sections/Approach}
\input{Sections/Evaluation}
\input{Sections/RelatedWork}
\input{Sections/Conclusion}

\bibliographystyle{ACM-Reference-Format}
\bibliography{refs}

\end{document}

%% file: macros.tex
\usepackage{microtype}% Recommended, but optional, packages for figures and better typesetting

\usepackage{amsmath,amssymb,amsfonts}
\usepackage{algorithmic}
\usepackage{graphicx}
\usepackage{textcomp}
\usepackage{xurl}
\usepackage[dvipsnames]{xcolor}
\usepackage{multirow}
\usepackage{booktabs} % For formal tables
\usepackage{comment}
\usepackage{subcaption}
\usepackage{rotating}
\usepackage{adjustbox}
\usepackage{array}
 \usepackage{float}
\usepackage{tabularx}
\usepackage{wrapfig}
\usepackage[super]{nth}
\usepackage{xfrac}
\usepackage{stmaryrd}
\usepackage{listings}
\usepackage{color, colortbl}
\usepackage{balance} % to balance the bibliography
\usepackage{lscape}
\usepackage{xspace}
\usepackage{framed}
\usepackage{syntax}
\usepackage{parcolumns}
\usepackage{pifont}
\usepackage{soul}
\usepackage[tikz]{bclogo}
\usepackage[ruled,vlined,linesnumbered,noend]{algorithm2e}
\SetKwInput{KwInput}{Input}
\SetKwInput{KwInputs}{Inputs}                % Set the Input
\SetKwInput{KwOutputs}{Outputs}              % set the Output
\SetKwInput{KwOutput}{Output}              % set the Output
\usepackage{enumitem}
\usepackage{makecell}
\usepackage[many]{tcolorbox}
\usepackage{cleveref}
\usepackage{capt-of}
\usepackage{mathtools}
\usepackage{amsthm}
\usepackage{minted}
\crefformat{section}{\S#2#1#3} % see manual of cleveref, section 8.2.1
\crefformat{subsection}{\S#2#1#3}
\crefformat{subsubsection}{\S#2#1#3}

\newcolumntype{R}[2]{%
    >{\adjustbox{angle=#1,lap=\width-(#2)}\bgroup}%
    l%
    <{\egroup}%
}
\usepackage{listings}
\usepackage{xcolor}
\usepackage{dirtree}
\usepackage{siunitx}
\newcommand{\reyhan}[1]{\textcolor{magenta}{\textbf{Reyhan:} #1}}

\newcommand{\approach}{\textsc{Rustine}\xspace}
\newcommand{\llmv}{DeepSeek-V3\xspace}
\newcommand{\llmr}{DeepSeek-R1\xspace}

\newcommand{\ollama}{Ollama\xspace}
\newcommand{\vllm}{vLLM\xspace}

\newcommand{\ctorust}{C2Rust\xspace}
\newcommand{\saferrust}{C2SaferRust\xspace}
\newcommand{\crown}{CROWN\xspace}
\newcommand{\laertes}{Laertes\xspace}
\newcommand{\rustmap}{RustMamp\xspace}
\newcommand{\syzygy}{Syzygy\xspace}

\newtcolorbox{boxK}{
    sharpish corners, % better drop shadow
    boxrule = 0pt,
    toprule = 4.5pt, % top rule weight
    enhanced,
    fuzzy shadow = {0pt}{-2pt}{-0.5pt}{0.5pt}{black!35} % {xshift}{yshift}{offset}{step}{options} 
}

\makeatletter
\newenvironment{btHighlight}[1][]
{\begingroup\tikzset{bt@Highlight@par/.style={#1}}\begin{lrbox}{\@tempboxa}}
{\end{lrbox}\bt@HL@box[bt@Highlight@par]{\@tempboxa}\endgroup}

\newcommand\btHL[1][]{%
  \begin{btHighlight}[#1]\bgroup\aftergroup\bt@HL@endenv%
}
\def\bt@HL@endenv{%
  \end{btHighlight}%   
  \egroup
}
\newcommand{\bt@HL@box}[2][]{%
  \tikz[#1]{%
    \pgfpathrectangle{\pgfpoint{1pt}{0pt}}{\pgfpoint{\wd #2}{\ht #2}}%
    \pgfusepath{use as bounding box}%
    \node[anchor=base west, fill=yellow,outer sep=0pt,inner xsep=1pt, inner ysep=0pt, rounded corners=3pt, minimum height=\ht\strutbox+1pt,#1]{\raisebox{1pt}{\strut}\strut\usebox{#2}};
  }%
}
\makeatother

\definecolor{codegreen}{rgb}{0,0.6,0}
\definecolor{codegray}{rgb}{0.5,0.5,0.5}
\definecolor{codepurple}{rgb}{0.58,0,0.82}
\definecolor{backcolour}{rgb}{0.95,0.95,0.92}

\lstdefinestyle{mystyle}{
language=Java,
 basicstyle=\linespread{1.1}\ttfamily\footnotesize,
     backgroundcolor=\color{backcolour}, commentstyle=\color{codegreen},
  keywordstyle=\color{magenta},
  numberstyle=\tiny\color{codegray},
  stringstyle=\color{codepurple}, xleftmargin=0.7cm,
    frame=tlbr, framesep=0.2cm, framerule=0pt, numbers=left,
    breaklines=true,
     moredelim=**[is][\btHL]{`}{`},
    moredelim=**[is][\color{orange}\bfseries]{@}{@}
}

\lstdefinestyle{stylewithcomment}{
language=Java,
 basicstyle=\linespread{0.9}\footnotesize,
     backgroundcolor=\color{backcolour}, commentstyle=\color{codegreen},
  keywordstyle=\color{magenta},
  numberstyle=\tiny\color{codegray},
  basicstyle=\scriptsize\ttfamily,
  stringstyle=\color{codepurple}, xleftmargin=0cm,
    frame=tlbr, framesep=0.1cm, framerule=0pt, numbers=none,
    breaklines=true,
    moredelim=**[is][\color{red}]{@}{@},
    moredelim=**[is][\bfseries]{`}{`}
}

\lstdefinestyle{stylewithcommentPy}{
language=Python,
 basicstyle=\linespread{0.9}\footnotesize,
     backgroundcolor=\color{backcolour}, commentstyle=\color{codegreen},
  keywordstyle=\color{magenta},
  numberstyle=\tiny\color{codegray},
  basicstyle=\scriptsize\ttfamily,
  stringstyle=\color{codepurple}, xleftmargin=0cm,
    frame=tlbr, framesep=0.1cm, framerule=0pt, numbers=none,
    breaklines=true,
    moredelim=**[is][\color{red}]{@}{@},
    moredelim=**[is][\bfseries]{`}{`}
}
\theoremstyle{plain}

\theoremstyle{definition}

\theoremstyle{remark}

\usepackage[textsize=tiny]{todonotes}

%% file: Sections/Introduction.tex
\section{Introduction}
\label{sec:introduction}

C is one of the most widely used programming languages\footnote{Second rank in TIOBE Index: https://www.tiobe.com/tiobe-index}. However, its permissive memory model and lack of built-in safety mechanisms make it prone to memory safety vulnerabilities~\cite{Nagarakatte2009SoftBound,Michael2022MSWasm,Hathhorn2015UndefinedC,Necula2005CCured,Wallach2024MemorySafetyManifesto,Serebryany2023GWPASan,Bernhard2022xTag}. Rust has emerged as a compelling alternative, offering strong memory safety guarantees. Manually migrating large-scale C projects to idiomatic Rust, however, is costly, time-consuming, and error-prone~\cite{Emre2021TranslatingSaferRust,Emre2023AliasingLimits,Li2025UserStudyC2Rust,Wu2025GenC2Rust,Zhang2023OwnershipGuided}. Automating this translation process at the repository level, while preserving correctness, idiomatic usage, and maintainability, has thus become a highly relevant research challenge. Existing approaches mark important progress in automating C to Rust translation~\cite{nitin2025c2saferrust,zhang2023ownership,emre2021translating,shetty2024syzygy,cai2025rustmap,yang2024vert}. However, they suffer from the following limitations: 

\begin{itemize}[leftmargin=*]
    \item \textbf{Non-Idiomatic Rust Output.} The majority of prior work~\cite{nitin2025c2saferrust,zhang2023ownership,emre2021translating} relies on \ctorust~\cite{ctorust} transpiler for initial translation. \ctorust translations frequently mirror the original C code structure and \emph{unsafe} semantics too closely, relying heavily on unsafe blocks~\cite{pan2024lost}. While these techniques improve some safety issues in the initial translation, their translation are not pure, failing to interact with the broader Rust ecosystem (e.g., crates or build tools). 

    \item \textbf{Scalability–Quality Tradeoff.} To overcome the previous limitation, recent techniques leverage large language models (LLMs) to translate C code to Rust from scratch~\cite{shetty2024syzygy,cai2025rustmap,yang2024vert}. These techniques rely on raw model power to overcome large-scale translation challenges, thereby, only work with frontier models. This can be prohibitively expensive, failing to scale to large projects\footnote{Such techniques either translated one or two relatively small projects, or a subset of the project.}. 
    For example, \syzygy~\cite{shetty2024syzygy} translated \emph{zopfli} project (\num{2937} LoC), costing around $\$800$. These techniques also require initial \emph{manual} effort, e.g., manual translation of structs in \syzygy, to help LLMs produce a \emph{compilable} translation. 

    \item \textbf{Insufficient Translation Validation.} Translation bugs are inevitable in LLM-based transpilation~\cite{pan2024lost}, and LLM-based techniques struggle to generate even a fully compilable code~\cite{khatry2025crust,sim2025large,yuan2025project,bai2025rustassure,transpilerrustify,wang2025evoc2rust,zhou2025llm}. Majority of compilation bugs are not just syntactic inaccuracy, but violation of Rust's safety guarantee---strict ownership, lifetime, and borrowing rules. This makes achieving $100\%$ compilability the \emph{bare minimum} validation requirement. The next level of validation is assessing functional equivalence, mostly done through testing in repository-level code translation\footnote{Formal methods can assess functional equivalence of a subset of program, but struggle to scale to entire repository~\cite{yang2024vert}.}. 
    Prior work barely reports details of test suites (e.g., number of tests, assertions, and coverage), and test suites of many evaluation subjects have low coverage or tests with no assertions, making them unsuitable to assess functional equivalence of translations.  

    \item \textbf{Lack of Comprehensive Evaluation Metrics.} Existing evaluation metrics focus only on a subset of safety aspects, e.g., the number of raw pointer declarations and dereferences or unsafe constructs, failing to assess the presence or absence of essential idiomaticity features in Rust translations, e.g., readability and usage pointer arithmetic/comparison.
    
    \item \textbf{Limited Error Reporting and Debugging Support.} Existing tools still cannot fully automate the translation of C projects to Rust and therefore require substantial human effort. However, they offer limited actionable diagnostics or tool support, reducing developer productivity when repairing partial translations. Notably, no prior C-to-Rust translation approach has conducted a human study to evaluate the maintainability or usability of the generated translations.

\end{itemize}

We propose \approach, an end-to-end pipeline to translate C repositories (application/test code and configuration/build files) to a \emph{maintainable}, \emph{idiomatic}, and \emph{functionally equivalent} Rust version:% \approach overcomes the mentioned limitations as follows:

\begin{itemize}[leftmargin=*]
    \item \textbf{Idiomatic Rust Translation.} \approach translates C projects from scratch. With all the translations produced by LLMs, the generated code is more readable and safer, hence better aligned with Rust idiomaticity.
    \approach success is not just from using LLM, but also employing sound program analysis techniques to \emph{automatically} (1) refactor the C code that would otherwise confuse the LLM for translation (resolving preprocessor directives, minimizing pointer operations, expanding macros, and maximizing constness); and (2) realize dependencies to guide the translation (caller-callee relationships, references in structures to other types, global state flows, and lifetime dependencies between variables).
    
    \item \textbf{Cost-effective Repository-level Translation.} \approach follows a two-tier prompting strategy in leveraging LLMs for code translation: an affordable LLM for the majority of the translation and a reasoning model in case of first model's failure. \approach considers several automated prompt crafting strategies to optimize the performance of LLMs, including RAG-based API translation and adaptive in-context learning example generation and usage.

    \item \textbf{Validation Beyond Syntactic Check.} \approach employs test translation and execution to evaluate functional correctness. This entails the existence of test suites with high coverage \emph{and} high-quality assertions. We manually augment the test suite of existing benchmarks with more tests and assertions, and report the functional equivalence based on the percentage of assertions passed, not just test passes. This transparency avoids bias against passing tests with no assertions, which, at best, indicate runtime correctness and not functional equivalence. 

    \item \textbf{Comprehensive Evaluation Metrics.} We evaluate different quality aspects of translations using metrics from prior work, as well as new ones such as \# pointer arithmetic/comparisons, idomaticity metric, readability metrics (cognitive complexity, Halstead, and SEI Maintenance Index), assertion pass rate, execution time, and cost.

    \item \textbf{Effective Debugging Procedure.} \approach implements a post-translation debugging. Inspired by the idea of delta debugging~\cite{zeller1999isolating,zeller2002simplifying,cleve2005cause,misherghi2006hdd}, it disables all failed assertions except one, to determine the culprits of non-equivalence in the translation corresponding to a specific failure. Narrowing down to a few translation units for each assertion enables \approach to effectively resolve many semantic mismatches, i.e., compilable code resulting in assertion failures.  
        
\end{itemize}

We evaluated the performance of \approach in translating $23$ C repositories into Rust, in comparison with \emph{six} leading prior approaches. The C repositories range from $27$ to \num{13200} lines of code, and are diverse in terms of using structs, unions, macros, raw pointer declarations/dereferences, and pointer arithmetic/comparisons. \approach generates \emph{fully compilable} Rust translations (both application and test code) for \emph{all the projects}, without any human intervention. In overall comparison with alternative approaches, \approach translations are \emph{safer} (less use of raw pointers and pointer arithmetic/comparison), \emph{more idiomatic} (fewer \texttt{\small{Clippy}}~\cite{clippy} warnings), and \emph{more readable} concerning three readability and maintainability metrics. For the subjects where translations, after \approach automated debugging, could not pass all assertions, a human developer resolved semantic mismatches in $4.5$ hours, on average, assisted by \approach debugging support. The final translations are $87\%$ \emph{functionally equivalent} (passing \num{1063099} assertions out of \num{1221192} in test suites of subjects, with average function and line coverage of $74.7\%$ and $72.2\%$). These notable advancements in automated translation of C repositories to Rust come with a \emph{low time and cost overhead}: on average, it takes $3.92$ hours for \approach to finish the translation, with the cost of $\$0.48$. 

Our contributions are: (1) an automated, cost-effective pipeline to translate large-scale C repositories into compilable, functionally equivalent, and idiomatic Rust; (2) the first approach to target and evaluate resolution of pointer arithmetic during translation; (3) comprehensive evaluation metrics to promote more transparent assessment of C to Rust translation; (4) enhancing benchmarking of C to Rust translation by augmenting the test suite of existing benchmarks with more tests and assertions; and (5) a comprehensive analysis of proposed technique compared to notable approaches, that can be used as baseline for future research (we discovered several issues in evaluation of other studied pipelines and recalculated metrics using our pipeline). 
Our artifacts are publicly available~\cite{website}.

%% file: Sections/Approach.tex
\section{Approach}
\label{sec:approach}

\begin{figure*}[t]
    %\vspace{-30pt}
    \includegraphics[width=0.8\textwidth]{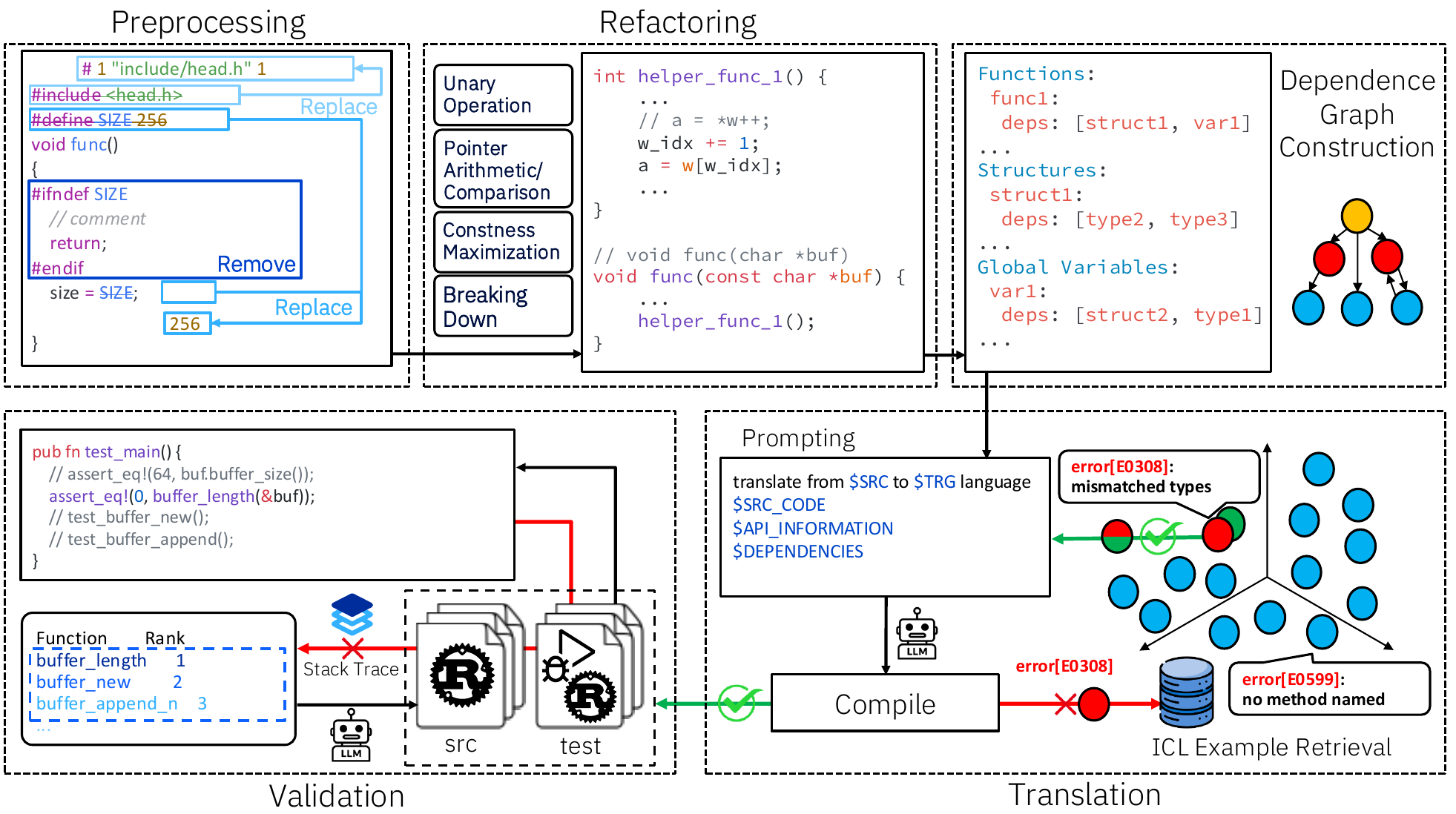}
    \vspace{-10pt}
    \caption{\approach framework consisting of five main components}
    \label{fig:framework}
\end{figure*}

\approach follows a five-stage architecture (Figure~\ref{fig:framework}) to progressively refine a C project into idiomatic Rust: \emph{Preprocessing} standardizes the code structure (\S \ref{subsec:preprocess}), \emph{Refactoring} resolves ambiguous, unsafe, anc challenging constructs in C code (\S \ref{subsec:refactoring}), \emph{Analysis} realizes intra- and inter-procedural dependencies of the C project for context engineering and determining translation order (\S \ref{subsec:analysis-decomposition}), an LLM \emph{translates} application and test code through an iterative feedback loop (\S \ref{subsec:translation}), and \emph{Validation} determines functional equivalence using translated tests, and debugs non-equivalent cases (\S \ref{subsec:debugging}). 

\vspace{-5pt}
\subsection{Automated Preprocessing of the Source Code}
\label{subsec:preprocess}

\approach uses the C preprocessor~\cite{gcccpp} to insert the contents of header files, substitute macro definitions, and resolve conditional compilation directives, e.g., \texttt{\small{\#ifdef}}, \texttt{\small{\#ifndef}}, and \texttt{\small{\#if defined()}}. This step standardizes the code and allows subsequent stages to focus solely on translating C syntax to Rust. Otherwise, the LLM must expand the macros itself, a process that can involve complex token concatenation~\cite{stallman1987c}, stringification~\cite{stallman1987c}, and recursive expansions~\cite{ernst2002empirical}, obscuring the underlying code logic. Some macros are context-dependent, i.e., the same macro produces different expansions based on prior definitions. Macros lack access to type definitions and function signatures from header files, making it difficult to generate correct Rust types and function calls.

\vspace{-5pt}
\subsection{Automated Refactoring of C Project}
\label{subsec:refactoring}

\begin{figure}
\centering
\begin{minipage}[c]{\linewidth}
\centering

\input{Figures/challenge}
\end{minipage}
\vspace{-0.5cm}
\caption{An illustrative example from \emph{zopfli} project (a),  refactored version with no pointer arithmetic (b), and corresponding translations of them (c) and (d)}
\label{fig:challenge1}
\end{figure}

Following preprocessing, \approach automatically refactors ambiguous or unsafe constructs that, if translated literally, would result in unidiomatic or unsafe Rust code. Specifically, \approach refactors usage of \emph{unary operators} (\S \ref{subsub-unary}) and \emph{pointer arithmetic} (\S \ref{subsub:pointer-arithmetic}), maximizes constness (\S \ref{subsub:maximization}), and breaks down large functions that do not fit into the context window of LLMs into smaller segments (\S \ref{subsub:breakdown}). This pre-translation step can significantly increase the likelihood of LLM generating correct, idiomatic Rust code. Figure \ref{fig:challenge1}-a shows an example from \emph{zopfoli} project, containing pointer arithmetic/comparisons (highlighted lines). Without refactoring, \approach results in the unsafe translation of Figure~\ref{fig:challenge1}-c, which contains unsafe lines and several raw pointer arithmetic. Figure~\ref{fig:challenge1}-b shows a refactored version, which LLM can translate into a more idiomatic Rust (Figure~\ref{fig:challenge1}-d). As a proxy for checking whether refactoring breaks functionality, \approach requires a successful post-refactoring test execution. Otherwise, it reverts the refactoring. 

\vspace{-5pt}
\subsubsection{Refactoring Unary Operators}
\label{subsub-unary}

Automated translation of unary operators is challenging since Rust requires explicit modification statements for clarity and error prevention. If done by an LLM, the translation may not consider the evaluation order and result in a semantic mismatch. \approach detects all pre- and post-increment/decrement operators. If the unary operator is a standalone expression, e.g., \texttt{\small{*var++;}}, \approach desugars it to a compound assignment, e.g., \texttt{\small{var+=1}}. Otherwise, it separates the expression into one compound assignment statement and one statement for the rest of the expression, preserving the order. When there are multiple unary operations in one statement, \approach skips refactoring to avoid semantic alternation. Line $7$ of Figure~\ref{fig:challenge1}-a shows such a case, where increments should only occur if the previous conditions succeed.

\vspace{-5pt}
\subsubsection{Refactoring Pointer Operations}
\label{subsub:pointer-arithmetic}

\begin{figure}
\centering
\vspace{-3pt}
  \begin{minipage}{\linewidth}
\input{Figures/refactoring}
 \vspace{-0.4cm}
\end{minipage}
\caption{Refactoring of the \emph{genann} project for resolving unary (middle) and pointer arithmetic (right) operations}
\label{fig:refactoring}
\end{figure}

\begin{wrapfigure}{R}{0.49\linewidth}
    \centering
    \scriptsize
    \begin{minipage}{\linewidth}
        \vspace{-12pt}
        \begin{algorithm}[H]
            \input{Algorithms/pointer}
        \end{algorithm}
        \vspace{-10pt}
    \end{minipage}
\end{wrapfigure}
Pointer operations (arithmetic/comparisons) enable efficient memory manipulation in C, but sacrifice safety (buffer overflows, dangling pointers, and other memory safety violations~\cite{One1996SmashingTS}) and readability. In contrast, Rust's ownership model prohibits pointer arithmetic or comparison~\cite{beingessner2016you}, requiring the LLM to find an equivalent safe translation. Such operations can be complex in C programs, making their translation a recipe for disaster. To alleviate this, \emph{following the refactoring of the unary operators}, \approach identifies the usages of pointer arithmetic/comparisons and transforms them into array-style indexing operations, using Algorithm~\ref{alg:pointer-refactor-detailed}. 

\approach starts by traversing the function's Abstract Syntax Tree (AST) to locate all pointers within the scope of the function (declared or passed) that are involved in arithmetic/comparison operations, e.g., addition or subtraction. For each such pointer, a corresponding integer index variable is declared at the beginning of the pointer declaration scope. This index variable tracks the pointer's position within the underlying array or memory block. For instance, for \texttt{\small{double const* w}} in \emph{genann} project (Figure~\ref{fig:refactoring}), which undergoes arithmetic operations inside the loop, an integer variable \texttt{\small{int w\_idx = 0}} is declared at the beginning of the function. 
\approach transforms the identified pointers depending on the type of pointer arithmetic: when pointers are used to access memory contents, \approach converts them into array indexing operations using the base array and the associated index variable (\texttt{\small{double sum = *w * -1.0}} in Figure~\ref{fig:refactoring} is transformed to \texttt{\small{double sum = w[w_idx] * -1.0}}); when arithmetic operations are performed directly on pointers, \approach applies index transformation on pointer and adds \texttt{\small{\&}} to access literal value of pointer (\texttt{\small{scan != end}} in Figure~\ref{fig:challenge1} transforms to \texttt{\small{\&scan[scan_idx] != end}}); when pointers are reintialized within the code, \approach resets their indices after re-initialization statements in the same scope. 

For \emph{alias-safe} transformations, \approach updates all transformed pointers that were passed to the function as an argument (regardless of them escaping or not). This ensures that changes in pointers are reflected outside of the function scope. In the examples of Figure~\ref{fig:challenge1}-a, two pointers \texttt{\small{scan}} and \texttt{\small{match}} are \emph{aliased} outside of the \texttt{\small{GetMatch}} scope. Both these pointers are transformed, and although they are passed by value, \approach updates them to ensure usage of them outside of \texttt{\small{GetMatch}} is not broken. 
This also ensures that, if only one of the aliased pointers is transformed, the semantics are not broken. In the example of Figure~\ref{fig:refactoring}, \texttt{\small{i}} and \texttt{\small{o}} point to \texttt{\small{output}} (although to non-overlapping sections). \approach only refactors \texttt{\small{o += 1}}, and since \texttt{\small{i}} is only accessed by index \texttt{\small{sum += *w * i[k]}}, the semantics are preserved. In addition to post-refactoring test execution, we also prove \approach pointer operation transformations are semantic preserving (details in Appendix).

Algorithm~\ref{alg:pointer-refactor-detailed} is \emph{bound-agnostic}, eliminating the need to specify object bounds explicitly, assuming safe pointer arithmetic. If this assumption does not hold, the translated Rust is either non-compilable or be corrected by the LLM into a safe implementation. 

\subsubsection{Constness Maximization.}
\label{subsub:maximization}

\approach analyzes data flow to identify opportunities for maximizing constness in function parameters that are never modified within their scope, and transforms them by adding \texttt{\small{const}} qualifiers to their declarations. For example, \texttt{\small{char *buffer}} and \texttt{\small{int}} are transformed to \texttt{\small{const char *buffer}} and \texttt{\small{const int}}, respectively. The modified C code with maximized constness directly guides the LLM to generate appropriate Rust constructs: \texttt{\small{const}} parameters translate naturally to immutable references (\texttt{\small{\&T}}) or owned immutable values, while non-const parameters indicate the need for mutable references (\texttt{\small{\&mut T}}) or owned mutable values. 

\vspace{-5pt}
\subsubsection{Large Function Breakdown}
\label{subsub:breakdown}

\begin{wrapfigure}{R}{0.54\linewidth}
    \centering
    \scriptsize
    \begin{minipage}{\linewidth}
        \vspace{-11pt}
        \begin{algorithm}[H]
            \input{Algorithms/breakdown}
        \end{algorithm}
        \vspace{-15pt}
    \end{minipage}
\end{wrapfigure}
Large functions may exceed the LLM context window, or even if not, the LLM may struggle to maintain attention across the entire function scope, leading to suboptimal translation quality~\cite{liu2024lost}. 
Furthermore, in case of inevitable translation bugs, the LLM is more likely to succeed in repair by focusing on smaller code rather than a large one. To account for this, \approach employs a flow-sensitive analysis to break down large functions---those larger than $50\%$ of LLM context window---prior to translation, using Algorithm~\ref{alg:function-breakdown}.  

\sloppy For a given function $F$ in C program that is larger than a breakdown threshold $\theta_f$, \approach determines basic blocks within control flow structures, i.e., \texttt{\small{if-else}} statements, \texttt{\small{while}}/\texttt{\small{for}} loops, and \texttt{\small{switch/case}}. The blocks \emph{larger} than threshold $\theta_b$ (to avoid excessive breakdown) with no early exit points, e.g., \texttt{\small{return}}, \texttt{\small{goto}}, or \texttt{\small{break}} statements are selected for breakdown. \approach moves candidate blocks from the previous step to new helper functions, and replaces each with a functional call. The return type of new helper functions is \texttt{\small{void}} and the input parameters are all variables accessed or modified within the block, passed by reference where necessary.

\subsection{Dependence Graph Construction}
\label{subsec:analysis-decomposition}

\approach performs a context-sensitive static analysis to construct a Dependence Graph (DG) for C projects. Such a data structure helps \approach track caller-callee relationships, references in structures to other types, global state flows, and lifetime dependencies between variables. \approach will use DG for three purposes: (1) to construct a project skeleton, (2) to guide the translation order, and (3) to provide a rich context to guide the LLM with translation (\S \ref{subsec:translation}).

\noindent \textbf{Definition 1.} Dependence Graph (DG) is a directed graph $DG = (V \coloneqq F \uplus S \uplus U \uplus G \uplus T, E)$, reflecting dependencies among program components ($v \in V$), including functions ($F$), structs ($S$), unions ($U$), global variables ($G$), and type definitions/enums ($T$). A directed edge $e_{i,j}=(v_i,v_j) \in E$ demonstrates an \emph{explicit} use-def or caller-callee relationship between $v_i$ and $v_j$. 

\begin{wrapfigure}[]{r}{0.47\textwidth}
    \centering
    \vspace{-10pt}
    \includegraphics[width=\linewidth]{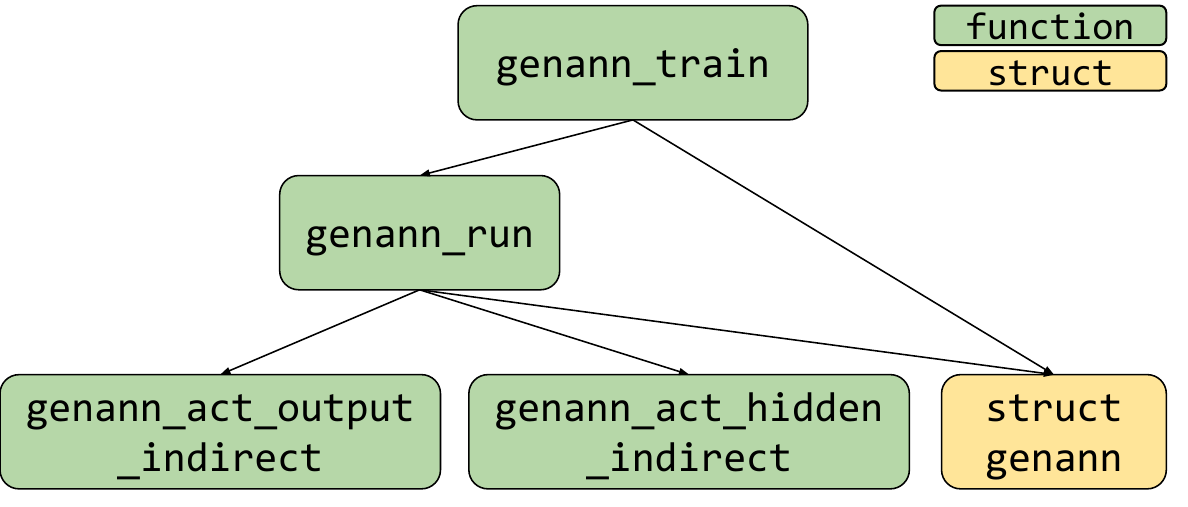}
    \vspace{-20pt}
    \caption{DG snapshot of Figure~\ref{fig:refactoring}}
    \vspace{-13pt}
    \label{fig:DG-example}
\end{wrapfigure}
Figure~\ref{fig:DG-example} shows a DG snapshot for \emph{genann} project, corresponding to Figure~\ref{fig:refactoring}.
%, showing dependencies between function \texttt{\small{genann_run}} and its related entities, like \texttt{\small{struct genann}}, which appears as an input argument (line 2), and \texttt{\small{genann_act_output}} represents a function call dependency (line 15). Although \texttt{\small{genann_run}} also depends on \texttt{\small{genann_act_hidden_indirect}}, this relationship is not visible in Figure \ref{fig:refactoring}. In \texttt{\small{genann}}, the function \texttt{\small{genan_train}} depends on \texttt{\small{genann_run}}. 
Cyclic DGs are inevitable in large C codebases due to the common use of mutual recursion, forward declarations, and complex header inclusion patterns. After construction of the initial DG, \approach detects strongly connected nodes (using Tarjan’s algorithm~\cite{tarjan1972depth}) and condenses each into a \emph{super translation unit}. This enables LLM to generate appropriate Rust constructs such as mutually recursive data structures or carefully designed module hierarchies.

\approach creates a compilable, memory-safe Rust project skeleton to support incremental translation. The structure includes a \texttt{\small{src/}} directory with \texttt{\small{*.rs}} files, aligning C units to Rust modules. The files contain placeholder functions with \emph{incomplete} Rust signatures (but with a complete commented signature in C for the context) and the \texttt{\small{todo!()}} body. \approach also consolidates global variables in \texttt{\small{src/globals.rs}} for maintainability, and identifies main functions, generating [bin] sections in \texttt{\small{Cargo.toml}}. Each struct/union will be placed in its own \texttt{\small{[struct_name].rs}} module for clarity. 

\subsubsection{Enhancing Cross-language API Translation}
\label{subsub:API-invocation}

An important consideration in real-world code translation is ensuring accurate translation of API calls (core libraries or external dependencies), which requires understanding the semantics of the original invocation and finding the equivalence in the target programming language, if any. This makes API calls in the source language essential dependency points that must be explicitly identified and analyzed to support translation. 

APIs invocations either belong to core libraries, e.g., \texttt{\small{std::atof}} and \texttt{\small{isatty}}, or external libraries, e.g., \texttt{\small{X11}}~\cite{xlib} and \texttt{\small{ncurses}}~\cite{ncurses}. \approach does not consider API invocation as a separate node in DG, but augments function node $F$ with invoked APIs listed as dependencies, accompanied by \emph{suggestions} about how to handle them during translation. To that end, \approach first locates all the API invocations.
%and determines whether they belong to core libraries. 
For core libraries, \approach retrieves API description and synopsis from Linux documentation~\cite{linux-man-pages}, helping the model better understand semantics of the API to find its equivalent in Rust, or write the semantics itself if no equivalent Rust API is found. For external libraries, \approach automatically generates Foreign Function Interface (FFI) using bindgen~\cite{rustbindgen}, and provides the corresponding FFI API as a suggestion during translation. 

In practice, such analysis proved to be useful for accurate API translation, as we will show in \S \ref{sec:evaluation}: most C core libraries now have an equivalent in Rust, and the provided context helped \approach using the equivalent Rust API in the translation. For example, \approach used \texttt{\small{f64::abs}} Rust API as an equivalent to \texttt{\small{std::fabs}} in C. \approach used FFI only in one case, where the library had no exact equivalence in Rust (\texttt{\small{X11}} library in \emph{xzoom} and \emph{grabc} projects). For other cases, \approach used equivalent API or corresponding wrappers in Rust, e.g., \texttt{\small{ncurses-rs}} crate~\cite{ncursesrs} for \texttt{\small{ncurses}} in C.

%it was able to translate with an equivalent third-party API in Rust, i.e., \texttt{\small{ncurses}}. 

%\reyhan{@Saman, please fill the placeholders. we have fabs and isatty but not sure should I use std:: for them or not. also, cite them}
%\reyhan{We need one figure here with three sub-figures, showing examples we discussed in the meeting}

\subsection{Translation}
\label{subsec:translation}

\begin{wrapfigure}{R}{0.54\linewidth}
    \centering
    \scriptsize
    \begin{minipage}{\linewidth}
        \vspace{-15pt}
        \begin{algorithm}[H]
            \input{Algorithms/main-loop}
        \end{algorithm}
        \vspace{-12pt}
    \end{minipage}
\end{wrapfigure}
\approach translates a project in the reverse topological order of the DG, using algorithm~\ref{alg:main-loop}. For each translation unit, it starts by prompting a general LLM. The initial prompt contains a translation unit in C, a fixed in-context learning (ICL)~\cite{brown2020language} example (to illustrate an idiomatic Rust translation of a sample C code to model), language-level hint rules, dependencies (e.g., global variables, used structs, function signatures of translated callees), and the context related to invoked C libraries as discussed in \S \ref{subsub:API-invocation}. \approach re-prompts the general LLM with a maximum budget of $A_{base}$, providing compilation errors as feedback to the model, to generate a compilable Rust translation. 

State-of-the-art LLMs can reliably produce syntactically correct Rust code. Most compilation errors, hence, stem from violations of Rust’s strict safety rules, particularly ownership and borrowing. Additional iterations, even when accompanied by error feedback, may not be helpful to resolve such complexities. Thereby, when reaching the $A_{base}$ budget, \approach switches to a reasoning model to analyze the nature of errors. \approach helps the model with \emph{adaptive} ICL examples~\cite{wu-etal-2023-self,rubin-etal-2022-learning,liu-etal-2022-makes} (\S \ref{subsub:adaptive-ICL}).

If the reasoning model also cannot generate a compilable code, \approach continues the translation loop with the latest feedback from the reasoning model. This design decision is motivated by a cost-quality trade-off: reasoning models are slower and more expensive, and using them multiple times increases the total runtime and cost. Reasoning models are also likely to resolve many major issues in the translation, leaving minor issues for the general model to resolve. Translation terminates after $A_{max}$ iterations, or when \approach generates a compilable translation. 

\begin{figure}
\centering
\begin{minipage}[c]{\linewidth}
\centering
\input{Figures/icl}
\end{minipage}
\vspace{-0.5cm}
\caption{An example of in-context learning from \textit{tulpindicator} project. 
(a) buggy translation; (b) error message; 
(c, d) ICL example: buggy code; 
(e) ICL example: fixed code; 
(f) compilable Rust translation}
\label{fig:icl}
\end{figure}

\subsubsection{Adaptive ICL Generation and Retrieval}
\label{subsub:adaptive-ICL}

In-context learning has been shown to effectively instruct LLMs with new tasks or help them solve challenging problems within known tasks. \approach uses a fixed ICL example for initial prompting to instruct the LLM with idiomatic Rust translation. Rust errors are diverse, i.e., $500$ official errors~\cite{rusterrorcode}, with some very complex, requiring more help for LLM than compilation feedback to resolve them. In the presence of persistent compilation errors, \approach switches to \emph{adaptive} ICL examples. Adaptivity means that, for each specific combination of translation unit and error message, \approach retrieves (from an ICL pool) or generates a similar ICL example to include in the prompt as a context. 

The ICL pool of \approach is indexed by Rust error codes~\cite{rusterrorcode}, e.g., \texttt{\small{E0308}} (type mismatch). When the ICL pool has examples corresponding to an error code, \approach retrieves one of the existing examples based on embedding similarity. If no example exists or existing examples are not similar, \approach prompts OpenAI o4-mini with the (1) buggy Rust code and (2) the corresponding error message to create a \emph{similar, yet simple} Rust code with the same issue, as well as a fix to it. Figure~\ref{fig:icl} shows an example from \emph{tulpindicator} project, where \approach produced an incorrect translation after $A_{base}$ attempts. The error stems from creating multiple mutable subslices of the same buffer, \texttt{\small{outref}}, within one scope, which leads to overlapping mutable borrows and triggers error \texttt{\small{E0499}}. The ICL example abstracts this buggy behavior on a two-dimensional array and offers an effective fix: it first partitions the slice using \texttt{\small{split\_at\_mut}} to obtain disjoint chunks, then derives each mutable reference from those chunks. Guided by the ICL example, \approach implements the same transformation to \texttt{\small{outref}} and produces three non-overlapping mutable slices.

\subsection{Validation and Debugging}
\label{subsec:debugging}

\begin{comment}
\begin{wrapfigure}{R}{0.5\linewidth}
  \centering
  \scriptsize
  \begin{minipage}{\linewidth}
  \vspace{-30pt}
    \begin{algorithm}[H]
    \input{Algorithms/debugging}
\end{algorithm}
\vspace{-0.5cm} % to remove the whitespace below algorithm
\end{minipage}
\end{wrapfigure}
\end{comment}
%Similar to the recent trend in repository-level code translation, 
\approach leverages test translation and execution for validation, and to determine an \emph{upper bound} for functional equivalence: after completing the translation of the entire project, 
%including the application and test code, 
\approach executes the translated tests. Even when the translation is compilable and safe, it may not be semantically equivalent to the source code. To account for post-translation validation issues, specifically test execution failures, \approach implements an automated debugging process.
%, illustrated by  Algorithm~\ref{alg:llm-debug}. 

When a failure occurs, \approach temporarily disables the failed assertions (corresponding to the reported panics) or the test (in case of other runtime errors or the absence of assertions) to complete test execution. It then enables failed assertions/tests, one by one, \emph{from the beginning}, and identifies the top $n$ functions on the stack trace as potential culprits. If a function in this shortlist is a \emph{focal method} examined by a \emph{passing assertion} in the test suite, \approach removes it from the shortlist and adds the next function in the stack trace to the shortlist. This exclusion avoids re-translating functions that have already demonstrated correct behavior under other test scenarios, following established principles of test-based behavioral consistency~\cite{pearson2017evaluating}. 

Next, \approach prompts a (reasoning) LLM to fix the semantic mismatch of the culprits, providing C source, Rust translation, and \texttt{\small stderr} context. If the result of re-prompting is not compilable, \approach reverts changes and disables the failing assertion again. Otherwise, it re-executes the test suite. If the re-execution results in a corresponding test/assertion pass without failing prior passing tests/assertions, \approach considers the semantic mismatch resolved. %In case of test/assertion failure, \approach re-checks the stack trace. If the current method is still at the top of the stack trace, the debugging terminates with reverting all changes and disabling the assertion/test. Otherwise, \approach continues with the next culprit from the shortlist. 
%\reyhan{note to self: edit the text based on Tianran comment}

The debugging approach is not perfect due to heuristics and looking at a subset of potential culprits, but it is lightweight in terms of time and cost. When automated debugging cannot fix a semantic mismatch (due to the limitations of LLM or incompleteness of heuristics), the debugging process can help a human expert in the loop to rescue: the human will see the full stack trace and re-translation, but use their expertise and reasoning to locate and fix the semantic mismatch.

%% file: Figures/challenge.tex
\definecolor{softgreen}{RGB}{200,255,200} 
\vspace{-4pt}
\begin{tabular}{c c}
% \toprule
% {\small Original C } & {\small unary-op refactor } & {\small pointer arithmetic refactor } \\
% \hline \\
 % \vspace{-1mm} &  \vspace{-1mm} A. genann project & \vspace{-1mm} \\
\begin{minipage}[t][][t]{0.47\textwidth}
\setlength{\fboxsep}{1pt}
\begin{lstlisting}[]
static const unsigned char *
GetMatch(const unsigned char *scan,const unsigned char *match,
 const unsigned char *end,const unsigned char *safe_end) {
 //... omitted code
 } else {
  while ((*@\colorbox{yellow}{scan < safe_end}@*) && (*@\colorbox{yellow}{*scan == *match}@*) &&
   (*@\colorbox{yellow}{*++scan}@*) == (*@\colorbox{yellow}{*++match}@*) && (*@\colorbox{yellow}{*++scan}@*) == (*@\colorbox{yellow}{*++match}@*) &&
   /** omitted code **/) {
      (*@\colorbox{yellow}{scan++;match++}@*);
    }
  }
  while ((*@\colorbox{yellow}{scan != end}@*) && (*@\colorbox{yellow}{*scan == *match}@*)) {
    (*@\colorbox{yellow}{scan++; match++}@*);
  }
  return scan;
}
\end{lstlisting}
\vspace{-10pt}
\end{minipage}
&
\begin{minipage}[t][][t]{0.47\textwidth}
\setlength{\fboxsep}{1pt}
\begin{lstlisting}[]
static const unsigned char *
GetMatch(const unsigned char *scan,const unsigned char *match,
 const unsigned char *end,const unsigned char *safe_end) {
 int scan_idx = 0, match_idx = 0;
  //... omitted code
 } else {
  while (&scan[scan_idx] < safe_end &&
  scan[scan_idx] == match[match_idx] &&
  scan[++scan_idx] == match[++match_idx] &&
  scan[++scan_idx] == match[++match_idx] &&
   /** omitted code **/) {
     scan_idx += 1; match_idx += 1;
  }
 }
 while (&scan[scan_idx] != end &&
 scan[scan_idx] == match[match_idx]) {
    scan_idx += 1; match_idx += 1;
 }
 scan = &scan[scan_idx];
 match = &match[match_idx];
 return scan;
\end{lstlisting}
\vspace{-10pt}
\end{minipage}
\\
\vspace{-4pt}\tiny (a) & \vspace{-4pt}\tiny (b)
\\
\begin{minipage}[t][][t]{0.47\textwidth}
\setlength{\fboxsep}{1pt}
\begin{lstlisting}[]
pub fn GetMatch<'a>(scan: &'a [u8],
match_: &'a [u8], end: &'a [u8], safe_end: &'a [u8])
-> Option<&'a [u8]> {
 //... omitted code
 } else {
  while (scan_ptr as usize) < (safe_end_ptr as usize) {
   (*@\colorbox{pink}{unsafe}@*) {
    (*@\colorbox{pink}{if *scan_ptr != *match_ptr ||}@*) 
    (*@\colorbox{pink}{*scan_ptr.add(1) != *match_ptr.add(1) ||}@*)
    (*@\colorbox{pink}{*scan_ptr.add(2) != *match_ptr.add(2) ||}@*)
     // ... omitted code {
      (*@\colorbox{pink}{break;}@*)
     }}}}
 while (scan_ptr as usize) < (end_ptr as usize) {
  (*@\colorbox{pink}{unsafe}@*) {
   (*@\colorbox{pink}{if *scan_ptr != *match_ptr \{ break; \} }@*)
   (*@\colorbox{pink}{scan_ptr = scan_ptr.add(1);}@*)
   (*@\colorbox{pink}{match_ptr = match_ptr.add(1);}@*)
\end{lstlisting}
\vspace{-7pt}
\end{minipage}
&
\begin{minipage}[t][][t]{0.47\textwidth}
\setlength{\fboxsep}{1pt}
\begin{lstlisting}[]
pub fn GetMatch<'a>(scan: &'a [u8], match_: &'a [u8],
end: &'a [u8], safe_end: &'a [u8]) -> Option<&'a [u8]> {
  let mut match_idx = 0;let mut scan_idx = 0;
  //... omitted code
  } else {
    while scan_idx < scan.len() && scan_idx < safe_end.len()
    && scan[scan_idx] == match_[match_idx]
    {
      (*@\colorbox{softgreen}{scan_idx += 1;match_idx += 1;}@*)
    }
  }
  while scan_idx < scan.len() && scan_idx < end.len()
   && scan[scan_idx] == match_[match_idx] {
     (*@\colorbox{softgreen}{scan_idx += 1; match_idx += 1;}@*)
  }
  Some(&scan[..scan_idx])
}
\end{lstlisting}
\vspace{-7pt}
\end{minipage}
\\
 \tiny (c) & \tiny (d)

\begin{comment}
\\
\begin{minipage}[t][][t]{0.3\textwidth}
\setlength{\fboxsep}{1pt}
\begin{lstlisting}[]
static unsigned long
CRC(const unsigned char* data,
    size_t size)
{
  unsigned long result = 0xFFFFFFFFUL;
  for (; size > 0; --size)
  {
    result = 
      crc32_table[(result ^ (*@\colorbox{pink}{*data++}@*))
        & 0xFF] ^ (result >> 8);
  }
  return result ^ 0xFFFFFFFFUL;
}
\end{lstlisting}
\vspace{-3mm}
\end{minipage}
&
\begin{minipage}[t][][t]{0.3\textwidth}
\setlength{\fboxsep}{1pt}
\begin{lstlisting}[]
static unsigned long
CRC(const unsigned char* data,
    size_t size) {
  unsigned long result = 0xFFFFFFFFUL;
  for (; size > 0; size -= 1) {
    result =
      crc32_table[(result ^ (*@\colorbox{softgreen}{*data}@*))
      & 0xFF] ^ (result >> 8);
      (*@\colorbox{yellow}{data += 1;}@*)
  }
  return result ^ 0xFFFFFFFFUL;
}
\end{lstlisting}
\vspace{-3mm}
\end{minipage}
&
\begin{minipage}[t][][t]{0.3\textwidth}
\setlength{\fboxsep}{1pt}
\begin{lstlisting}[]
static unsigned long
CRC(const unsigned char* data,
    size_t size){
  (*@\colorbox{softgreen}{int data_idx = 0}@*);
  unsigned long result = 0xFFFFFFFFUL;
  for (; size > 0; size -= 1) {
    result =
      crc32_table[
       (result ^ (*@\colorbox{softgreen}{data[data_idx]}@*))
       & 0xFF] ^ (result >> 8);
      (*@\colorbox{softgreen}{data_idx += 1;}@*)
  }
  return result ^ 0xFFFFFFFFUL;
}
\end{lstlisting}
\vspace{-3mm}
\end{minipage}
\\
\ &  \tiny(c) & 
\end{comment}
\\
\end{tabular}

%% file: Figures/refactoring.tex
\definecolor{softgreen}{RGB}{200,255,200} 
\vspace{-4pt}
\begin{tabular}{c c c}
% \toprule
% {\small Original C } & {\small unary-op refactor } & {\small pointer arithmetic refactor } \\
% \hline \\
 % \vspace{-1mm} &  \vspace{-1mm} A. genann project & \vspace{-1mm} \\
\begin{minipage}[t][][t]{0.3\textwidth}
\setlength{\fboxsep}{1pt}
\begin{lstlisting}[]
double const*
genann_run(genann* ann, double* inputs){
  double const* w = ann->weight;
  double* o = ann->output + ann->inputs;
  double const* i = ann->output;
  // ... omitted code
  for (j = 0; j < ann->outputs; (*@\colorbox{pink}{++j}@*)){
    double sum = (*@\colorbox{pink}{*w++}@*) * -1.0;    
    
    for (k = 0; k < ann->inputs; (*@\colorbox{pink}{++k}@*)
    {
      sum += (*@\colorbox{pink}{*w++}@*) * i[k];

    }
    (*@\colorbox{pink}{*o++}@*) = genann_act_output(ann, sum);
  }
\end{lstlisting}
\vspace{-8pt}
\end{minipage}
&
\begin{minipage}[t][][t]{0.3\textwidth}
\setlength{\fboxsep}{1pt}
\begin{lstlisting}[]
double const*
genann_run(genann* ann, double* inputs){
  double const* w = ann->weight;
  double* o = ann->output + ann->inputs;
  double const* i = ann->output;
  // ... omitted code
  for (j=0;j < ann->outputs;(*@\colorbox{softgreen}{j += 1}@*)){
    double sum = *w * -1.0;
    (*@\colorbox{yellow}{w += 1}@*);
    for (k = 0;k < ann->inputs;(*@\colorbox{softgreen}{k += 1}@*))
    {
      sum += *w * i[k];
      (*@\colorbox{yellow}{w += 1}@*);
    }
    *o = genann_act_output(ann, sum);
    (*@\colorbox{yellow}{o += 1}@*);
  }
\end{lstlisting}
\vspace{-8pt}
\end{minipage}
&
\begin{minipage}[t][][t]{0.3\textwidth}
\setlength{\fboxsep}{1pt}
\begin{lstlisting}[]
double const*
genann_run(genann* ann, double* inputs){
  (*@\colorbox{softgreen}{int w_idx = 0;}@*)
  (*@\colorbox{softgreen}{int o_idx = 0;}@*)
  double const* w = ann->weight;
  double* o = ann->output + ann->inputs;
  double const* i = ann->output;
  // ... omitted code
  for (j = 0; j < ann->outputs; j += 1){
    (*@\colorbox{softgreen}{double sum = w[w_idx] * -1.0;}@*)
    (*@\colorbox{softgreen}{w_idx += 1;}@*)
    for (k = 0; k < ann->inputs;k+=1){
      (*@\colorbox{softgreen}{sum += w[w_idx] * i[k];}@*)
      (*@\colorbox{softgreen}{w_idx += 1;}@*)
    }
    (*@\colorbox{softgreen}{o[o_idx]=genann_act_output(ann,sum);}@*)
    (*@\colorbox{softgreen}{o_idx += 1;}@*)
\end{lstlisting}
\vspace{-8pt}
\end{minipage}
\vspace{-4pt}
\\
\\
\end{tabular}

%% file: Algorithms/pointer.tex
\caption{\small{Pointer Operation Refactoring}}
\label{alg:pointer-refactor-detailed}

\KwInput{Function $F$}
\KwOutput{Modified $F$ with pointer arithmetic converted to array indexing}

%\tcp{Stage 1:}% Identify pointers with arithmetic operations}
$\mathit{all\_pointers} \leftarrow \textsc{LocatePtrs}(F)$\;
$\mathit{possible\_pointers} \leftarrow \emptyset$\;

\ForEach{pointer $p \in \mathit{all\_pointers}$}{
    \If{$\textsc{HasArithmCompOperations}(p)$}{
        $\mathit{possible\_pointers} \leftarrow \mathit{possible\_pointers} \cup \{p\}$\;
        $\mathit{name} \leftarrow p.\mathit{name} + \text{``\_idx''}$\;
        $\mathit{decl} \leftarrow \textsc{CreateDeclaration}(\text{``int''}, \mathit{name}, 0)$\;
        $\textsc{DeclareBeforScope}(F, p, \mathit{decl})$\;
        %$\textsc{ResetAfterPtrInitialization}(F, \mathit{decl})$\;
        \uIf{$\textsc{AccessContents}(p)$}{
            %$p \leftarrow \textsc{ReplaceWithIndex}(F, p)$\;
            $\textsc{ReplaceWithIndex}(F, p)$\;
            % \tcp{e.g., $p++$ becomes $p\_idx++$}
        }
        \ElseIf{$\textsc{PtrUsed}(p)$}{
            %$p \leftarrow \textsc{ReplaceWithPtrAccessWithIndex}(F, p)$\;
            $\textsc{ReplaceWithPtrAccessWithIndex}(F, p)$\;
            % \tcp{e.g., $*p$ becomes $p[p\_idx]$}
        }
        \ElseIf{$\textsc{PtrReInitialized}(p)$}{
            $\textsc{ResetIndex}(F, p)$\;
        }
    }
}

\begin{comment}
\BlankLine

%\tcp{Stage 2}%: Transform pointer operations to index operations}
\ForEach{$p \in \textsc{TraverseAST}(F)$}{
    \If{$n \in \mathit{possible\_pointers}$}{
        \uIf{$\textsc{IsPtrArithmetic}(n)$}{
            $n \leftarrow \textsc{ReplaceWithIndex}(n)$\;
            % \tcp{e.g., $p++$ becomes $p\_idx++$}
        }
        \ElseIf{$\textsc{IsPtrUsage}(n)$}{
            $n \leftarrow \textsc{ReplaceWithPtrAccessWithIndex}(n)$\;
            % \tcp{e.g., $*p$ becomes $p[p\_idx]$}
        }
    }
}
\end{comment}

% \tcp{Stage 3: update non-const input pointers before exit points}
\ForEach{$p \in \mathit{possible\_pointers}$}{
    \If{$\textsc{PtrTransformed}(p) \& \textsc{PtrIsArgument}(p)$}{
        $\textsc{UpdatePtrBeforeReturn}(F, p)$\;
    }
}

\Return{$F$}\;

%% file: Algorithms/breakdown.tex
\caption{\small{Large Function Breakdown}}
\label{alg:function-breakdown}

\KwInputs{Function $F$, Breakdown threshold $\theta_{f}$, Block threshold $\theta_{b}$}
%\KwData{Function size threshold $\theta_{f}$}
%\KwData{Block size threshold $\theta_{b}$}
\KwOutput{Updated code with breakdown}
\If{$|F| < \theta_{f}$}{
    \Return{$F$}
}
%\tcp{Stage 1: Candidate Identification}
$\mathit{candidates,helpers} \leftarrow \emptyset$\;
\ForEach{$B$ in $F$}{
    \If{$|B| > \theta_{b}$ \textbf{and} $\neg\textsc{HasEarlyExit}(B)$}{
        $\mathit{candidates} \leftarrow \mathit{candidates} \cup \{B\}$\;
    }
}
%\tcp{Stage 2: Dependency Analysis}
\ForEach{$B \in \mathit{candidates}$}{
    $V_{read},V_{write} \leftarrow \textsc{ExtractReadWriteVariables}(B)$\;
    %$V_{write} \leftarrow \textsc{ExtractWriteVariables}(B)$\;
    $B.\mathit{dependencies} \leftarrow V_{read} \cup V_{write}$\;
    %$B.\mathit{dependencies} \leftarrow V_{deps}$\;
     $\mathit{name} \leftarrow \text{``helper\_''} + F.\mathit{name} + \text{``\_''} + B.\mathit{identifier}$\;
     $\mathit{params} \leftarrow \emptyset$\;
     \ForEach{$v \in B.\mathit{dependencies}$}{
        \eIf{$v \in V_{write}$}{
            $\mathit{params} \leftarrow \mathit{params} \cup \{v^*\}$ \tcp{Pass by reference}
        }{
            $\mathit{params} \leftarrow \mathit{params} \cup \{v\}$ \tcp{Pass by value}
        }
    }
    
    $H \leftarrow \textsc{CreateFunction}(\mathit{name}, \mathit{params}, \text{void}, B.\mathit{body})$\;
    $\mathit{helpers} \leftarrow \mathit{helpers} \cup \{H\}$\;
}
$\textsc{UpdateCode}(F,\mathit{helpers})$\;
%\tcp{Stage 4: Call Site Replacement}
\begin{comment}
$F' \leftarrow F$ \tcp*{Create modified version of original function}
\ForEach{candidate block $B \in \mathit{candidates}$}{
    $H \leftarrow \textsc{GetCorrespondingHelper}(B, \mathit{helpers})$\;
    $\mathit{args} \leftarrow \textsc{PrepareArguments}(B.\mathit{dependencies})$\;
    $\mathit{call} \leftarrow \textsc{CreateFunctionCall}(H.\mathit{name}, \mathit{args})$\;
    $\textsc{Replace}(F', B, \mathit{call})$\;
}
\Return{$\{F'\} \cup \mathit{helpers}$}\;
\end{comment}

\begin{comment}
    %\tcp{Stage 3: Function Generation}
\ForEach{candidate block $B \in \mathit{candidates}$}{
    $\mathit{name} \leftarrow \text{``helper\_''} + F.\mathit{name} + \text{``\_''} + B.\mathit{identifier}$\;
    $\mathit{params} \leftarrow \emptyset$\;
    
    \ForEach{variable $v \in B.\mathit{dependencies}$}{
        \eIf{$v \in V_{write}$}{
            $\mathit{params} \leftarrow \mathit{params} \cup \{v^*\}$ \tcp*{Pass by reference}
        }{
            $\mathit{params} \leftarrow \mathit{params} \cup \{v\}$ \tcp*{Pass by value}
        }
    }
    
    $H \leftarrow \textsc{CreateFunction}(\mathit{name}, \mathit{params}, \text{void}, B.\mathit{body})$\;
    $\mathit{helpers} \leftarrow \mathit{helpers} \cup \{H\}$\;
}
\end{comment}

%% file: Algorithms/main-loop.tex
\caption{\small{\approach Translation Loop}}
    \label{alg:main-loop}
    \SetKwProg{Fn}{Function}{}{}
    \SetKwInOut{Input}{Inputs}
    \SetKwInOut{Output}{Output}
    \SetKwFunction{GenPrompt}{generatePrompt}
    \SetKwFunction{Translate}{translateFragment}
    \SetKwFunction{Compile}{compileCheck}
    \SetKwFunction{HasUnres}{hasUnresolvedImports}
    \SetKwFunction{FixImports}{fixImports}
    \SetKwFunction{RetrieveICL}{\textsc{retrieveICL}}
    %\SetKwFunction{Fix}{fix}
    \SetKwFunction{Fix}{Translate}
    \SetKw{Return}{return}
    
    \Input{Translation Unit $TU$, Base Model $M$, Reasoning Model $R$, Max Attempts $A_{max}$, Base Attempts $A_{base}$, Reasoning Attempts $A_{reason}$, ICL Pool $\mathcal{I}$}
    \Output{Rust Translation $T$}
    
    $T, compOut$ $\gets \emptyset$; 
    
    \For{$i \gets 1$ \KwTo $A_{max}$}{%
      \For{$k \gets 1$ \KwTo $A_{base}$}{%
        $T, compOut \gets$ \textsc{Translate($M, TU, compOut, ICL_f$)}\;
        \If{\textsc{IsCompilable($compOut$)}}
        {
            \Return $T$;
        }
      }
      \For{$k \gets 1$ \KwTo $A_{reason}$}{%
        $ICL_{TU}$ $\gets$ \textsc{RetrieveGenerateICL($\mathcal{I}, TU, compOut$)}\;
        $T, compOut \gets$ \textsc{Translate($R, TU, compOut, ICL_{TU}$)}\;
        \If{\textsc{IsCompilable($compOut$)}}
        {
            \Return $T$;
        }
      }
    }
    
    \Return $T$;
    
    \BlankLine

    \Fn{\textsc{Translate($LLM, TU, compOut, ICL$)}}{
        $prompt \gets$ \textsc{GeneratePrompt($TU, compOut, ICL$)}\;
        $T \gets$ \textsc{TranslateUnit($LLM, prompt$)}\;
        $compOut \gets$ \textsc{CompilationCheck($T$)}\;
        \Return $T, compOut$\;
    }

    \begin{comment}
    \BlankLine
    \textbf{Auxiliary:} \textsc{RetrieveGenerateICL($\mathcal{I}, TU, compOut$)}
    \Begin{
      $E \gets \text{ErrorCodes}(compOut)$\;
      $\mathcal{C} \gets \{x \in \mathcal{I} \mid \text{ErrorCodes}(x)\cap E \neq \varnothing\}$\;
      \For{$x \in \mathcal{C}$}{
        $\mathrm{score}(x) \gets \mathrm{Sim}(\mathrm{Emb}(TU), \mathrm{Emb}(x))$\;
      }
      \Return $\arg\max_{x \in \mathcal{C}} \mathrm{score}(x)$\;
    }
    \end{comment}

%% file: Figures/icl.tex
\lstdefinelanguage{Rust}{
    keywords={fn,let,mut,pub,match,Some,Option,Result,ref,return,impl,for,in,while,loop,if,else},
    keywordstyle=\color{blue}\bfseries,
    sensitive=true,
    comment=[l]{//},
    morecomment=[s]{/*}{*/},
    commentstyle=\color{green!40!black},
    stringstyle=\color{red},
    morestring=[b]{"},
}
\definecolor{iclgreen}{RGB}{227,246,230}
\definecolor{iclblue}{RGB}{219,230,253}
\definecolor{iclred}{RGB}{255,220,220}
\vspace{-8pt}
\setlength{\fboxsep}{1pt}
\begin{center}
\begin{tabular}[t]{@{}p{0.46\textwidth}@{\hspace{0.04\textwidth}}p{0.5\textwidth}@{}}
\begin{minipage}[t]{\linewidth}
\begin{lstlisting}[language=Rust]
pub fn run_ti_ref(info: &TiIndicatorInfo,
                  options: &[f64], goal: i32) -> i32 {
  // ...
  let mut outref = OUTREF.lock().unwrap();
  let row0 = (*@\colorbox{iclred}{\&mut outref[0][..]}@*);
  let row1 = (*@\colorbox{iclred}{\&mut outref[1][..]}@*);
  let row2 = (*@\colorbox{iclred}{\&mut outref[2][..]}@*);

  let outputs_ref: &mut [&mut [f64]]
      = (*@\colorbox{iclred}{\&mut [row0, row1, row2]}@*);
  let ret = (info.indicator_ref)(
    4000, Some(inputs_ref), Some(options), Some(outputs_ref)
  );
}
\end{lstlisting}
\vspace{-18pt}\centering{\scriptsize{(a)}}
\vspace{-9pt}
\end{minipage}
&
\begin{minipage}[t]{\linewidth}
\begin{lstlisting}[language=Rust]
error[E0499]: cannot borrow `outref` as (*@\colorbox{iclblue}{mutable more than once}@*)
(*@\colorbox{iclblue}{at a time}@*)
   --> src/benchmark.rs:6:19
   |
5  | let row0 = (*@\colorbox{iclblue}{\&mut outref[0][..]}@*);
   |                 ------ first mutable borrow occurs here
6  | let row1 = (*@\colorbox{iclblue}{\&mut outref[1][..]}@*);
   |                 ^^^^^^ second mutable borrow occurs here
8  | let outputs_ref: &mut [&mut [f64]] = &mut [row0, row1, row2];
   |                                            ---- first borrow
   | later used here
\end{lstlisting}
\vspace{-13pt}\centering{\scriptsize{(b)}}
\vspace{-15pt}
\end{minipage}
\end{tabular}

\begin{tabular}[t]{@{}p{0.63\textwidth}@{\hspace{0.04\textwidth}}p{0.33\textwidth}@{}}
\begin{minipage}[t]{\linewidth}
\begin{tabular}[t]{@{}p{0.34\linewidth}@{\hspace{0.02\linewidth}}p{0.64\linewidth}@{}}
\begin{minipage}[t]{\linewidth}
\begin{lstlisting}[language=Rust]
fn process_data(
  mut data: [[f64; 100]; 3]
) {
  let outputs = &mut [
    &mut data[0][..],
    &mut data[1][..],
    &mut data[2][..],
  ];
  some_function(outputs);
}
\end{lstlisting}
\vspace{-10pt}\centering{\scriptsize{(c)}}
\end{minipage}
&
\begin{minipage}[t]{\linewidth}
\begin{lstlisting}[language=Rust]
error[E0499]: cannot borrow `data` as (*@\colorbox{iclblue}{mutable more}@*)
(*@\colorbox{iclblue}{than once at a time}@*)
   --> src/main.rs:5:14
 4 |  let outputs = &mut [
   |  _______________-
 5 | |  (*@\colorbox{iclblue}{\&mut data[0][..]}@*),
   | |       ---- first mutable borrow occurs here
 6 | |  (*@\colorbox{iclblue}{\&mut data[1][..]}@*),
   | |       ^^^^ second mutable borrow occurs here
 7 | |  &mut data[2][..],
 8 | |     ];
   | |_____- first borrow later used here
\end{lstlisting}
\vspace{-9pt}\centering{\scriptsize{(d)}}
\end{minipage}
\end{tabular}
\vspace{6pt}
\begin{minipage}[t]{\linewidth}
\begin{lstlisting}[language=Rust]
fn process_data(mut data: [[f64; 100]; 3]) {
  (*@\colorbox{iclgreen}{let (first, rest) = data.split_at_mut(1);}@*)
  (*@\colorbox{iclgreen}{let (second, third) = rest.split_at_mut(1);}@*)
  let outputs = &mut [&mut first[0][..], &mut second[0][..], &mut third[0][..]];
  some_function(outputs);
}
\end{lstlisting}
\vspace{-19pt}\centering{\scriptsize{(e)}}
\end{minipage}
\end{minipage}
&
\begin{minipage}[t]{\linewidth}
\begin{lstlisting}[language=Rust]
pub fn run_ti_ref(info: &TiIndicatorInfo,
                  options: &[f64],
                  goal: i32
) -> i32 {
  // ...
  let mut outref = OUTREF.lock().unwrap();
  let rows = &mut *outref;
  (*@\colorbox{iclgreen}{let (first, rest) = rows.split_at_mut(1);}@*)
  (*@\colorbox{iclgreen}{let (second, third) = rest.split_at_mut(1);}@*)
  let row0 = (*@\colorbox{iclgreen}{\&mut first[0][..]}@*);
  let row1 = (*@\colorbox{iclgreen}{\&mut second[0][..]}@*);
  let row2 = (*@\colorbox{iclgreen}{\&mut third[0][..]}@*);
  let outputs_ref: &mut [&mut [f64]]
    = &mut [row0, row1, row2];
  let ret = (info.indicator_ref)(
    4000, Some(inputs_ref),
    Some(options), Some(outputs_ref)
  );
}
\end{lstlisting}
\vspace{-13pt}\centering{\scriptsize{(f)}}
\end{minipage}
\end{tabular}
\end{center}

%% file: Sections/Evaluation.tex
\section{Empirical Evaluation}
\label{sec:evaluation}

We evaluate contributions of \approach concerning the following research questions: 

\begin{enumerate}[label=\bfseries RQ\arabic*.]
    
    \item \textit{Effectiveness of \approach and Comparison with Alternative Approaches}. To what extent can \approach facilitate translation of the real-world C projects to idiomatic Rust? How much effort is required by a human developer to fix semantic mismatches in the translations? How does the performance of \approach in code translation compare with other techniques?

    \item \textit{Effectiveness of the Refactoring.} To what extent refactoring is successful in resolving pointer arithmetic in \approach translations? 

    \item \textit{Translation Bugs}. What are the most common translation issues produced by \approach?

    \item \textit{Performance and Cost}. How long does it take for \approach to translate subject projects from C to Rust? How much does it cost to use \approach for translation?

\end{enumerate}

\input{Sections/Evaluation-Setup}
\input{Sections/Evaluation-RQ1}

\input{Sections/Evaluation-RQ2}

\input{Sections/Evaluation-RQ3}
\input{Sections/Evaluation-RQ4}

%% file: Sections/Evaluation-Setup.tex
\subsection{Experimental Setup}
\label{subsec:eval-setup}

%\reyhan{add a paragraph here to explain which specific numbers are used for the parameters in algorithms.}
%After every eight runs, I switch to using r1. Actually, I implemented a loop pattern: running 8 times with the regular deepseek-v3, then 2 times with deepseek-r1, and repeating this cycle continuously. total of 30 reprompting

\subsubsection{Alternative Approaches and Subject Programs}
\label{subsubsec:eval-setup-subjects}

We compare \approach with leading prior work that attempted repository-level translation of C projects to Rust. Given new evaluation metrics that were not considered by prior work, but are necessary for proper evaluation of translation quality (\S \ref{subsubsec:eval-setup-metrics}), we focused on the tools that either made their translations publicly available or we were able to run them with no runtime exception to generate translations\footnote{We did not include techniques such as Fluorine~\cite{eniser2024towards} or VERT~\cite{yang2024vert}, since they only translate a few program slices of real-world projects.}. Considering these criteria, we compare \approach with the following techniques (alphabetically ordered):

\textbf{\ctorust}~\cite{ctorust} is a rule-based transpiler that converts C code into Rust while preserving semantics. 
\textbf{\saferrust}~\cite{nitin2025c2saferrust} improves the \emph{idiomaticity} of \ctorust translations by breaking down the translations by \ctorust into smaller translation units, and prompting GPT-4o mini with static analysis-provided contexts to improve the translation. \textbf{\crown}~\cite{zhang2023ownership} builds on top of \ctorust and leverages ownership analysis to resolve unsafe usages in the translation. 
\textbf{\laertes}~\cite{emre2021translating} also relies on \ctorust for initial Rust translation and lifts raw pointers into safe Rust references through an iterative, compiler-in-the-loop approach. 

\textbf{\rustmap}~\cite{cai2025rustmap} and \textbf{\syzygy}~\cite{shetty2024syzygy} perform a neurosymbolic compositional approach~\cite{ibrahimzada2025alphatrans} to translate C projects into idiomatic Rust \emph{from scratch}. Both techniques leverage GPT-4o for translation and, due to the high cost, only translate one (\rustmap) and two (\syzygy) repositories. 

\begin{table*}[t]
    \scriptsize
    \centering
    \caption{Subject programs and corresponding properties. Subjects are sorted by size (LoC).}
    \vspace{-10pt}
    \input{Tables/Subjects}
    %\vspace{-10pt}
    \label{table:subjects}
\end{table*}

Table~\ref{table:subjects} shows the list of $23$ programs from the evaluation of the prior technique that met our subject selection criteria. The programs range from $27$ to \num{13200} lines of code, and are diverse in terms of using structs, unions, macros, and raw pointer declarations/dereferences, with $17$ and $22$ of them containing $452$ and \num{2636} instances of pointer operations and macros, respectively\footnote{%We were able to access the translations in 
We did not compare with CRUST-Bench~\cite{khatry2025crust} as it evaluates the power of specific LLMs rather than the pipeline.}. Another 

\vspace{-3pt}
\subsubsection{Choice of LLM}
\label{subsubsec:eval-setup-LLM}

To balance cost and effectiveness, we use two LLMs in \approach pipeline: \llmv for most of the translation.
%, as it offers comparable frontier-level performance with considerably lower cost compared to OpenAI, Claude, and Gemini models. 
For cases that \llmv fails in translation due to compilation errors after \emph{eight} re-prompting, \approach uses \llmr, a reasoning model, to better analyze the failure reasons with a re-prompting budget of \emph{two}. 
%Due to lower number of re-prompting efforts, the higher cost of \llmr is overall negligible (more details in \S \ref{sec:translaterepair}). 
%\approach uses \llmo for generating in-context examples due to its overall superiority, including more comprehensive knowledge about Rust features. 
%\llmo is more expensive than others, but its usage by \approach is negligible, resulting in minimal cost. 
%Prior research has shown that using stronger models improves the translation performance \cite{ibrahimzada2025alphatrans}, making the results in this paper a lower bound of \approach capabilities. 
Users can configure \approach to use any LLM depending on their budget by replacing the API key\footnote{API- or open-access models hosted using \ollama~\cite{ollama}/\vllm~\cite{vllm} and exposed via an API key-protected endpoint.}. 

\vspace{-3pt}
\subsubsection{Metrics}
\label{subsubsec:eval-setup-metrics}

We measure the following metrics for both C and Rust code. 
%The metrics marked with $^*$ are new, i.e., none of alternative approaches measured or reported numbers corresponding to these metrics. 
We identified several bugs or incorrect assumptions in calculation of some metrics by existing techniques (marked with $^\dagger$) and report recalculated numbers using our pipeline.
%, which leverages reliable tools to measure those metrics. 

\begin{itemize}[leftmargin=*]
    \item \textbf{\# Raw Pointer Declarations/Dereferences$^\dagger$\footnote{Some pipelines used a regex-based approach to calculate these metrics, making them unreliable and error-prone.}.} 
    %The number of raw pointer declarations and dereferences directly reflects the extent of unsafe memory manipulation in translated Rust code. 
    For raw pointer declaration in C, \approach uses \texttt{\small{pycparser}}~\cite{pyparser} to find \texttt{\small{Decl}} nodes in AST with \texttt{\small{PtrDecl}} type. It also looks for \texttt{\small{UnaryOp}} nodes with the \texttt{\small{*}} operator for determining pointer dereferences.    
    For Rust, \approach uses (\texttt{\small{rustc_hir}} crate~\cite{hir}) to identify nodes corresponding to raw pointer types, e.g., \texttt{\small{*const T}} and \texttt{\small{*mut T}}, declared within variable bindings, function signatures, and type annotations. For dereference, \approach locates \texttt{\small{ExprKind::Unary}} expressions with \texttt{\small{Deref}} as operator and raw pointer operand. 

    \item \textbf{\# Pointer Operations.} 
    %Use of pointer arithmetic implies a low-level memory manipulation that is discouraged in Rust. 
    \sloppy None of the prior work addresses or measures pointer arithmetic/comparisons in translations. \approach uses \texttt{\small{rustc_hir}} crate to locates function calls on pointer types (\texttt{\small{ExprKind::MethodCall}}) that correspond to arithmetic operations, e.g., \texttt{\small{add}}, and \texttt{\small{byte_add}}~\cite{rustpointer}. 
    %, and \texttt{\small{wrapping_add}}~\cite{rustpointer}. 
    
    \item \textbf{\# Unsafe Constructs$^\dagger$\footnote{Similar to previous metrics, some techniques followed an incomplete regex-based approach for counting unsafe constructs. While not using regex, \saferrust does not support newer versions of Rust (1.81 and above) and cannot calculate these metrics inside struct methods or within complex statements such as \texttt{\footnotesize{sum+=unsafe \{*w\} * unsafe \{*i.add(k as usize)\}}}. It marks the entire functions as unsafe rather than using unsafe blocks for specific statements, reporting \emph{fewer} unsafe issues.}.} 
    %Counting the number of unsafe lines, type casts, and function calls provides insight into how the translation aligns with Rust's safety guarantees. 
    \approach uses Rust compiler (\texttt{\footnotesize{rustc_hir}} crate~\cite{hir}) to calculate the number of unsafe \emph{lines}, \emph{type casts}, and \emph{calls}.
    
    \item \sloppy \textbf{Idiomaticity\footnote{\ctorust default pipeline suppresses \texttt{\footnotesize{Clippy}} warnings. We updated the config of transpiler-based techniques to ensure proper calculation of warnings.}.} 
    %This metric reflects how well the output adheres to Rust’s best practices, abstractions, and language conventions. 
    \approach uses \emph{Clippy}~\cite{clippy} for measuring idimaticity.
    %, a widely used Rust linter that checks $750$ lints categorized into nine groups: \texttt{\small{cargo}}, \texttt{\small{complexity}}, \texttt{\small{correctness}}, \texttt{\small{nursery}}, \texttt{\small{pedantic}}, \texttt{\small{perf}}, \texttt{\small{restriction}}, \texttt{\small{style}}, \texttt{\small{suspicious}}. 
    \approach excludes \texttt{\small{nursery}} and \texttt{\small{restriction}} as they either concern new lints that are still under development or may have false positives. \texttt{\small{Clippy}} document also suggests against using \texttt{\small{restriction}} by default due to the subjectiveness of the lints. 
    Except for \texttt{\small{clippy::style}} that will be used to measure readability, \approach computes the total number of raised flags as $Clippy_{flags}=\sum \texttt{\small{clippy::flag}}_i$, where $i \in \{\texttt{\small{cargo}}, \texttt{\small{complexity}}, \texttt{\small{correctness}}, \texttt{\small{perf}}, \texttt{\small{suspicious}}\}$. 
    %Prior research has shown that idiomatic Rust projects have an average warning density of $21$ Clippy warnings per KLoC~\cite{li2024unleashing}. As a result, a translation with $Clippy_{flags} \le 21$ can be considered idiomatic. At the same time, 
    Transpiler-based techniques can increase code size $1.8\times$ when converting C to Rust~\cite{emre2021translating}. Without normalizing the number of Clippy warnings concerning inflation size, aggregating the number of raised flags favors
    %will be lower for those technique, favoring 
    unnecessary verbosity, redundancy, or boilerplate~\cite{pan2024lost}. \approach measures Translation Size Inflation\footnote{\approach relies on Tokei~\cite{tokei} to measure LoC, which does not include blank lines and comments when measuring LoC.} as  $TSI=\sfrac{LoC_{Rust}}{LoC_{C}}$ and reports idiomaticity as $TSI \times Clippy_{flags}$ per KLoC. 
    
    \item \textbf{Readability.} 
    %Except for \rustmap that only measures \emph{cognitive complexity}~\cite{campbell2018cognitive}, none of the prior techniques concern readability. 
    \approach uses rust-code-analysis~\cite{ardito2020rust,rustcodeanalysis} 
    %to compute code quality metrics for the Rust translations. It 
    and reports \emph{Cognitive Complexity}, \emph{Halstead}~\cite{curtis1979measuring}, and \emph{SEI Maintainability Index}~\cite{oman1992metrics}, as well as the number of \texttt{\small{clippy::style}} violations. These metrics highlight different aspects of readability, making the evaluation of translation more transparent. Except for the \emph{SEI}, the lower number for each metric indicates more readability. 
    
    \item \textbf{Assertion Pass Rate.} 
    %None of the prior work provides clear statistics about the number of tests per project and corresponding coverage. At best, they mention that reported results are for the translations that pass the translated tests. As we will discuss (\S \ref{subsec:rq-effectiveness}), several projects had no tests, and some tests were only for the runtime execution of code without any assertions. Even when tests exist, artifacts of prior techniques did not contain test translations, or test translation was unsuccessful with compilation errors. 
    %For a proper evaluation, w
    We augmented test suites of subjects with more tests and assertions, resulting in average function and line coverage of $74.7\%$ and $72.2\%$. %The test suites of C programs 
    They include \num{1221192} assertions and cover $93.3\%$ of the refactored code, on average. 
    %To assess the success of preserving semantics through translations, 
    \approach reports the number of executed, passed, and failed assertions in the Rust translation.
    %, i.e., executing translated tests on the translated code. %1,221,192
    
    \item \textbf{Execution Time and Cost.} \approach reports the time and cost of
    %measures the execution time of translation and computes the cost of 
    translating projects. It also presents the number of tokens involved in translating each project\footnote{The numbers reported for \texttt{\footnotesize{usage.prompt\_tokens}} and \texttt{\footnotesize{usage.completion\_tokens}} properties by the LLM API.}, which practitioners can use to approximate the cost of using commercial models with \approach (\S \ref{subsec:rq-performance}). 
\end{itemize}

%As we discuss, superiority in one metric does not imply overall superiority. For example, while \texttt{\small{Clippy}} warnings are useful to highlight some aspects of \emph{idiomaticity}, they are not exhaustive: \texttt{\small{Clippy}} does not flag many \emph{C-shaped} patterns that \ctorust tends to emit, such as heavy use of \texttt{\small{libc::c_int}}, raw pointers, manual index loops, and unsafe blocks, even though the \ctorust authors themselves describe the output as non-idiomatic and call for further cleanup tools. As a result, the ultimate quality should be determined based on considering performance in all dimensions. 

%% file: Tables/Subjects.tex
\centering
\setlength{\tabcolsep}{1pt}
\footnotesize
\resizebox{\textwidth}{!}{
\begin{tabular}{|c|c|c|c|c|c|c|c|cc|cc|cc|ccc|}
\hline

\multirow{3}{*}{\textbf{ID}} & 
\multirow{3}{*}{\textbf{Subject}} & 
\multirow{3}{*}{\textbf{LoC}} & 
\multirow{3}{*}{\textbf{\# Files}} & 
\multirow{3}{*}{\textbf{\# Functions}} &
\multirow{3}{*}{\textbf{\# Structs}} &
\multirow{3}{*}{\textbf{\# Unions}} &
\multirow{3}{*}{\begin{tabular}[c]{@{}c@{}}\textbf{\# Global}\\ \textbf{Variables}\end{tabular}} &
\multicolumn{2}{c|}{\multirow{2}{*}{\textbf{\# Macros}}} & 
\multicolumn{2}{c|}{\multirow{2}{*}{\textbf{\# Pointer Arithmetic}}} &
\multicolumn{2}{c|}{\multirow{2}{*}{\textbf{\# Raw Pointers}}} & 

\multicolumn{3}{c|}{\textbf{Tests}} 
\\ \cline{15-17} 

& 
& 
& 
& 
\multicolumn{1}{c|}{} &
& 
& 
& 
\multicolumn{2}{c|}{} & \multicolumn{2}{c|}{} &
\multicolumn{2}{c|}{} &

\multicolumn{2}{c|}{\textbf{\% Coverage}} &
\multirow{2}{*}{\textbf{\# Tests (Assertions)}} 
\\ \cline{9-14} \cline{15-16}

& 
& 
& 
& 
&
& 
& 
& 
\multicolumn{1}{c|}{\textbf{Config}} &
\textbf{Function} &
\multicolumn{1}{c|}{\textbf{Before}} &
\textbf{After} &
\multicolumn{1}{c|}{\textbf{Declarations}} &
\textbf{Dereferences} &

\multicolumn{1}{c|}{\textbf{Func}} &
\multicolumn{1}{c|}{\textbf{Line}} & 
\\ \hline

% needs to be repeated per program
$1$&
\multicolumn{1}{l|}{qsort}&
$27$&                           
$1$&                                                  
$3$&                             
$0$&                            
$0$&          
$0$&     
$0$& 
$0$&            
$0$&                                         
$0$&                                      
$2$&                
$4$&                               
                       
$100$& 
\multicolumn{1}{c|}{$100$}&       
$5(21)$
\\ 

$2$&
\multicolumn{1}{l|}{bst}&
$65$&                           
$1$&                                     
$5$&              
$3$&                             
$0$&                                
$0$&                                    
$2$&            
$0$&                                         
$0$&                   
$0$&                                      
$9$&                
$0$&                               
                          
$100$& 
\multicolumn{1}{c|}{$100$}&       
$6(6)$
\\ 

$3$&
\multicolumn{1}{l|}{rgba}&
$411$&                           
$3$&                                     
$19$&              
$1$&                             
$0$&                            
$1$&                                                        
$12$&            
$1$&                                         
$20$&                   
$0$&                                      
$14$&                
$61$&                               
                         
$92$& 
\multicolumn{1}{c|}{$85$}&           
$5(20)$
\\ 

$4$&
\multicolumn{1}{l|}{quadtree}&
$437$&                           
$7$&                                     
$36$&              
$8$&                             
$0$&                            
$2$&                                    
$17$&            
$1$&                                         
$0$&                   
$0$&                                      
$55$&                
$7$&                               
                           
$100$& 
\multicolumn{1}{c|}{$95$}&      
$4(34)$
\\ 

$5$&
\multicolumn{1}{l|}{buffer}&
$452$&                           
$3$&                                     
$42$&              
$1$&                             
$0$&                            
$0$&                                                                
$17$&            
$2$&                                         
$0$&                   
$0$&                                      
$61$&                
$1$&                               
                         
$86$& 
\multicolumn{1}{c|}{$83$}&   
$17(54)$
\\ 

$6$&
\multicolumn{1}{l|}{grabc}&
$490$&                           
$1$&                                     
$10$&              
$0$&                             
$0$&                            
$11$&                                    
$24$&            
$0$&                                         
$1$&                   
$0$&                                      
$19$&                
$15$&                               
                    
$33$& 
\multicolumn{1}{c|}{$18$}&     
$3(2)$
\\ 

$7$&
\multicolumn{1}{l|}{urlparser}&
$563$&                           
$5$&                                     
$27$&              
$3$&                             
$0$&                            
$5$&                                    
$23$&            
$3$&                                         
$17$&                   
$0$&                                      
$82$&                
$49$&                               
                          
$83$& 
\multicolumn{1}{c|}{$80$}&   
$1(46)$
\\ 

$8$&
\multicolumn{1}{l|}{xzoom}&
$659$&                           
$1$&                                     
$8$&              
$0$&                             
$0$&                            
$32$&                                    
$48$&            
$2$&                                         
$40$&                   
$0$&                                      
$13$&                
$24$&                               
                        
$-$& 
\multicolumn{1}{c|}{$-$}&          
$-$
\\ 

$9$&
\multicolumn{1}{l|}{genann}&
$690$&                           
$8$&                                     
$30$&              
$2$&                             
$0$&                            
$12$&                                    
$51$&            
$10$&                                         
$28$&                   
$4$&                                      
$62$&                
$93$&                               
                          
$96$& 
\multicolumn{1}{c|}{$94$}&  
$9(521556)$
\\ 

$10$&
\multicolumn{1}{l|}{ht}&
$699$&                           
$10$&                                     
$23$&              
$4$&                             
$0$&                            
$3$&                                    
$50$&            
$0$&                                         
$7$&                   
$0$&                                      
$68$&                
$28$&                               
% $30$&                           
$60$& 
\multicolumn{1}{c|}{$56$}&    
$8(1)$
\\ 

$11$&
\multicolumn{1}{l|}{robotfindskitten}&
$838$&                           
$3$&                                     
$17$&              
$1$&                             
$0$&                            
$8$&                                    
$49$&            
$5$&                                         
$0$&                   
$0$&                                      
$1$&                
$0$&                               
% $26$&                           
$53$& 
\multicolumn{1}{c|}{$63$}&    
$7(47)$
\\ 

$12$&
\multicolumn{1}{l|}{libcsv}&
$1035$&                           
$7$&                                     
$31$&              
$2$&                             
$0$&                            
$5$&                                    
$64$&            
$5$&                                         
$5$&                   
$0$&                                      
$65$&                
$8$&                               
% $48$&                           
$75$& 
\multicolumn{1}{c|}{$92$}&    
$45(7406)$
\\ 

$13$&
\multicolumn{1}{l|}{avl-tree}&
$1170$&                           
$7$&                                     
$50$&              
$6$&                             
$0$&                            
$3$&                                    
$41$&            
$7$&                                         
$0$&                   
$0$&                                      
$122$&                
$4$&                               
% $61$&                           
$40$& 
\multicolumn{1}{c|}{$45$}&   
$11(12)$
\\ 

$14$&
\multicolumn{1}{l|}{libopenaptx}&
$1333$&                           
$4$&                                     
$44$&              
$15$&                             
$0$&                            
$41$&                                    
$29$&            
$3$&                                         
$2$&                   
$0$&                                      
$65$&                
$20$&                               
% $107$&                           
$95$& 
\multicolumn{1}{c|}{$84$}&         
$4(9)$
\\ 

$15$&
\multicolumn{1}{l|}{libtree}&
$1412$&                           
$4$&                                     
$30$&              
$17$&                             
$0$&                            
$1$&                                    
$69$&            
$0$&                                         
$17$&                   
$3$&                                      
$91$&                
$31$&                               
% $63$&                           
$100$& 
\multicolumn{1}{c|}{$64$}&    
$57(121)$
\\ 

$16$&
\multicolumn{1}{l|}{opl}&
$1642$&                           
$2$&                                     
$117$&              
$20$&                             
$0$&                            
$7$&                                    
$23$&            
$4$&                                         
$7$&                   
$0$&                                      
$148$&                
$28$&                               
% $146$&                           
$63$& 
\multicolumn{1}{c|}{$56$}&        
$10(14)$
\\ 

$17$&
\multicolumn{1}{l|}{libzahl}&
$2575$&                           
$52$&                                     
$59$&              
$0$&                             
$0$&                            
$35$&                                    
$98$&            
$28$&                                         
$26$&                   
$0$&                                      
$15$&                
$43$&                               
% $99$&                           
$98$& 
\multicolumn{1}{c|}{$93$}&    
$13(1570)$
\\ 

$18$&
\multicolumn{1}{l|}{zopfli}&
$2937$&                           
$26$&                                     
$105$&              
$10$&                             
$0$&                               
$1$&                                    
$254$&            
$1$&                                         
$26$&                   
$0$&                                      
$289$&                
$859$&                               
% $158$&                           
$95$& 
\multicolumn{1}{c|}{$88$}&   
$11(40)$
\\ 

$19$&
\multicolumn{1}{l|}{snudown}&
$5271$&                           
$19$&                                     
$153$&              
$89$&                             
$0$&                            
$23$&                                    
$138$&            
$18$&                                         
$5$&                   
$0$&                                      
$441$&                
$44$&                               
% $302$&                           
$35$& 
\multicolumn{1}{c|}{$30$}&    
$-$
\\ 

$20$&
\multicolumn{1}{l|}{lodepng}&
$5606$&                           
$6$&                                     
$235$&              
$19$&                             
$0$&                            
$16$&                                    
$168$&            
$9$&                                         
$31$&                   
$0$&                                      
$543$&                
$222$&                               
% $341$&                           
$78$& 
\multicolumn{1}{c|}{$80$}&       
$28(683994)$
\\ 

$21$&
\multicolumn{1}{l|}{bzip2}&
$5861$&                           
$15$&                                     
$118$&              
$8$&                             
$0$&                            
$36$&                                    
$251$&            
$83$&                                         
$10$&                   
$0$&                                      
$222$&                
$129$&                               
% $426$&                           
$10$& 
\multicolumn{1}{c|}{$6$}&  
$1(36)$
\\ 

$22$&
\multicolumn{1}{l|}{binn}&
$6361$&                           
$10$&                                     
$315$&              
$3$&                             
$1$&                            
$18$&                                    
$188$&            
$16$&                                         
$74$&                   
$0$&                                      
$625$&                
$220$&                               
% $341$&                           
$65$& 
\multicolumn{1}{c|}{$87$}&    
$14(1949)$
\\ 

$23$&
\multicolumn{1}{l|}{tulpindicator}&
$13200$&                           
$152$&                                     
$326$&              
$9$&                             
$0$&                            
$32$&                                    
$673$&            
$149$&                                         
$136$&                   
$7$&                                      
$1073$&                
$155$&                               
% $534$&                           
$85$& 
\multicolumn{1}{c|}{$90$}&    
$9(4252)$
\\ 

% $24$&
% \multicolumn{1}{l|}{heman}&
% $...$&                           
% $...$&                                     
% $...$&              
% $...$&                             
% $...$&                            
% $...$&                                    
% $...$&            
% $...$&                                         
% $...$&                   
% $...$&                                      
% $...$&                
% $...$&                               
% $...$&                           
% $...$& 
% \multicolumn{1}{c|}{$...$}&    
% $...$
% \\ 

\hline
\end{tabular}
}

%% file: Sections/Evaluation-RQ1.tex
\subsection{RQ1. Effectiveness of \approach and Comparison with Alternative Approaches}
\label{subsec:rq-effectiveness}

\begin{table*}[t]
    \scriptsize
    \centering
    \caption{Validation of translations by \approach and alternative approaches. Abbreviations: \ctorust (C2R), \saferrust (C2SR), \crown (C), \laertes (L), \rustmap (R), and \syzygy (S).}
    \vspace{-10pt}
    \input{Tables/Effectiveness-Validation}
    %\vspace{-10pt}
    \label{table:effectiveness-validation}
\end{table*}

\begin{table*}[t]
    \scriptsize
    \centering
    \caption{Safety characteristics of translations by \approach and alternative approaches. Abbreviations: \ctorust (C2R), \saferrust (C2SR), \crown (C), \laertes (L), \rustmap (R), and \syzygy (S).}
    \vspace{-10pt}
    \input{Tables/Effectiveness-Safety}
    %\vspace{-10pt}
    \label{table:effectiveness-safety}
\end{table*}

\textbf{\approach successfully translates all the programs and tests, achieving $100\%$ application and test code compilation success} (gray rows under column \emph{Compilation Success} of Table~\ref{table:effectiveness-validation}). \ctorust, \crown, and \laertes generated a non-compilable application code for \emph{libzahl} (ID=$17$). \crown also resulted in a non-compilable application and test code in Rust for \emph{zopfli} (ID=$18$). \saferrust test code translation for \emph{urlparser} (ID=$7$) and \emph{gennan} (ID=$9$) was non-compilable. \ctorust, \saferrust, \crown, and \laertes all generated non-compilable test code translation for \emph{tulipindicator} (ID=$23$). We found no test translation by \syzygy for \emph{zopfli} (ID=$18$). The execution of translated tests by \rustmap also stuck in a loop and never terminates for \emph{bzip2} (ID=$21$). These cases are marked with "-" in Table~\ref{table:effectiveness-validation}. The $100\%$ compilation success of \approach is achieved through \textbf{full automation, without any human intervention} as in \syzygy and \rustmap.

Table~\ref{table:effectiveness-validation} shows the validation results of \approach, compared to alternative techniques. Overall, \approach resulted in $100\%$ functional equivalence in translation of $10$ projects \textbf{without any human intervention}. That is, executing the translated tests passed all the assertions. We mark the ID of those projects with $*$ in Table~\ref{table:effectiveness-validation}. For the remaining subjects, with at least one assertion failure, we asked a developer with more than ten years of relevant industry experience---proficiency in C and Rust---and no familiarity\footnote{No familiarity with projects determines an upper bound for the amount of time required to fix semantic mismatches in \approach translations, compared to the developers of these projects.} with the subjects to repair the translations within a budget of $10$ hours per project. The developer used \approach debugging 
%\approach provided the developer with dynamic debugging information 
to facilitate the process. 

\begin{wrapfigure}{r}{0.46\columnwidth}
    \vspace{-5pt}
    \includegraphics[width=0.45\columnwidth]{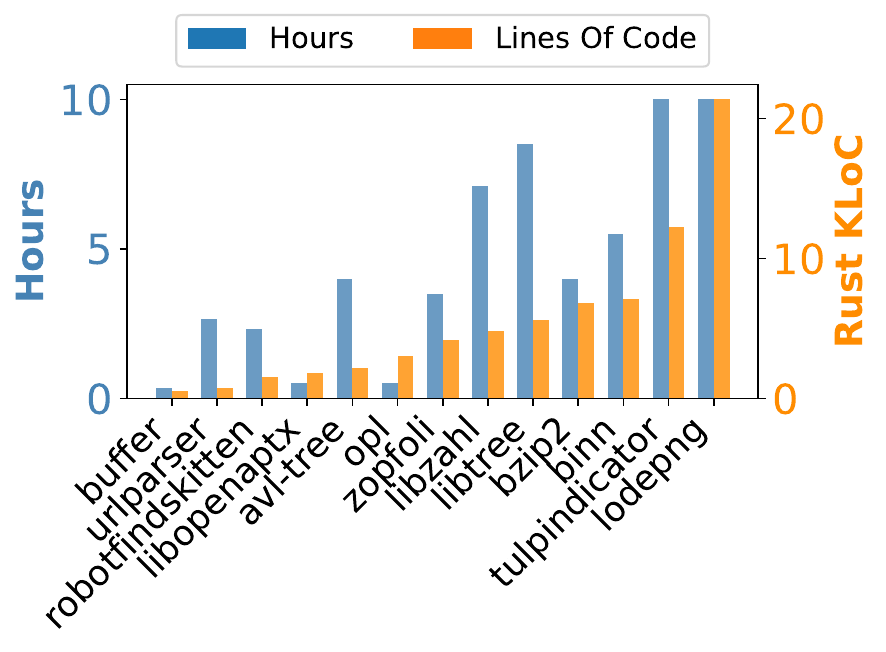}
    \vspace{-15pt}
    \caption{Manual debugging efforts for projects with at least one assertion failure}
    \vspace{-10pt}
    \label{fig:manual-debugging}
\end{wrapfigure}

The human developer 
%repaired translations for the $11$ remaining projects, achieving a 
achieved $100\%$ assertion pass rate on $11$ remaining projects, within $4.5$ hours on average (min=$0.33$ and max=$10$ hours). For \emph{tulipindicator} (ID=$23$) and \emph{lodepng} (ID=$20$), the assertion pass rate after manual effort was $94\%$ and $76\%$, respectively. Figure~\ref{fig:manual-debugging} shows the effort of the human developer.
%in fixing the semantic mismatch in translations. 
%The orange bars indicate the size of Rust translation (KLoC), and the blue bars show the effort (Hours). 
The Spearman Rank Order Correlation~\cite{inbook} shows that there is a strong correlation between the size of the project and the required effort to fix semantic mismatches in translations ($\rho=0.86$ with strong statistical significance, p-value$=1.59e-4<0.05$). The human developer reported the provided debugging information by \approach \textbf{useful to pinpoint the semantic mismatch}. They also noted \textbf{the nature of the semantic mismatch} as an important factor in the amount of debugging and effort, in addition to the size.  

\vspace{-5pt}
Table~\ref{table:effectiveness-validation} shows that \approach, fully automated or with human in the loop, \textbf{generates semantically equivalent translations similar to transpiler-based techniques}, which preserve semantics by construct. Some 
%of the 
test translations by these techniques were not compilable to 
%run and 
evaluate functional equivalence. 
%We also noticed that some 
Some techniques remove 
%a subset of 
assertions during translation. For example, the number of executed assertions in translations of \emph{qsort} (ID=$1$) with \ctorust, \saferrust, \crown, and \laertes is $10$, 
%which is 
less than the original $21$ assertions.
%in the C program. 
%We could not run tests on 
\saferrust 
%translations: 
either did not have test translations% did not exist 
%for projects, or if they existed, 
or they were not compilable in \emph{bin} mode. 

We further evaluate the quality of the translations across the following metrics (results in Tables~\ref{table:effectiveness-safety}, gray rows for \approach and white rows for alternative approaches): 

\subsubsection{Raw Pointer Declarations/Dereferences.}
\label{subsubsection:safety-rpdeclaration}

\approach effectively reduces the usage of raw pointers in translations: \textbf{the number of raw pointer declarations/derefenreces is reduced from \num{4080}/\num{2045} in C to $198$/$126$ in Rust, respectively} (\emph{Raw Pointers} columns in Table~\ref{table:subjects}; cf. Table~\ref{table:effectiveness-safety}). Translations of transpiler-based techniques contain drastically more raw pointer declarations and dereferences. \ctorust, \saferrust, \crown, and \laertes have \num{4904}/\num{20169}, \num{1400}/\num{5510}, \num{3027}/\num{12298}, and \num{2255}/\num{9685} \textbf{more raw pointer declarations and dereferences}, concerning overlapping projects. 
%LLM-based techniques that translate C projects to Rust contain considerably fewer raw pointer declarations and dereferences. 
\approach translations contain $5$/$49$ \textbf{fewer raw pointer declarations and dereferences} compared to \rustmap in \emph{bzip2}, and only one more raw pointer declarations compared to \syzygy. %This is due to \approach careful prompt crafting, i.e., instructing LLMs through natural language and in-context examples to translate without use of raw pointers. We believe the remaining few cases of raw pointers in translations are a form of compensation that LLMs provide for prioritizing compilation success or even functional equivalence.

\subsubsection{Pointer Arithmetic/Comparison.}
\label{subsubsection:safety-arithmetic}

\begin{figure}[t]
    \centering
    \input{Listings/rustinepointerinflation}
    \vspace{-3mm}
    \caption{Examples of introducing pointer arithmetic in Rust translation by \approach, which did not exist in the original C code, for \emph{bzip2} (left) and \emph{libtree} (right)}
    \label{code:safety}
\end{figure}

\approach \textbf{effectively resolved many pointer operation cases} through refactoring (\emph{Before} and \emph{After} of \emph{Pointer Arithmetic} columns under Table~\ref{table:subjects}), reducing them from $452$ to $14$. During translation, \approach resolves \emph{four} additional instances of pointer arithmetic in project \emph{gennan} (ID=$9$), but surprisingly, \textbf{adds $243$ new cases of pointer arithmetic} in translations of $14$ projects (column \emph{Pointer Arithmetic} in Table~\ref{table:effectiveness-safety}). We carefully investigated these cases and identified the following two reasons for this: (1) interacting with C libraries with no equivalence in Rust, e.g., operating system libraries, and (2) manipulating strings. 

Figure~\ref{code:safety} shows such a case. For \emph{bzip2} (left side), the C program casts the opaque pointer \texttt{\small{BZFILE}} into structure type \texttt{\small{bzFile}}, and directly accesses its inner handle field. Rust forbids such casting and field access on opaque types. 
%Between the choices of duplicating the entire struct (as done by \ctorust) and using pointer arithmetic, \approach LLM chooses the latter, although non exists in the original C code: 
The translation maintains a separate array \texttt{\small{bz_fp}} of file pointers and computes the correct \texttt{\small{FILE}} via an offset derived from the opaque pointer’s address. For \emph{libtree} (right side), the C program checks whether the variable \texttt{\small{soname}} is a valid offset not equal to \texttt{\small{MAX_OFFSET_T}}, and whether the string at position \texttt{\small{soname_buf_offset}} in \texttt{\small{s->string_table.arr}} appears in the program’s exclude list, storing the result in \texttt{\small{in_exclude_list}}.
Rust does not allow indexing raw bytes with arbitrary offsets, and \texttt{\small{\&[u8]}} or \texttt{\small{\&str}} slicing requires valid UTF-8 and known bounds. To avoid complex lifetime tracking and unsafe transmutes that could violate aliasing rules, the LLM %resorts to pointer arithmetic: it 
computes the pointer to the start of the target string with \texttt{\small{.add(soname\_buf\_offset)}} and constructs a \texttt{\small{CStr}} directly from that address.

\begin{wrapfigure}{r}{0.5\columnwidth}
\vspace{-20pt}
\input{Listings/c2rustpointerinflation}
\vspace{-2mm}
\caption{An example of pointer arithmetic inflation in the translation of \emph{xzoom} project by \ctorust}
    \label{c2rustsafety}
    \vspace{-3mm}
\end{wrapfigure}
\sloppy Compared to \approach, translations of transpiler-based techniques contain \textbf{considerably higher} pointer operations. Such techniques not only move existing pointer arithmetic in the C program to translation, but also drastically inflate their usage. The primary reason for pointer arithmetic inflation in transpiler-based techniques, however, is different: these tools convert array subscript accesses (\texttt{\small{[]}}) to explicit pointer arithmetic, even in cases where the original C code used clean, readable subscript notation. Figure \ref{c2rustsafety} shows an example of such cases in \ctorust translation. The C code computes the address of a pixel in an \texttt{\small{XImage}}
structure using two nested subscript expressions that account for pixel format, row padding (\texttt{\small{bytes_per_line}}), and per-component offset (\texttt{\small{xoffset}}). For implementing the same logic in Rust, \ctorust mechanically replaces every \texttt{\small{[]}} with a chain of \texttt{\small{.offset()}} and \texttt{\small{.wrapping_mul()}} calls, repeatedly recomputing the base address of the \texttt{\small{ximage}} array and the size of \texttt{\small{T}}. The resulting expression spans $17$ lines and contains seven separate pointer arithmetic operations, with none in C code. 
The \textbf{margin with LLM-based approaches is smaller}: \rustmap translations of \emph{bzip2} (ID=$22$) include $69$ pointer arithmetic, compared to $20$ in \approach; \syzygy and \approach both generate translations with \emph{zero} pointer arithmetic for \emph{zopfli} (ID=$18$).

\subsubsection{Unsafe Constrcuts.}
\label{subsubsection:safety-unsafe-constrcuts}

\approach translations contain \textbf{considerably fewer unsafe constructs compared to transpiler-based approaches}, as shown in Table~\ref{table:effectiveness-safety}. The number of unsafe $\langle$lines,type casts,calls$\rangle$ in \approach translations ($\langle 114,47,130 \rangle$) are \textbf{notably less} compared to \rustmap ($\langle 974,70,372 \rangle$). \syzygy translations contain no unsafe constructs, compared to $14$ in \approach translations. We believe the higher cost of translation of \emph{zopfli} by \syzygy ($\$800$) compared to $<\$0.71$ cost of translation by \approach, can explain the difference: re-prompting the LLM more can be helpful in reducing the safety issues, but can also increase the cost of translation. 

\subsubsection{Idiomaticity.} 
\label{subsubsection:idiomaticity}

%\texttt{\small{Clippy}} is widely-used for evaluating the idiomaticity of Rust code~\cite{li2024unleashing}. 
\begin{wrapfigure}{r}{0.4\columnwidth}
    \vspace{-15pt}
    \includegraphics[width=0.4\columnwidth]{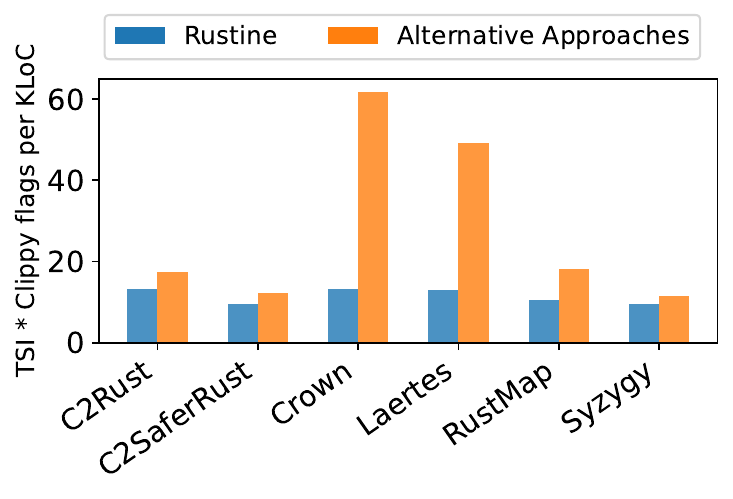}
    \vspace{-23pt}
    \caption{$TSI \times Clippy_{flags}$ per KLoC comparison of \approach with alternative approaches}
    \vspace{-10pt}
    \label{fig:idiomacity-comp}
\end{wrapfigure}
\begin{figure}[t]
    \centering
    \vspace{-10pt}
    \includegraphics[width=\linewidth]{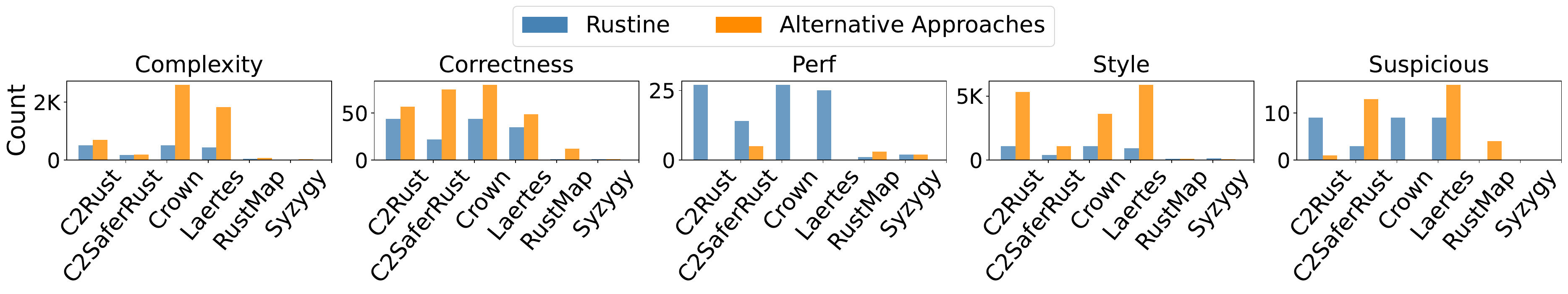}
    \vspace{-20pt}
    \caption{Pairwise comparison of \approach with others concerning raised \texttt{\small{Clippy}} flags on mutual projects}
    \label{fig:clippy}
\end{figure}

Figure~\ref{fig:idiomacity-comp} shows the TSI-normalized number of all raised \texttt{\small{Clippy}} flags for \approach, compared to alternative approaches concerning mutual projects. Figure~\ref{fig:clippy} also shows the mutual comparison per each \texttt{\small{Clippy}} flag. These figures show that \textbf{\approach consistently raises fewer flags per normalized KLoC compared to other approaches}, i.e., it better adheres to Rust's idiomatic patterns compared to existing approaches. Concerning individual flag categories, \approach consistently raises fewer \texttt{\small{complexity}} and \texttt{\small{correctness}} flags compared to other approaches. 

\begin{wrapfigure}{r}{0.5\columnwidth}
\vspace{-20pt}
\input{Listings/perf-warning-example}
\vspace{-2mm}
\caption{An example of \texttt{\small{clippy::perf}} warning for \approach, while not complaining about \ctorust unsafe block}
    \label{fig:perf-example}
    \vspace{-8pt}
\end{wrapfigure}
The \texttt{\small{correctness}} warnings indicate opportunities to improve idiomaticity rather than functional incorrectness. Our investigation confirms the \textbf{majority of \texttt{\small{correctness}} warnings for \approach are \emph{false positives} or acceptable trade-offs}: (1) C code usually uses \texttt{\small{int}} for lengths, requiring bound checking. \approach translates \texttt{\small{int}} types to \texttt{\small{usize}}, which are always positive. However, it preserves the C structure and keeps the bound check, resulting in \texttt{\small{Clippy::correctness}} warning. (2) Some C libraries have no equivalence in Rust, e.g., \texttt{\small{X11}}~\cite{xlib} lib in \emph{xzoom}, making the use of FFI in Rust to interact with them inevitable (\S \ref{subsub:API-invocation}). \texttt{\small{Clippy}} may raise \texttt{\small{correctness}} warnings on bindgen-generated FFI bindings rather than translated application logic. (3) Rust provides built-in named constants, e.g., \texttt{\small{f64::consts::LOG2_E}}, while C does not. When translating a C code containing similar but hard-coded variables, the LLM typically preserves the C structure and redefines them, raising \texttt{\small{correctness}} warning. 

Transpiler-based approaches, surprisingly, raised almost \emph{zero} \texttt{\small{perf}} warnings. A deeper investigation reveals such warnings are mostly improvements to \emph{closely} idiomatic Rust code. The transpiler-based techniques suffer from serious safety issues discussed before, far from being idiomatic, and escaping linter checks (false negatives). The example of Figure~\ref{fig:perf-example} shows such a case: \ctorust translations is unsafe, using C API and raw pointers to copy the buffer. \approach translation is safe with no pointers, but \texttt{\small{Clippy}} suggests using \texttt{\small{copy_from_slide}} rather than a \texttt{\small{loop}}. This observation suggests that, \textbf{\texttt{\small{Clippy}} warnings should be not be used alone to determine the idiomaticity of translations, without comparison concerning different safety aspects and readability}. 

\subsubsection{Readability}
\label{subsubsection:readability}

\begin{figure}[t]
    \vspace{-0.6cm}
    \centering
    \scriptsize 
    \includegraphics[width=\linewidth]{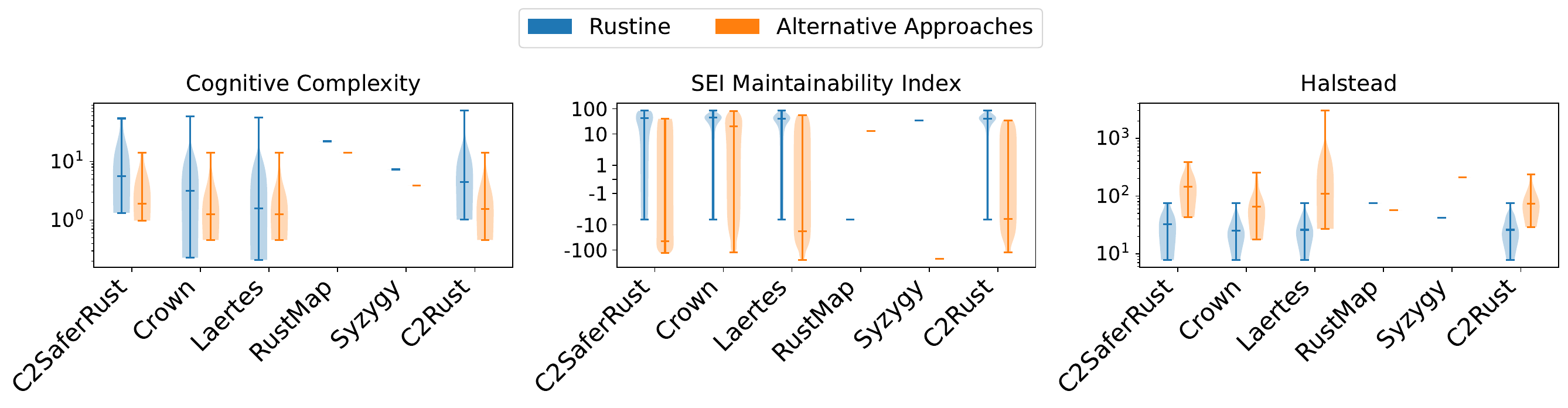}
    \vspace{-15pt}
    \caption{Comparing the readability of translations generated by \approach and alternative approaches. For Cognitive Complexity and Halstead, lower values indicate more readability. For SEI, a higher value is better}
    \label{fig:codeqq}
    %\vspace{-0.4cm}
\end{figure}

\approach generates readable and well-structured Rust code that \textbf{significantly outperforms existing translation tools across multiple metrics}. Figure~\ref{fig:codeqq} compares the readability of \approach translations with others on the overlapping projects. \approach translations consistently score lower cognitive complexity, lower Halstead, and higher SEI. The only exception is comparison with \rustmap, where \approach translations achieve a lower SEI score and slightly higher Halstead. As shown in Figure~\ref{fig:clippy}, \approach raises considerably fewer \texttt{\small{clippy::style}} flags, demonstrating its superior adherence to Rust coding conventions. These results confirm that \textbf{\approach preserves functionality and actively enhances code structure and maintainability during translation, producing output that approaches hand-written Rust quality}.

%% file: Tables/Effectiveness-Validation.tex
\centering
\setlength{\tabcolsep}{3pt}
\footnotesize
\resizebox{\textwidth}{!}{
\begin{tabular}{|c|c|c|cc|ccc|}
\hline
\multirow{2}{*}{\textbf{ID}} 
& \multirow{2}{*}{\textbf{Tool}} 
& \multirow{2}{*}{\begin{tabular}[c]{@{}c@{}}\% \textbf{Compilation}\\ \textbf{Success}\end{tabular}} 
& \multicolumn{2}{c|}{\% \textbf{Coverage}} 
& \multicolumn{3}{c|}{\# \textbf{Assertions}} \\ \cline{4-8}
& & & \textbf{Function} & \textbf{Line} 
& \textbf{Executed} & \textbf{Passed} & \textbf{Failed} \\ \hline

% Should be repeated per each subject
% Rustify
\multirow{2}{*}{1*} & 
\approach \cellcolor{gray!25} & 
\multicolumn{1}{c|}{\cellcolor{gray!25} $100$}&
\cellcolor{gray!25} $100$&  
\multicolumn{1}{c|}{\cellcolor{gray!25}$92$} &
\cellcolor{gray!25} \cellcolor{gray!25} $21$&   
\cellcolor{gray!25} $21$&  
\cellcolor{gray!25} $0$\\ \cline{2-8} 
% Others
& 
$\langle$ C2R,C2SR,C,L $\rangle$ & 
\multicolumn{1}{c|}{$\langle$ $100$,$100$,$100$,$100$ $\rangle$} & 
$\langle$ $100$,$100$,$100$,$100$ $\rangle$& 
\multicolumn{1}{c|}{$\langle$ $100$,$-$,$100$,$100$ $\rangle$} & 
$\langle$ $10$,$-$,$10$,$10$ $\rangle$& 
$\langle$ $10$,$-$,$10$,$10$ $\rangle$& 
$\langle$ $0$,$-$,$0$,$0$ $\rangle$\\ \hline

% Rustify
\multirow{2}{*}{2*} & 
\approach \cellcolor{gray!25} & 
\multicolumn{1}{c|}{\cellcolor{gray!25}$100$} &
\cellcolor{gray!25} $92$&  
\multicolumn{1}{c|}{\cellcolor{gray!25}$95$} &
\cellcolor{gray!25} \cellcolor{gray!25} $6$&   
\cellcolor{gray!25} $6$&  
\cellcolor{gray!25} $0$
\\ \cline{2-8} 
% Others
& 
$\langle$ C2R,C,L $\rangle$ & 
\multicolumn{1}{c|}{$\langle$ $100$,$100$,$100$ $\rangle$} & 
$\langle$ $87$,$85$,$74$ $\rangle$& 
\multicolumn{1}{c|}{$\langle$ $73$,$70$,$63$ $\rangle$} & 
$\langle$ $6$,$6$,$6$ $\rangle$& 
$\langle$ $6$,$6$,$6$ $\rangle$& 
$\langle$ $0$,$0$,$0$ $\rangle$
\\ \hline

% Rustify
\multirow{2}{*}{3*} & 
\approach \cellcolor{gray!25} & 
\multicolumn{1}{c|}{\cellcolor{gray!25}$100$} &
\cellcolor{gray!25} $99$&  
\multicolumn{1}{c|}{\cellcolor{gray!25}$93$} &
\cellcolor{gray!25} \cellcolor{gray!25} $20$&   
\cellcolor{gray!25} $20$&  
\cellcolor{gray!25} $0$
\\ \cline{2-8} 
% Others
& 
$\langle$ C2R,C,L $\rangle$ & 
\multicolumn{1}{c|}{$\langle$ $100$,$100$,$100$ $\rangle$} & 
$\langle$ $83$,$83$,$23$ $\rangle$& 
\multicolumn{1}{c|}{$\langle$ $62$,$69$,$22$ $\rangle$} & 
$\langle$ $20$,$20$,$20$ $\rangle$& 
$\langle$ $20$,$20$,$18$ $\rangle$& 
$\langle$ $0$,$0$,$2$ $\rangle$
\\ \hline

% Rustify
\multirow{2}{*}{4*} & 
\approach \cellcolor{gray!25} & 
\multicolumn{1}{c|}{\cellcolor{gray!25}$100$} &
\cellcolor{gray!25} $91$&  
\multicolumn{1}{c|}{\cellcolor{gray!25}$83$} &
\cellcolor{gray!25} \cellcolor{gray!25} $34$&   
\cellcolor{gray!25} $34$&  
\cellcolor{gray!25} $0$\\ \cline{2-8} 
% Others
& 
$\langle$ C2R,C,L $\rangle$ & 
\multicolumn{1}{c|}{$\langle$ $100$,$100$,$100$ $\rangle$} & 
$\langle$ $81$,$76$,$27$ $\rangle$& 
\multicolumn{1}{c|}{$\langle$ $78$,$63$,$48$ $\rangle$} & 
$\langle$ $34$,$34$,$34$ $\rangle$& 
$\langle$ $34$,$34$,$34$ $\rangle$& 
$\langle$ $0$,$0$,$0$ $\rangle$\\ \hline

% Rustify
\multirow{2}{*}{5} & 
\approach \cellcolor{gray!25} & 
\multicolumn{1}{c|}{\cellcolor{gray!25}$100$} &
\cellcolor{gray!25} $91$&  
\multicolumn{1}{c|}{\cellcolor{gray!25}$88$} &
\cellcolor{gray!25} \cellcolor{gray!25} $54$&   
\cellcolor{gray!25} $54$&  
\cellcolor{gray!25} $0$
\\ \cline{2-8} 
% Others
& 
$\langle$ C2R,C,L $\rangle$ & 
\multicolumn{1}{c|}{$\langle$ $100$,$100$,$100$ $\rangle$} & 
$\langle$ $87$,$92$,$30$ $\rangle$& 
\multicolumn{1}{c|}{$\langle$ $57$,$66$,$42$ $\rangle$} & 
$\langle$ $54$,$54$,$54$ $\rangle$& 
$\langle$ $54$,$50$,$50$ $\rangle$& 
$\langle$ $0$,$4$,$4$ $\rangle$
\\ \hline

% Rustify
\multirow{2}{*}{6*} & 
\approach \cellcolor{gray!25} & 
\multicolumn{1}{c|}{\cellcolor{gray!25}$100$} &
\cellcolor{gray!25} $11$&  
\multicolumn{1}{c|}{\cellcolor{gray!25}$10$} &
\cellcolor{gray!25} \cellcolor{gray!25} $4$&   
\cellcolor{gray!25} $4$&  
\cellcolor{gray!25} $0$
\\ \cline{2-8} 
% Others
& 
$\langle$ C2R,C2SR,C,L $\rangle$ & 
\multicolumn{1}{c|}{$\langle$ $100$,$100$,$100$,$100$ $\rangle$} & 
$\langle$ $10$,$-$,$10$,$10$ $\rangle$& 
\multicolumn{1}{c|}{$\langle$ $11$,$-$,$10$,$9$ $\rangle$} & 
$\langle$ $4$,$-$,$4$,$4$ $\rangle$& 
$\langle$ $4$,$-$,$4$,$4$ $\rangle$& 
$\langle$ $0$,$-$,$0$,$0$ $\rangle$
\\ \hline

% Rustify
\multirow{2}{*}{7} & 
\approach \cellcolor{gray!25} & 
\multicolumn{1}{c|}{\cellcolor{gray!25}$100$} &
\cellcolor{gray!25} $75$&  
\multicolumn{1}{c|}{\cellcolor{gray!25}$74$} &
\cellcolor{gray!25} \cellcolor{gray!25} $46$&   
\cellcolor{gray!25} $46$&  
\cellcolor{gray!25} $0$
\\ \cline{2-8} 
% Others
& 
$\langle$ C2R,C2SR,C,L $\rangle$ & 
\multicolumn{1}{c|}{$\langle$ $100$,$100$,$100$,$100$ $\rangle$} & 
$\langle$ $61$,$-$,$60$,$60$ $\rangle$& 
\multicolumn{1}{c|}{$\langle$ $41$,$-$,$39$,$35$ $\rangle$} & 
$\langle$ $46$,$-$,$46$,$46$ $\rangle$ & 
$\langle$ $46$,$-$,$46$,$46$ $\rangle$ & 
$\langle$ $0$,$-$,$0$,$0$ $\rangle$ 
\\ \hline

% Rustify
\multirow{2}{*}{8*} & 
\approach \cellcolor{gray!25} & 
\multicolumn{1}{c|}{\cellcolor{gray!25}$100$} &
\cellcolor{gray!25} $-$&  
\multicolumn{1}{c|}{\cellcolor{gray!25}$-$} &
\cellcolor{gray!25} \cellcolor{gray!25} $-$&   
\cellcolor{gray!25} $-$&  
\cellcolor{gray!25} $-$
\\ \cline{2-8} 
% Others
& 
$\langle$ C2R,C2SR,C,L $\rangle$ & 
\multicolumn{1}{c|}{$\langle$ $100$,$100$,$100$,$100$ $\rangle$} & 
$-$& 
\multicolumn{1}{c|}{$-$} & 
$-$& 
$-$& 
$-$
\\ \hline

% Rustify
\multirow{2}{*}{9*} & 
\approach \cellcolor{gray!25} & 
\multicolumn{1}{c|}{\cellcolor{gray!25}$100$} &
\cellcolor{gray!25} $84$&  
\multicolumn{1}{c|}{\cellcolor{gray!25}$79$} &
\cellcolor{gray!25} \cellcolor{gray!25} $521556$&   
\cellcolor{gray!25} $521556$&  
\cellcolor{gray!25} $0$
\\ \cline{2-8} 
% Others
& 
$\langle$ C2R,C2SR,C,L $\rangle$ & 
\multicolumn{1}{c|}{$\langle$ $100$,$100$,$100$,$100$ $\rangle$} & 
$\langle$ $83$,$-$,$81$,$81$ $\rangle$& 
\multicolumn{1}{c|}{$\langle$ $68$,$-$,$67$,$68$ $\rangle$} & 
$\langle$ $521556$,$-$,$521556$,$521556$ $\rangle$& 
$\langle$ $521556$,$-$,$521556$,$521556$ $\rangle$& 
$\langle$ $0$,$-$,$0$,$0$ $\rangle$
\\ \hline

% Rustify
\multirow{2}{*}{10*} & 
\approach \cellcolor{gray!25} & 
\multicolumn{1}{c|}{\cellcolor{gray!25}$100$} &
\cellcolor{gray!25} $61$&  
\multicolumn{1}{c|}{\cellcolor{gray!25}$67$} &
\cellcolor{gray!25} \cellcolor{gray!25} $1$&   
\cellcolor{gray!25} $1$&  
\cellcolor{gray!25} $0$
\\ \cline{2-8} 
% Others
& 
$\langle$ C2R,C,L $\rangle$ & 
\multicolumn{1}{c|}{$\langle$ $100$,$100$,$100$ $\rangle$} & 
$\langle$ $55$,$56$,$54$ $\rangle$ & 
\multicolumn{1}{c|}{$\langle$ $40$,$41$,$39$ $\rangle$} & 
$\langle$ $1$,$1$,$1$ $\rangle$ & 
$\langle$ $1$,$1$,$1$ $\rangle$ & 
$\langle$ $0$,$0$,$0$ $\rangle$ 
\\ \hline

% Rustify
\multirow{2}{*}{11} & 
\approach \cellcolor{gray!25} & 
\multicolumn{1}{c|}{\cellcolor{gray!25}$100$} &
\cellcolor{gray!25} $63$&  
\multicolumn{1}{c|}{\cellcolor{gray!25}$61$} &
\cellcolor{gray!25} \cellcolor{gray!25} $47$&   
\cellcolor{gray!25} $47$&  
\cellcolor{gray!25} $0$
\\ \cline{2-8} 
% Others
& 
$\langle$ C2R,C,L $\rangle$ & 
\multicolumn{1}{c|}{$\langle$ $100$,$100$,$100$ $\rangle$} & 
$\langle$ $51$,$49$,$43$ $\rangle$ & 
\multicolumn{1}{c|}{$\langle$ $59$,$55$,$51$ $\rangle$} & 
$\langle$ $47$,$47$,$47$ $\rangle$ & 
$\langle$ $47$,$47$,$47$ $\rangle$ & 
$\langle$ $0$,$0$,$0$ $\rangle$ 
\\ \hline

% Rustify
\multirow{2}{*}{12*} & 
\approach \cellcolor{gray!25} & 
\multicolumn{1}{c|}{\cellcolor{gray!25}$100$} &
\cellcolor{gray!25} $45$&  
\multicolumn{1}{c|}{\cellcolor{gray!25}$52$} &
\cellcolor{gray!25} \cellcolor{gray!25} $7406$&   
\cellcolor{gray!25} $7406$&  
\cellcolor{gray!25} $0$
\\ \cline{2-8} 
% Others
& 
$\langle$ C2R,C,L $\rangle$ & 
\multicolumn{1}{c|}{$\langle$ $100$,$100$,$100$ $\rangle$} & 
$\langle$ $48$,$45$,$47$ $\rangle$ & 
\multicolumn{1}{c|}{$\langle$ $84$,$79$,$77$ $\rangle$} & 
$\langle$ $7406$,$7406$,$7406$ $\rangle$ & 
$\langle$ $7406$,$7406$,$7406$ $\rangle$ & 
$\langle$ $0$,$0$,$0$ $\rangle$ 
\\ \hline

% Rustify
\multirow{2}{*}{13} & 
\approach \cellcolor{gray!25} & 
\multicolumn{1}{c|}{\cellcolor{gray!25}$100$} &
\cellcolor{gray!25} $29$&  
\multicolumn{1}{c|}{\cellcolor{gray!25}$31$} &
\cellcolor{gray!25} \cellcolor{gray!25} $12$&   
\cellcolor{gray!25} $12$&  
\cellcolor{gray!25} $0$
\\ \cline{2-8} 
% Others
& 
$\langle$ C2R,C,L $\rangle$ & 
\multicolumn{1}{c|}{$\langle$ $100$,$100$,$100$ $\rangle$} & 
$\langle$ $35$,$32$,$33$ $\rangle$ & 
\multicolumn{1}{c|}{$\langle$ $33$,$31$,$32$ $\rangle$} & 
$\langle$ $12$,$12$,$12$ $\rangle$ & 
$\langle$ $10$,$10$,$10$ $\rangle$ & 
$\langle$ $2$,$2$,$2$ $\rangle$ 
\\ \hline

% Rustify
\multirow{2}{*}{14} & 
\approach \cellcolor{gray!25} & 
\multicolumn{1}{c|}{\cellcolor{gray!25}$100$} &
\cellcolor{gray!25} $95$&  
\multicolumn{1}{c|}{\cellcolor{gray!25}$81$} &
\cellcolor{gray!25} \cellcolor{gray!25} $9$&   
\cellcolor{gray!25} $9$&  
\cellcolor{gray!25} $0$
\\ \cline{2-8} 
% Others
& 
$\langle$ C2R,C $\rangle$ & 
\multicolumn{1}{c|}{$\langle$ $100$,$100$ $\rangle$} & 
$\langle$ $81$,$75$ $\rangle$ & 
\multicolumn{1}{c|}{$\langle$ $87$,$74$ $\rangle$} & 
$\langle$ $9$,$9$ $\rangle$ & 
$\langle$ $9$,$9$ $\rangle$ & 
$\langle$ $0$,$0$ $\rangle$  
\\ \hline

% Rustify
\multirow{2}{*}{15} & 
\approach \cellcolor{gray!25} & 
\multicolumn{1}{c|}{\cellcolor{gray!25}$100$} &
\cellcolor{gray!25} $82$&  
\multicolumn{1}{c|}{\cellcolor{gray!25}$75$} &
\cellcolor{gray!25} \cellcolor{gray!25} $121$&   
\cellcolor{gray!25} $121$&  
\cellcolor{gray!25} $0$
\\ \cline{2-8} 
% Others
& 
$\langle$ C2R,C,L $\rangle$ & 
\multicolumn{1}{c|}{$\langle$ $100$,$100$,$100$ $\rangle$} & 
$\langle$ $90$,$89$,$90$ $\rangle$ & 
\multicolumn{1}{c|}{$\langle$ $61$,$60$,$61$ $\rangle$} & 
$\langle$ $121$,$121$,$121$ $\rangle$ & 
$\langle$ $121$,$121$,$121$ $\rangle$ & 
$\langle$ $0$,$0$,$0$ $\rangle$ 
\\ \hline

% Rustify
\multirow{2}{*}{16} & 
\approach \cellcolor{gray!25} & 
\multicolumn{1}{c|}{\cellcolor{gray!25}$100$} &
\cellcolor{gray!25} $45$&  
\multicolumn{1}{c|}{\cellcolor{gray!25}$44$} &
\cellcolor{gray!25} \cellcolor{gray!25} $14$&   
\cellcolor{gray!25} $14$&  
\cellcolor{gray!25} $0$
\\ \cline{2-8} 
% Others
& 
$\langle$ C2R,C $\rangle$ & 
\multicolumn{1}{c|}{$\langle$ $100$,$100$ $\rangle$} & 
$\langle$ $45$,$41$ $\rangle$ & 
\multicolumn{1}{c|}{$\langle$ $40$,$41$ $\rangle$} & 
$\langle$ $14$,$14$ $\rangle$ & 
$\langle$ $14$,$14$ $\rangle$ & 
$\langle$ $0$,$0$ $\rangle$ 
\\ \hline

% Rustify
\multirow{2}{*}{17} & 
\approach \cellcolor{gray!25} & 
\multicolumn{1}{c|}{\cellcolor{gray!25}$100$} &
\cellcolor{gray!25} $87$&  
\multicolumn{1}{c|}{\cellcolor{gray!25}$60$} &
\cellcolor{gray!25} \cellcolor{gray!25} $1570$&   
\cellcolor{gray!25} $1570$&  
\cellcolor{gray!25} $0$
\\ \cline{2-8} 
% Others
& 
$\langle$ C2R,C,L $\rangle$ & 
\multicolumn{1}{c|}{$\langle$ $0$,$0$,$0$ $\rangle$} & 
$\langle$ $-$,$-$,$-$ $\rangle$ & 
\multicolumn{1}{c|}{$\langle$ $-$,$-$,$-$ $\rangle$} & 
$\langle$ $-$,$-$,$-$ $\rangle$ & 
$\langle$ $-$,$-$,$-$ $\rangle$ & 
$\langle$ $-$,$-$,$-$ $\rangle$ 
\\ \hline

% Rustify
\multirow{2}{*}{18} & 
\approach \cellcolor{gray!25} & 
\multicolumn{1}{c|}{\cellcolor{gray!25}$100$} &
\cellcolor{gray!25} $92$&  
\multicolumn{1}{c|}{\cellcolor{gray!25}$84$} &
\cellcolor{gray!25} \cellcolor{gray!25} $40$&   
\cellcolor{gray!25} $40$&  
\cellcolor{gray!25} $0$
\\ \cline{2-8} 
% Others
& 
$\langle$ C2R,C,S $\rangle$ & 
\multicolumn{1}{c|}{$\langle$ $100$,$0$,$100$ $\rangle$} & 
$\langle$ $86$,$-$,$-$ $\rangle$ & 
\multicolumn{1}{c|}{$\langle$ $80$,$-$,$-$ $\rangle$} & 
$\langle$ $40$,$-$,$-$ $\rangle$ & 
$\langle$ $40$,$-$,$-$ $\rangle$ & 
$\langle$ $0$,$-$,$-$ $\rangle$ 
\\ \hline

% Rustify
\multirow{2}{*}{19*} & 
\approach \cellcolor{gray!25} & 
\multicolumn{1}{c|}{\cellcolor{gray!25}$100$} &
\cellcolor{gray!25} $-$&  
\multicolumn{1}{c|}{\cellcolor{gray!25}$-$} &
\cellcolor{gray!25} \cellcolor{gray!25} $-$&   
\cellcolor{gray!25} $-$&  
\cellcolor{gray!25} $-$
\\ \cline{2-8} 
% Others
& 
$\langle$ C2R,C2SR,C,L $\rangle$ & 
\multicolumn{1}{c|}{$\langle$ $100$,$100$,$100$,$100$ $\rangle$} & 
$-$ & 
\multicolumn{1}{c|}{$-$} & 
$-$ & 
$-$ & 
$-$
\\ \hline

% Rustify
\multirow{2}{*}{20} & 
\approach \cellcolor{gray!25} & 
\multicolumn{1}{c|}{\cellcolor{gray!25}$100$} &
\cellcolor{gray!25} $61$&  
\multicolumn{1}{c|}{\cellcolor{gray!25}$60$} &
\cellcolor{gray!25} \cellcolor{gray!25} $683994$&   
\cellcolor{gray!25} $526159$&  
\cellcolor{gray!25} $157835$
\\ \cline{2-8} 
% Others
& 
$\langle$ C2R,C,L $\rangle$ & 
\multicolumn{1}{c|}{$\langle$ $100$,$100$,$100$ $\rangle$} & 
$\langle$ $72$,$-$,$-$ $\rangle$ & 
\multicolumn{1}{c|}{$\langle$ $73$,$-$,$-$ $\rangle$} & 
$\langle$ $683994$,$-$,$-$ $\rangle$ & 
$\langle$ $683994$,$-$,$-$ $\rangle$ & 
$\langle$ $0$,$-$,$-$ $\rangle$ 
\\ \hline

% Rustify
\multirow{2}{*}{21} & 
\approach \cellcolor{gray!25} & 
\multicolumn{1}{c|}{\cellcolor{gray!25}$100$} &
\cellcolor{gray!25} $13$&  
\multicolumn{1}{c|}{\cellcolor{gray!25}$12$} &
\cellcolor{gray!25} \cellcolor{gray!25} $36$&   
\cellcolor{gray!25} $36$&  
\cellcolor{gray!25} $0$
\\ \cline{2-8} 
% Others
& 
$\langle$ C2R,C2SR,R,C,L $\rangle$ & 
\multicolumn{1}{c|}{$\langle$ $100$,$100$,$100$,$100$,$100$ $\rangle$} & 
$\langle$ $5$,$-$,$-$,$5$.$5$ $\rangle$ & 
\multicolumn{1}{c|}{$\langle$ $5$,$-$,$-$,$4$,$4$ $\rangle$} & 
$\langle$ $36$,$-$,$-$,$36$,$36$ $\rangle$ & 
$\langle$ $36$,$-$,$-$,$36$,$36$ $\rangle$ & 
$\langle$ $0$,$-$,$-$,$0$,$0$ $\rangle$ 
\\ \hline

% Rustify
\multirow{2}{*}{22} & 
\approach \cellcolor{gray!25} & 
\multicolumn{1}{c|}{\cellcolor{gray!25}$100$} &
\cellcolor{gray!25} $55$&  
\multicolumn{1}{c|}{\cellcolor{gray!25}$60$} &
\cellcolor{gray!25} \cellcolor{gray!25} $1949$&   
\cellcolor{gray!25} $1949$&  
\cellcolor{gray!25} $0$
\\ \cline{2-8} 
% Others
& 
$\langle$ C2R,C,L $\rangle$ & 
\multicolumn{1}{c|}{$\langle$ $100$,$100$,$100$ $\rangle$} & 
$\langle$ $60$,$60$,$61$ $\rangle$ & 
\multicolumn{1}{c|}{$\langle$ $79$,$72$,$75$ $\rangle$} & 
$\langle$ $1949$,$1949$,$1949$ $\rangle$ & 
$\langle$ $1949$,$1949$,$1949$ $\rangle$ & 
$\langle$ $0$,$0$,$0$ $\rangle$ 
\\ \hline

% Rustify
\multirow{2}{*}{23} & 
\approach \cellcolor{gray!25} & 
\multicolumn{1}{c|}{\cellcolor{gray!25}$100$} &
\cellcolor{gray!25} $65$&  
\multicolumn{1}{c|}{\cellcolor{gray!25}$60$} &
\cellcolor{gray!25} \cellcolor{gray!25} $4252$&   
\cellcolor{gray!25} $3994$&  
\cellcolor{gray!25} $258$
\\ \cline{2-8} 
% Others
& 
$\langle$ C2R,C2SR,C,L $\rangle$ & 
\multicolumn{1}{c|}{$\langle$ $100$,$100$,$100$,$100$ $\rangle$} & 
$-$ & 
\multicolumn{1}{c|}{$-$} & 
$-$ & 
$-$ & 
$-$ 
\\ \hline

\end{tabular}
}

%% file: Tables/Effectiveness-Safety.tex
\centering
\setlength{\tabcolsep}{2pt}
\footnotesize
\resizebox{\textwidth}{!}{
\begin{tabular}{|c|c|c|c|c|c|c|c|}
\hline
\multirow{2}{*}{\textbf{ID}} 
& \multirow{2}{*}{\textbf{Tool}} 
& \multirow{2}{*}{\begin{tabular}[c]{@{}c@{}}\# \textbf{Pointer}\\ \textbf{Arithmetic}\end{tabular}} 
& \multicolumn{2}{c|}{\# \textbf{Raw Pointers}} 
& \multicolumn{3}{c|}{\# \textbf{Unsafe Constructs}} \\ \cline{4-8}
& & & \textbf{Declarations} & \textbf{Dereferences} 
& \textbf{Lines} & \textbf{Type Casts} & \textbf{Calls} \\ \hline

% Should be repeated per each subject
% Rustify
\multirow{2}{*}{1} & 
\approach \cellcolor{gray!25} &  
\multicolumn{1}{c|}{\cellcolor{gray!25}$\mathbf{0}$} &
\cellcolor{gray!25} $\mathbf{0}$&  
\multicolumn{1}{c|}{\cellcolor{gray!25}$\mathbf{0}$} &
\cellcolor{gray!25} $\mathbf{0}$&  
\cellcolor{gray!25} $\mathbf{0}$&
\cellcolor{gray!25} $\mathbf{0}$\\ \cline{2-8} 
% Others
& 
$\langle$ C2R,C2SR,C,L $\rangle$ & 
\multicolumn{1}{c|}{$\langle$ $5$,$5$,$5$,$5$ $\rangle$} & 
$\langle$ $4$,$1$,$4$,$2$ $\rangle$& 
\multicolumn{1}{c|}{$\langle$ $10$,$6$,$10$,$6$ $\rangle$} & 
$\langle$ $39$,$19$,$39$,$32$ $\rangle$& 
$\langle$ $12$,$13$,$12$,$12$ $\rangle$& 
$\langle$ $11$,$10$,$11$,$23$ $\rangle$\\ \hline

% Rustify
\multirow{2}{*}{2} & 
\approach \cellcolor{gray!25} & 
\multicolumn{1}{c|}{\cellcolor{gray!25}$0$} &
\cellcolor{gray!25} $\mathbf{0}$&  
\multicolumn{1}{c|}{\cellcolor{gray!25}$\mathbf{0}$} &
\cellcolor{gray!25} $\mathbf{0}$&  
\cellcolor{gray!25} $\mathbf{0}$&
\cellcolor{gray!25} $\mathbf{0}$
\\ \cline{2-8} 
% Others
& 
$\langle$ C2R,C,L $\rangle$ & 
\multicolumn{1}{c|}{$\langle$ $0$,$0$,$0$ $\rangle$} & 
$\langle$ $9$,$4$,$9$ $\rangle$ & 
\multicolumn{1}{c|}{$\langle$ $29$,$7$,$29$ $\rangle$} & 
$\langle$ $60$,$61$,$68$ $\rangle$ & 
$\langle$ $8$,$6$,$8$ $\rangle$ & 
$\langle$ $21$,$92$,$21$ $\rangle$ 
\\ \hline

% Rustify
\multirow{2}{*}{3} & 
\approach \cellcolor{gray!25} & 
\multicolumn{1}{c|}{\cellcolor{gray!25}$\mathbf{0}$} &
\cellcolor{gray!25} $\mathbf{0}$&  
\multicolumn{1}{c|}{\cellcolor{gray!25}$\mathbf{0}$} &
\cellcolor{gray!25} $\mathbf{0}$&  
\cellcolor{gray!25} $\mathbf{0}$&
\cellcolor{gray!25} $\mathbf{0}$
\\ \cline{2-8} 
% Others
& 
$\langle$ C2R,C,L $\rangle$ & 
\multicolumn{1}{c|}{$\langle$ $33$,$33$,$33$ $\rangle$} & 
$\langle$ $21$,$16$,$16$ $\rangle$ & 
\multicolumn{1}{c|}{$\langle$ $78$,$66$,$66$ $\rangle$} & 
$\langle$ $706$,$604$,$1621$ $\rangle$ & 
$\langle$ $525$,$562$,$974$ $\rangle$ & 
$\langle$ $155$,$148$,$187$ $\rangle$ 
\\ \hline

% Rustify
\multirow{2}{*}{4} & 
\approach \cellcolor{gray!25} & 
\multicolumn{1}{c|}{\cellcolor{gray!25}$\mathbf{0}$} &
\cellcolor{gray!25} $\mathbf{4}$&  
\multicolumn{1}{c|}{\cellcolor{gray!25}$\mathbf{2}$} &
\cellcolor{gray!25} $\mathbf{26}$&  
\cellcolor{gray!25} $\mathbf{0}$&
\cellcolor{gray!25} $\mathbf{8}$\\ \cline{2-8} 
% Others
& 
$\langle$ C2R,C,L $\rangle$ & 
\multicolumn{1}{c|}{$\langle$ $1$,$1$,$1$ $\rangle$} & 
$\langle$ $50$,$33$,$47$ $\rangle$& 
\multicolumn{1}{c|}{$\langle$ $210$,$108$,$206$ $\rangle$} & 
$\langle$ $980$,$759$,$819$ $\rangle$& 
$\langle$ $419$,$443$,$376$ $\rangle$& 
$\langle$ $294$,$509$,$266$ $\rangle$\\ \hline

% Rustify
\multirow{2}{*}{5} & 
\approach \cellcolor{gray!25} & 
\multicolumn{1}{c|}{\cellcolor{gray!25}$\mathbf{1}$} &
\cellcolor{gray!25} $\mathbf{0}$&  
\multicolumn{1}{c|}{\cellcolor{gray!25}$\mathbf{0}$} &
\cellcolor{gray!25} $\mathbf{0}$&  
\cellcolor{gray!25} $\mathbf{0}$&
\cellcolor{gray!25} $\mathbf{0}$
\\ \cline{2-8} 
% Others
& 
$\langle$ C2R,C,L $\rangle$ & 
\multicolumn{1}{c|}{$\langle$ $23$,$23$,$23$ $\rangle$} & 
$\langle$ $61$,$19$,$57$ $\rangle$ & 
\multicolumn{1}{c|}{$\langle$ $82$,$19$,$73$ $\rangle$} & 
$\langle$ $1008$,$780$,$871$ $\rangle$ & 
$\langle$ $575$,$620$,$557$ $\rangle$ & 
$\langle$ $334$,$518$,$327$ $\rangle$ 
\\ \hline

% Rustify
\multirow{2}{*}{6} & 
\approach \cellcolor{gray!25} & 
\multicolumn{1}{c|}{\cellcolor{gray!25}$\mathbf{0}$} &
\cellcolor{gray!25} $15$&  
\multicolumn{1}{c|}{\cellcolor{gray!25}$\mathbf{7}$} &
\cellcolor{gray!25} $112$&  
\cellcolor{gray!25} $15$&
\cellcolor{gray!25} $51$
\\ \cline{2-8} 
% Others
& 
$\langle$ C2R,C2SR,C,L $\rangle$ & 
\multicolumn{1}{c|}{$\langle$ $8$,$1$,$8$,$2$ $\rangle$} & 
$\langle$ $17$,$\mathbf{7}$,$17$,$8$ $\rangle$ & 
\multicolumn{1}{c|}{$\langle$ $28$,$\mathbf{11}$,$29$,$12$ $\rangle$} & 
$\langle$ $590$,$\mathbf{84}$,$590$,$213$ $\rangle$ & 
$\langle$ $247$,$\mathbf{10}$,$247$,$122$ $\rangle$ & 
$\langle$ $100$,$\mathbf{33}$,$106$,$71$ $\rangle$ 
\\ \hline

% Rustify
\multirow{2}{*}{7} & 
\approach \cellcolor{gray!25} & 
\multicolumn{1}{c|}{\cellcolor{gray!25}$\mathbf{0}$} &
\cellcolor{gray!25} $\mathbf{0}$&  
\multicolumn{1}{c|}{\cellcolor{gray!25}$\mathbf{0}$} &
\cellcolor{gray!25} $\mathbf{4}$&  
\cellcolor{gray!25} $\mathbf{2}$&
\cellcolor{gray!25} $\mathbf{3}$
\\ \cline{2-8} 
% Others
& 
$\langle$ C2R,C2SR,C,L $\rangle$ & 
\multicolumn{1}{c|}{$\langle$ $52$,$1$,$5$,$10$ $\rangle$} & 
$\langle$ $85$,$81$,$78$,$74$ $\rangle$ & 
\multicolumn{1}{c|}{$\langle$ $133$,$51$,$42$,$2$ $\rangle$} & 
$\langle$ $1490$,$702$,$932$,$1002$ $\rangle$ & 
$\langle$ $804$,$450$,$706$,$1167$ $\rangle$ & 
$\langle$ $441$,$439$,$397$,$572$ $\rangle$ 
\\ \hline

% Rustify
\multirow{2}{*}{8} & 
\approach \cellcolor{gray!25} & 
\multicolumn{1}{c|}{\cellcolor{gray!25}$\mathbf{21}$} &
\cellcolor{gray!25} $\mathbf{21}$&  
\multicolumn{1}{c|}{\cellcolor{gray!25}$\mathbf{17}$} &
\cellcolor{gray!25} $\mathbf{453}$&  
\cellcolor{gray!25} $\mathbf{57}$&
\cellcolor{gray!25} $\mathbf{274}$
\\ \cline{2-8} 
% Others
& 
$\langle$ C2R,C2SR,C,L $\rangle$ & 
\multicolumn{1}{c|}{$\langle$ $125$,$115$,$125$,$117$ $\rangle$} & 
$\langle$ $29$,$29$,$29$,$29$ $\rangle$ & 
\multicolumn{1}{c|}{$\langle$ $172$,$168$,$172$,$172$ $\rangle$} & 
$\langle$ $1473$,$1467$,$1473$,$1530$ $\rangle$ & 
$\langle$ $877$,$631$,$877$,$897$ $\rangle$ & 
$\langle$ $329$,$325$,$329$,$321$ $\rangle$ 
\\ \hline

% Rustify
\multirow{2}{*}{9} & 
\approach \cellcolor{gray!25} & 
\multicolumn{1}{c|}{\cellcolor{gray!25}$\mathbf{0}$} &
\cellcolor{gray!25} $\mathbf{1}$&  
\multicolumn{1}{c|}{\cellcolor{gray!25}$\mathbf{4}$} &
\cellcolor{gray!25} $\mathbf{43}$&  
\cellcolor{gray!25} $\mathbf{8}$&
\cellcolor{gray!25} $\mathbf{21}$
\\ \cline{2-8} 
% Others
& 
$\langle$ C2R,C2SR,C,L $\rangle$ & 
\multicolumn{1}{c|}{$\langle$ $127$,$82$,$113$,$246$ $\rangle$} & 
$\langle$ $88$,$63$,$71$,$73$ $\rangle$ & 
\multicolumn{1}{c|}{$\langle$ $355$,$273$,$317$,$339$ $\rangle$} & 
$\langle$ $2140$,$994$,$1775$,$1650$ $\rangle$ & 
$\langle$ $1385$,$362$,$1317$,$1329$ $\rangle$ & 
$\langle$ $597$,$578$,$643$,$581$ $\rangle$ 
\\ \hline

% Rustify
\multirow{2}{*}{10} & 
\approach \cellcolor{gray!25} & 
\multicolumn{1}{c|}{\cellcolor{gray!25}$\mathbf{1}$} &
\cellcolor{gray!25} $\mathbf{1}$&  
\multicolumn{1}{c|}{\cellcolor{gray!25}$\mathbf{0}$} &
\cellcolor{gray!25} $\mathbf{7}$&  
\cellcolor{gray!25} $\mathbf{0}$&
\cellcolor{gray!25} $\mathbf{9}$
\\ \cline{2-8} 
% Others
& 
$\langle$ C2R,C,L $\rangle$ & 
\multicolumn{1}{c|}{$\langle$ $66$,$22$,$22$ $\rangle$} & 
$\langle$ $74$,$13$,$13$ $\rangle$ & 
\multicolumn{1}{c|}{$\langle$ $126$,$24$,$31$ $\rangle$} & 
$\langle$ $1083$,$178$,$176$ $\rangle$ & 
$\langle$ $521$,$72$,$72$ $\rangle$ & 
$\langle$ $339$,$117$,$102$ $\rangle$ 
\\ \hline

% Rustify
\multirow{2}{*}{11} & 
\approach \cellcolor{gray!25} & 
\multicolumn{1}{c|}{\cellcolor{gray!25}$\mathbf{0}$} &
\cellcolor{gray!25} $\mathbf{1}$&  
\multicolumn{1}{c|}{\cellcolor{gray!25}$\mathbf{0}$} &
\cellcolor{gray!25} $20$&  
\cellcolor{gray!25} $1$&
\cellcolor{gray!25} $8$
\\ \cline{2-8} 
% Others
& 
$\langle$ C2R,C,L $\rangle$ & 
\multicolumn{1}{c|}{$\langle$ $20$,$21$,$16$ $\rangle$} & 
$\langle$ $3$,$3$,$2$ $\rangle$ & 
\multicolumn{1}{c|}{$\langle$ $24$,$24$,$24$ $\rangle$} & 
$\langle$ $556$,$513$,$475$ $\rangle$ & 
$\langle$ $346$,$340$,$331$ $\rangle$ & 
$\langle$ $164$,$160$,$162$ $\rangle$ 
\\ \hline

% Rustify
\multirow{2}{*}{12} & 
\approach \cellcolor{gray!25} & 
\multicolumn{1}{c|}{\cellcolor{gray!25}$4$} &
\cellcolor{gray!25} $6$&  
\multicolumn{1}{c|}{\cellcolor{gray!25}$8$} &
\cellcolor{gray!25} $113$&  
\cellcolor{gray!25} $2$&
\cellcolor{gray!25} $41$
\\ \cline{2-8} 
% Others
& 
$\langle$ C2R,C,L $\rangle$ & 
\multicolumn{1}{c|}{$\langle$ $115$,$63$,$60$ $\rangle$} & 
$\langle$ $57$,$23$,$19$ $\rangle$ & 
\multicolumn{1}{c|}{$\langle$ $185$,$31$,$26$ $\rangle$} & 
$\langle$ $2857$,$799$,$873$ $\rangle$ & 
$\langle$ $1794$,$423$,$26$ $\rangle$ & 
$\langle$ $513$,$496$,$19$ $\rangle$ 
\\ \hline

% Rustify
\multirow{2}{*}{13} & 
\approach \cellcolor{gray!25} & 
\multicolumn{1}{c|}{\cellcolor{gray!25}$0$} &
\cellcolor{gray!25} $15$&  
\multicolumn{1}{c|}{\cellcolor{gray!25}$12$} &
\cellcolor{gray!25} $93$&  
\cellcolor{gray!25} $5$&
\cellcolor{gray!25} $45$
\\ \cline{2-8} 
% Others
& 
$\langle$ C2R,C,L $\rangle$ & 
\multicolumn{1}{c|}{$\langle$ $21$,$0$,$0$ $\rangle$} & 
$\langle$ $117$,$5$,$13$ $\rangle$ & 
\multicolumn{1}{c|}{$\langle$ $366$,$8$,$53$ $\rangle$} & 
$\langle$ $2019$,$84$,$91$ $\rangle$ & 
$\langle$ $1296$,$30$,$18$ $\rangle$ & 
$\langle$ $503$,$202$,$37$ $\rangle$ 
\\ \hline

% Rustify
\multirow{2}{*}{14} & 
\approach \cellcolor{gray!25} & 
\multicolumn{1}{c|}{\cellcolor{gray!25}$7$} &
\cellcolor{gray!25} $1$&  
\multicolumn{1}{c|}{\cellcolor{gray!25}$0$} &
\cellcolor{gray!25} $5$&  
\cellcolor{gray!25} $2$&
\cellcolor{gray!25} $2$
\\ \cline{2-8} 
% Others
& 
$\langle$ C2R,C $\rangle$ & 
\multicolumn{1}{c|}{$\langle$ $236$,$236$ $\rangle$} & 
$\langle$ $84$,$84$ $\rangle$ & 
\multicolumn{1}{c|}{$\langle$ $361$,$361$ $\rangle$} & 
$\langle$ $2077$,$2077$ $\rangle$ & 
$\langle$ $1389$,$6$ $\rangle$ & 
$\langle$ $580$,$8$ $\rangle$ 
\\ \hline

% Rustify
\multirow{2}{*}{15} & 
\approach \cellcolor{gray!25} & 
\multicolumn{1}{c|}{\cellcolor{gray!25}$22$} &
\cellcolor{gray!25} $8$&  
\multicolumn{1}{c|}{\cellcolor{gray!25}$4$} &
\cellcolor{gray!25} $162$&  
\cellcolor{gray!25} $25$&
\cellcolor{gray!25} $131$
\\ \cline{2-8} 
% Others
& 
$\langle$ C2R,C,L $\rangle$ & 
\multicolumn{1}{c|}{$\langle$ $198$,$199$,$199$ $\rangle$} & 
$\langle$ $95$,$78$,$76$ $\rangle$ & 
\multicolumn{1}{c|}{$\langle$ $358$,$201$,$246$ $\rangle$} & 
$\langle$ $2202$,$2134$,$2199$ $\rangle$ & 
$\langle$ $1178$,$1231$,$1241$ $\rangle$ & 
$\langle$ $625$,$1010$,$899$ $\rangle$ 
\\ \hline

% Rustify
\multirow{2}{*}{16} & 
\approach \cellcolor{gray!25} & 
\multicolumn{1}{c|}{\cellcolor{gray!25}$12$} &
\cellcolor{gray!25} $0$&  
\multicolumn{1}{c|}{\cellcolor{gray!25}$0$} &
\cellcolor{gray!25} $5$&  
\cellcolor{gray!25} $0$&
\cellcolor{gray!25} $5$
\\ \cline{2-8} 
% Others
& 
$\langle$ C2R,C $\rangle$ & 
\multicolumn{1}{c|}{$\langle$ $170$,$177$ $\rangle$} & 
$\langle$ $150$,$150$ $\rangle$ & 
\multicolumn{1}{c|}{$\langle$ $566$,$569$ $\rangle$} & 
$\langle$ $3839$,$3839$ $\rangle$ & 
$\langle$ $2375$,$1$ $\rangle$ & 
$\langle$ $622$,$2$ $\rangle$ 
\\ \hline

% Rustify
\multirow{2}{*}{17} & 
\approach \cellcolor{gray!25} & 
\multicolumn{1}{c|}{\cellcolor{gray!25}$12$} &
\cellcolor{gray!25} $6$&  
\multicolumn{1}{c|}{\cellcolor{gray!25}$1$} &
\cellcolor{gray!25} $1731$&  
\cellcolor{gray!25} $8$&
\cellcolor{gray!25} $658$
\\ \cline{2-8} 
% Others
& 
$\langle$ C2R,C,L $\rangle$ & 
\multicolumn{1}{c|}{$\langle$ $281$,$-$,$-$ $\rangle$} & 
$\langle$ $193$,$-$,$-$ $\rangle$ & 
\multicolumn{1}{c|}{$\langle$ $650$,$-$,$-$ $\rangle$} & 
$\langle$ $2903$,$-$,$-$ $\rangle$ & 
$\langle$ $1107$,$-$,$-$ $\rangle$ & 
$\langle$ $1240$,$-$,$-$ $\rangle$ 
\\ \hline

% Rustify
\multirow{2}{*}{18} & 
\approach \cellcolor{gray!25} & 
\multicolumn{1}{c|}{\cellcolor{gray!25}$3$} &
\cellcolor{gray!25} $1$&  
\multicolumn{1}{c|}{\cellcolor{gray!25}$0$} &
\cellcolor{gray!25} $8$&  
\cellcolor{gray!25} $1$&
\cellcolor{gray!25} $5$
\\ \cline{2-8} 
% Others
& 
$\langle$ C2R,C,S $\rangle$ & 
\multicolumn{1}{c|}{$\langle$ $819$,$-$,$8$ $\rangle$} & 
$\langle$ $291$,$-$,$0$ $\rangle$ & 
\multicolumn{1}{c|}{$\langle$ $1359$,$-$,$0$ $\rangle$} & 
$\langle$ $8085$,$-$,$0$ $\rangle$ & 
$\langle$ $3703$,$-$,$0$ $\rangle$ & 
$\langle$ $1924$,$-$,$0$ $\rangle$ 
\\ \hline

% Rustify
\multirow{2}{*}{19} & 
\approach \cellcolor{gray!25} & 
\multicolumn{1}{c|}{\cellcolor{gray!25}$2$} &
\cellcolor{gray!25} $8$&  
\multicolumn{1}{c|}{\cellcolor{gray!25}$0$} &
\cellcolor{gray!25} $52$&  
\cellcolor{gray!25} $17$&
\cellcolor{gray!25} $38$
\\ \cline{2-8} 
% Others
& 
$\langle$ C2R,C2SR,C,L $\rangle$ & 
\multicolumn{1}{c|}{$\langle$ $916$,$1873$,$1460$,$922$ $\rangle$} & 
$\langle$ $244$,$231$,$432$,$224$ $\rangle$ & 
\multicolumn{1}{c|}{$\langle$ $842$,$747$,$1468$,$765$ $\rangle$} & 
$\langle$ $5842$,$5583$,$10835$,$7213$ $\rangle$ & 
$\langle$ $3252$,$2867$,$69$,$5940$ $\rangle$ & 
$\langle$ $1587$,$1533$,$57$,$1772$ $\rangle$ 
\\ \hline

% Rustify
\multirow{2}{*}{20} & 
\approach \cellcolor{gray!25} & 
\multicolumn{1}{c|}{\cellcolor{gray!25}$90$} &
\cellcolor{gray!25} $27$&  
\multicolumn{1}{c|}{\cellcolor{gray!25}$17$} &
\cellcolor{gray!25} $117$&  
\cellcolor{gray!25} $11$&
\cellcolor{gray!25} $39$
\\ \cline{2-8} 
% Others
& 
$\langle$ C2R,C,L $\rangle$ & 
\multicolumn{1}{c|}{$\langle$ $2055$,$2119$,$2122$ $\rangle$} & 
$\langle$ $552$,$450$,$400$ $\rangle$ & 
\multicolumn{1}{c|}{$\langle$ $2781$,$2352$,$1886$ $\rangle$} & 
$\langle$ $13030$,$13000$,$13127$ $\rangle$ & 
$\langle$ $8632$,$8929$,$8985$ $\rangle$ & 
$\langle$ $3452$,$5069$,$5601$ $\rangle$ 
\\ \hline

% Rustify
\multirow{2}{*}{21} & 
\approach \cellcolor{gray!25} & 
\multicolumn{1}{c|}{\cellcolor{gray!25}$7$} &
\cellcolor{gray!25} $19$&  
\multicolumn{1}{c|}{\cellcolor{gray!25}$20$} &
\cellcolor{gray!25} $114$&  
\cellcolor{gray!25} $47$&
\cellcolor{gray!25} $130$
\\ \cline{2-8} 
% Others
& 
$\langle$ C2R,C2SR,R,C,L $\rangle$ & 
\multicolumn{1}{c|}{$\langle$ $1149$,$849$,$24$,$1105$,$1111$ $\rangle$} & 
$\langle$ $230$,$199$,$24$,$194$,$206$ $\rangle$ & 
\multicolumn{1}{c|}{$\langle$ $4338$,$2691$,$69$,$3674$,$3699$ $\rangle$} & 
$\langle$ $13008$,$10328$,$974$,$12592$,$13115$ $\rangle$ & 
$\langle$ $7421$,$4097$,$70$,$7672$,$8699$ $\rangle$ & 
$\langle$ $2611$,$2455$,$372$,$2929$,$2788$ $\rangle$ 
\\ \hline

% Rustify
\multirow{2}{*}{22} & 
\approach \cellcolor{gray!25} &   
\multicolumn{1}{c|}{\cellcolor{gray!25}$56$} &
\cellcolor{gray!25} $57$&  
\multicolumn{1}{c|}{\cellcolor{gray!25}$20$} &
\cellcolor{gray!25} $331$&  
\cellcolor{gray!25} $140$&
\cellcolor{gray!25} $335$
\\ \cline{2-8} 
% Others
& 
$\langle$ C2R,C,L $\rangle$ & 
\multicolumn{1}{c|}{$\langle$ $110$,$110$,$116$ $\rangle$} & 
$\langle$ $369$,$317$,$302$ $\rangle$ & 
\multicolumn{1}{c|}{$\langle$ $498$,$329$,$339$ $\rangle$} & 
$\langle$ $4163$,$4139$,$4105$ $\rangle$ & 
$\langle$ $1635$,$1572$,$1574$ $\rangle$ & 
$\langle$ $631$,$1060$,$1036$ $\rangle$ 
\\ \hline

% Rustify
\multirow{2}{*}{23} & 
\approach \cellcolor{gray!25} & 
\multicolumn{1}{c|}{\cellcolor{gray!25}$\mathbf{15}$} &
\cellcolor{gray!25} $\mathbf{7}$&  
\multicolumn{1}{c|}{\cellcolor{gray!25}$\mathbf{14}$} &
\cellcolor{gray!25} $\mathbf{34}$&  
\cellcolor{gray!25} $\mathbf{16}$&
\cellcolor{gray!25} $\mathbf{22}$
\\ \cline{2-8} 
% Others
& 
$\langle$ C2R,C2SR,C,L $\rangle$ &  
\multicolumn{1}{c|}{$\langle$ $4678$,$1215$,$1283$,$2572$ $\rangle$} & 
$\langle$ $2279$,$860$,$865$,$866$ $\rangle$ & 
\multicolumn{1}{c|}{$\langle$ $6744$,$1625$,$1847$,$1847$ $\rangle$} & 
$\langle$ $35228$,$17314$,$17769$,$25594$ $\rangle$ & 
$\langle$ $22861$,$12274$,$13195$,$13050$ $\rangle$ & 
$\langle$ $8518$,$2638$,$2419$,$2595$ $\rangle$ 
\\ \hline

\end{tabular}
}

%% file: Listings/rustinepointerinflation.tex
\begin{tabular}{c  c }
\begin{minipage}[t][][t]{0.47\textwidth}
\begin{minted}[frame=lines,framesep=1.1mm,baselinestretch=0.47, fontsize=\scriptsize,
        breaklines, breakanywhere, linenos,numbersep=2pt,
        highlightlines={5}]{c}
              -- bzip2 Source Code in C --
         
void BZ_API(BZ2_bzclose) (BZFILE* b) {
    // ... omitted code
    fp = ((bzFile *)b)->handle;
    // ... omitted code
}
    \end{minted}
\end{minipage}
&
\begin{minipage}[t][][t]{0.47\textwidth}
\begin{minted}[frame=lines,framesep=1.1mm,baselinestretch=0.47, fontsize=\scriptsize,
        breaklines, breakanywhere, linenos,numbersep=2pt,
        highlightlines={3}]{c}
            -- libtree Source Code in C --
            
int in_exclude_list = soname != MAX_OFFSET_T && is_in_exclude_list(s->string_table.arr[soname_buf_offset]);
    \end{minted}
\vspace{-3pt}
\end{minipage}
\\
\begin{minipage}[t][][t]{0.47\textwidth}
\vspace{-5pt}
   \begin{minted}[frame=lines,framesep=1mm,baselinestretch=0.47, fontsize=\scriptsize,
        breaklines, breakanywhere, linenos,numbersep=2pt,
        highlightlines={5,6,7}]{rust}
        -- Rustine Translation of bzip2 --

pub fn BZ2_bzclose(b: Option<&mut BZFILE>) {
    // ... omitted code
    let fp = unsafe { 
        *bz_fp.offset(fp_idx as isize)
    };
    // ... omitted code
}
    \end{minted}
\end{minipage}
&
\begin{minipage}[t][][t]{0.47\textwidth}
\vspace{-16pt}
       \begin{minted}[frame=lines,framesep=1mm,baselinestretch=0.47, fontsize=\scriptsize,
        breaklines, breakanywhere, linenos,numbersep=2pt,
        highlightlines={4,5,8}]{rust}
        -- Rustine Translation of libtree --       
let in_exclude_list = soname != u64::MAX && {
    let soname_str = unsafe {
        CStr::from_ptr(s.string_table.arr.as_ref()
        .unwrap().as_ptr()
        .add(soname_buf_offset) as *const i8)
    };
    is_in_exclude_list(Some(soname_str.to_str().unwrap())) != 0
}
    \end{minted}
\end{minipage}

\end{tabular}

%% file: Listings/c2rustpointerinflation.tex
\begin{tabular}{c}
% \toprule
% {\tiny c2rust translation of xzoom code. \vspace{-1mm}} \\
% \hline \\
\begin{minipage}[t][][t]{0.48\textwidth}
\begin{minted}[frame=lines,framesep=1.1mm,baselinestretch=0.47, fontsize=\scriptsize,
        breaklines, breakanywhere, linenos,numbersep=2pt,
        highlightlines={4,5}]{c}
             -- xzoom Source Code in C --
             
(T *)(&ximage[t]
    ->data[(ximage[t]->xoffset + (x)) * sizeof(T) +
        (y)*ximage[t]->bytes_per_line])
    \end{minted}
\vspace{-8pt}
\end{minipage}
\\
\begin{minipage}[t][][t]{0.48\textwidth}
    \begin{minted}[frame=lines,framesep=1mm,baselinestretch=0.47, fontsize=\scriptsize,
        breaklines, breakanywhere, linenos,numbersep=2pt,
        highlightlines={6}]{rust}
           -- C2Rust Translation of xzoom --

&mut *((**ximage.as_mut_ptr().offset(
            1 as libc::c_int as isize))
           .data)
     .offset((((**ximage.as_mut_ptr().offset(
            1 as libc::c_int as isize))
            .xoffset +
        0 as libc::c_int) as libc::c_ulong)
            .wrapping_mul(
                ::core::mem::size_of::<libc::c_uchar>()
                    as libc::c_ulong, )
            .wrapping_add(
                (j * magy *
                (**ximage.as_mut_ptr().offset(
                    1 as libc::c_int as isize))
                    .bytes_per_line)
                    as libc::c_ulong, ) as isize, )
    as *mut libc::c_char as *mut libc::c_uchar;
    \end{minted}
% \begin{lstlisting}[]
% // RUST SOURCE CODE
% &mut *((**ximage.as_mut_ptr().offset(
%             1 as libc::c_int as isize))
%            .data)
%      .offset((((**ximage.as_mut_ptr().offset(
%                     1 as libc::c_int as isize))
%                    .xoffset +
%                0 as libc::c_int) as libc::c_ulong)
%                  .wrapping_mul(
%                      ::core::mem::size_of::<libc::c_uchar>()
%                          as libc::c_ulong, )
%                  .wrapping_add(
%                      (j * magy *
%                       (**ximage.as_mut_ptr().offset(
%                            1 as libc::c_int as isize))
%                           .bytes_per_line)
%                          as libc::c_ulong, ) as isize, )
%          as *mut libc::c_char as *mut libc::c_uchar;
% \end{lstlisting}
\end{minipage}

\end{tabular}

%% file: Listings/perf-warning-example.tex
\begin{tabular}{c}
\begin{minipage}[t][][t]{0.48\textwidth}
\begin{minted}[frame=lines,framesep=1.1mm,baselinestretch=0.47, fontsize=\scriptsize,
        breaklines, breakanywhere, linenos,numbersep=2pt,
        highlightlines={4}]{c}
              -- buffer Source Code in C --         
buffer_t *buffer_new_with_copy(char *str) {
  // ... omitted code
  memcpy(self->alloc, str, len);
    \end{minted}
\vspace{-10pt}
\end{minipage}
\\
\begin{minipage}[t][][t]{0.48\textwidth}
    \begin{minted}[frame=lines,framesep=1mm,baselinestretch=0.47, fontsize=\scriptsize,
        breaklines, breakanywhere, linenos,numbersep=2pt,
        highlightlines={5}]{rust}
         -- C2Rust Translation of buffer --
pub unsafe extern "C" fn buffer_new_with_copy(
    mut str: *mut libc::c_char, ) -> *mut buffer_t {
  // ... omitted code
  memcpy((*self_0).alloc as * mut libc::c_void, str as *const libc::c_void, len);
    \end{minted}
    \vspace{-10pt}
\end{minipage}
\\
\begin{minipage}[t][][t]{0.48\textwidth}
    \begin{minted}[frame=lines,framesep=1mm,baselinestretch=0.47, fontsize=\scriptsize,
        breaklines, breakanywhere, linenos,numbersep=2pt,
        highlightlines={6-10}]{rust}
           -- Rustine Translation of buffer --           
pub fn buffer_new_with_copy(str: Option<&str>)
    -> Option<BufferT> {
  // ... omitted code
  let to_copy = std::cmp::min(alloc.len(), src.len());
  for i in 0..to_copy { 
    alloc[i] = src[i]; 
  }
CLIPPY: ^ help: try replacing the loop by: 
"alloc[..to_copy].copy_from_slice(&src[..to_copy]);"
    \end{minted}
\end{minipage}
\end{tabular}

%% file: Sections/Evaluation-RQ2.tex
\subsection{RQ2. Effectiveness of the Refactoring}
\label{subsec:rq-ablation}

An important contribution of \approach is considering raw pointers, and specifically pointer operations, as first-class citizens.
%in the translation pipeline. 
As we mentioned, even when pointer operations do not exist in the C program, both transpiler- and LLM-based techniques result in Rust translations with pointer arithmetic/comparisons. But when they exist, the inflation explodes in translations. To investigate this observation in a more controlled setting, we excluded the pointer arithmetic refactoring component of \approach and retranslated the translation units involving pointer operations.

\begin{figure}[t]
    \centering
    \scriptsize 
    \includegraphics[width=\linewidth]{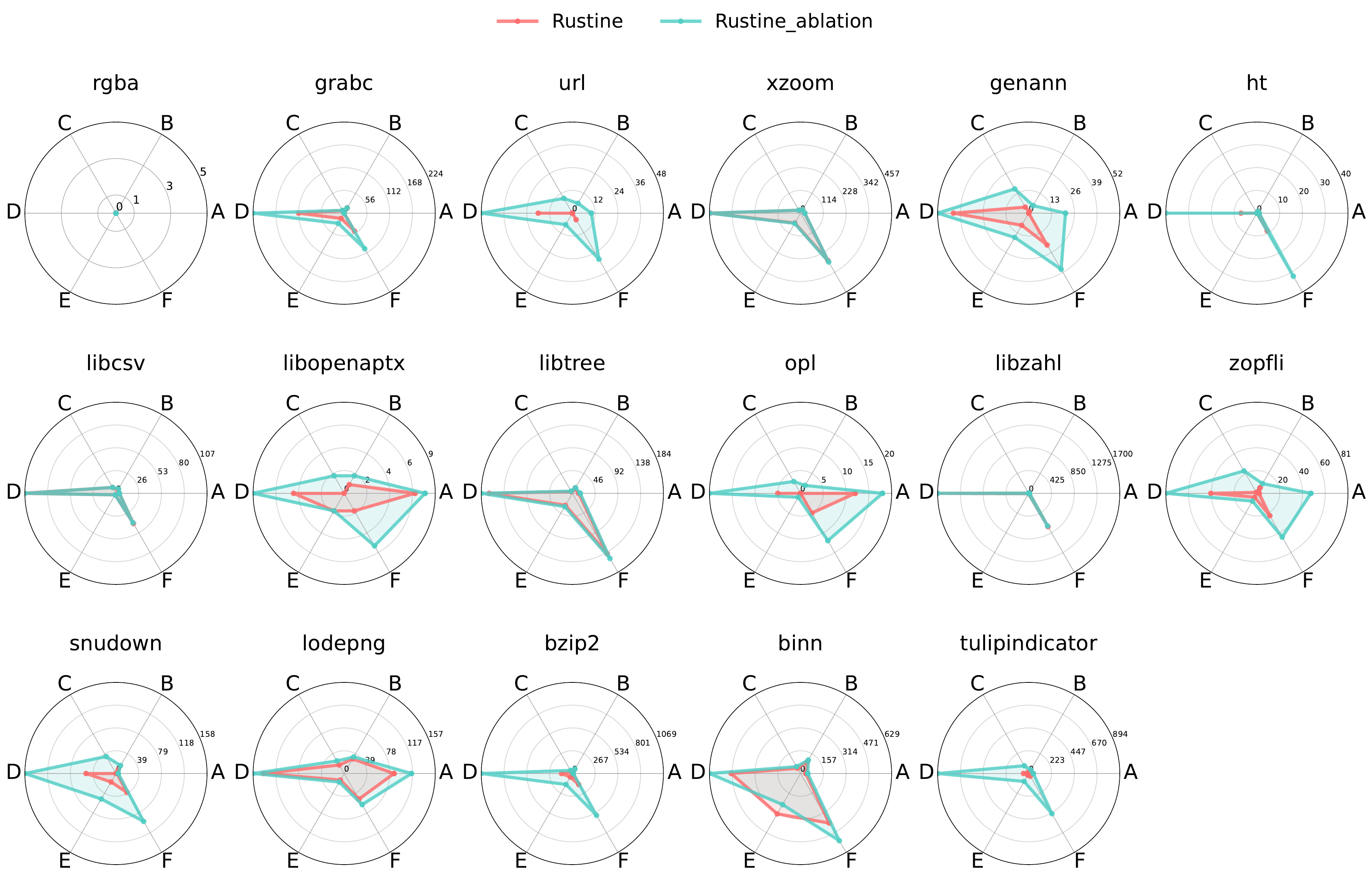}
    \vspace{-0.4cm}
    \caption{Impact of removing pointer arithmetic resolution before translation. Comparison dimenstions are: A=Ptr arithmetic, B=Ptr decleration, C=Ptr derefrence, D=Unsafe lines, E=Unsafe type cast, F=Unsafe calls}
    \label{fig:ablation}
    %\vspace{-0.8cm}
\end{figure}

Figure~\ref{fig:ablation} corroborates 
\textbf{a significant rise in not only pointer operation cases, but also raw pointer declarations/dereferences, and unsafe constructs}. On average, the number of pointer arithmetic increases by $83.01\%$, raw pointer declarations/dereferences by $52.6\%$/$154\%$, and unsafe $\langle$lines,type casts,calls$\rangle$ by $\langle 64.42\%,31.67\%,61.46\% \rangle$. These results, along with examples shown before (Figure~\ref{fig:challenge1}), confirm that the \textbf{refactoring component is critical in \approach to minimize safety issues} in Rust translation: in the absence of pointer arithmetic refactoring, direct pointer manipulations are preserved, or even worse, will be inflated during translation. 

%77.12\%.

%The results of this experiment are presented in Figure \ref{fig:codeq}. The 12 selected subjects in Figure \ref{fig:codeq} visually compare the output of \approach (red polygon) against the ablation version (teal polygon) across several key metrics.
%A clear and consistent pattern emerges: the ablation results are substantially larger across all unsafe-related axes, indicating a marked regression in translation quality when refactoring is disabled. 
%Figure\ref{fig:refactoring} illustrates impact of this refactoring with a concrete example.
%These results unequivocally demonstrate the effectiveness of our pointer arithmetic refactoring. 
%\tianran{Do we need the figure for ablation? We already have the table...}

\begin{comment}
\begin{table*}[t]
    \scriptsize
    \centering
    \caption{Importance of pointer arithmetic refactoring in \approach. Abbreviations in the table stand for \textbf{RP}: Raw Pointers. 
    \reyhan{this table and figure 13 show the same information. I prefer figure over table, but that figure is not readable. You need to move it to drawings and adjust the font there}
    }
    \vspace{-10pt}
    \input{Tables/Ablation-Pointer}
    \vspace{-10pt}
    \label{table:ablation}
\end{table*}
\end{comment}

%% file: Sections/Evaluation-RQ3.tex
\subsection{RQ3. Translation Bugs}
\label{subsec:rq-taxonomy}

We discuss examples of translation bugs due to improper treatment of \emph{dynamic memory management}, \emph{global variables}, and \emph{Rust APIs}, which were resolved through automated or manual debugging. 

\subsubsection{Dynamic Memory Management.}

In C, manual dynamic memory management is unavoidable due to the absence of high-level abstractions. Developers must explicitly track allocation sizes, alignments, and pointer validity.
\approach often mimics this low-level pattern in translations. However, due to the inherent complexity of manual memory management, translations are often incorrect or unsafe (pointer lifetime after reallocation or failing to update references to the data).

The code below is an example from \emph{buffer} project, where the C code uses \texttt{\small{realloc}} to double the size of a dynamically growing buffer. \approach follows the same pattern, but fails to correctly handle memory alignment and pointer validity across the program.
Due to the complexity of Rust’s memory management mechanisms, automated debugging cannot fix the issue. The developer resolved the issue by replacing raw pointers and \texttt{realloc} with \texttt{\small{Vec<T>}} directly in the struct layout, eliminating the need for manual memory management.

\vspace{3pt}
\input{Listings/dyn}
\vspace{3pt}

In many cases, when provided with sufficient contextual prompting, \approach defines \texttt{\small{Vec<T>}} in struct layouts and APIs or avoids manual allocation, resulting in compilable, idiomatic, and logically correct Rust code.
However, even when using \texttt{\small{Vec<T>}}, \approach may struggle to translate the C memory allocation into proper Rust equivalent, especially when Rust-specific APIs are involved. As shown in the example below, \approach initially translated the C \texttt{\small{realloc}} usage to \texttt{\small{resize}}, which directly increases the vector length, resulta ing in semantic mismatch and assertion failure. During the automated debugging process, \approach successfully patched the semantic mismatch by using \texttt{\small{reserve}}, which expands the capacity without changing the length.

\vspace{5pt}
\noindent
\input{Listings/memresize}
\vspace{3pt}

\subsubsection{Global Variables.}

The LLM may struggle with initialization and the use of global variables due to strict ownership rules. Despite explicit guidance (context about global variables and in-context examples) to use \texttt{\small{lazy\_static}} for handling global state in Rust, the LLM sometimes failed to perform proper initialization.
Consider the example below from \emph{tulipindicator}, where the C code has a global array of \texttt{\small{struct ti_indicator_info}}. \approach initial translation uses unsafe zero-initialization instead of properly populating the structure fields. The manually debugged version properly initializes each field with appropriate values.

\vspace{3pt}

\input{Listings/global}
\vspace{3pt}

\approach may fail to properly handle locking when accessing global variables. In Rust, locks acquired through \texttt{\small{Mutex::lock()}} are held until the guard goes out of scope, which can lead to deadlocks if not managed carefully. \approach sometimes failed to recognize when locks should be explicitly released, particularly in scenarios when variable accessed by multiple function calls within the same scope. The example below shows such a case in \emph{robotfindskitten} prohect, where the global variable \texttt{\small{robot}} is accessed but never released for subsequent function calls reading it. In the manually fixed version, the developer explicitly drops the lock and solves the deadlock issue.

\vspace{3pt}

\input{Listings/lock}
\vspace{3pt}

% \begin{wrapfigure}{r}{0.45\columnwidth}
% \vspace{-10pt}
% \input{Listings/lock}
% \vspace{-2mm}
% \caption{An example of lock mishandling when accessing global variables }
%     \label{code:lock}
%     \vspace{-3mm}
% \end{wrapfigure}

%\reyhan{we need examples for both initialization failure and removing lock handling mechanisms. I also don't understand the last sentence: what do you mean that it omits them? Do you mean they exist in C code but LLM ignores them in translation? One important thing here is difference between concurrency management in Rust. How this can relate to it?}
%\reyhan{note to self: add referecnes to the approach section when we discuss global variables and handling of them}

\subsubsection{API Misuse.}
API invocations in translations are either due to mapping the C APIs into their equivalents in Rust, or introducing a Rust API to implement a C code with no API, likely to generate a more idiomatic Rust. \approach is pretty successful in the former, due to pre-translation analysis (\S \ref{subsub:API-invocation}) and use of adaptive ICL examples (\S \ref{subsub:adaptive-ICL}). However, we observed several cases where the LLM's creativity in using APIs was incorrect in the initial translation. The example below shows such a case, where \approach implemented the reverse byte copying in C by combining \texttt{\small{to\_be\_bytes()}}, to convert number to \emph{big-endian} bytes, followed by \texttt{\small{from\_be\_bytes()}}, to convert \emph{big-endian} bytes to number, which, in fact, cancels the conversion. Automated debugging of \approach resolves the issue with a proper API usage, \texttt{\small{to\_ne\_bytes()}} and \texttt{\small{from\_ne\_bytes()}}.

\vspace{3pt}
\input{Listings/binn-debug}

%% file: Listings/dyn.tex
\begin{minipage}{.47\linewidth}
\begin{minted}[frame=lines,framesep=1mm,baselinestretch=0.47, fontsize=\scriptsize,
    breaklines, breakanywhere, linenos,numbersep=2pt,
    highlightlines={10}]{rust}
              -- Rustine Translation --

v.capacity *= 2;
let layout = Layout::array::<u64>(v.capacity).unwrap();
let old_ptr = match & v.p {
    Some(boxed) = > Box::into_raw(boxed.clone()) as *mut u64,
    None = > unreachable !(),
};

let ptr = unsafe{std::alloc::realloc(old_ptr as * mut u8, layout, v.capacity *std::mem::size_of::<u64>()) as * mut u64};
\end{minted}
\end{minipage}\hfill
\begin{minipage}{.46\linewidth}
\begin{minted}[frame=lines,framesep=1.1mm,baselinestretch=0.47, fontsize=\scriptsize,
        breaklines, breakanywhere, linenos,numbersep=2pt,
        highlightlines={4}]{c}
         -- buffer Source Code in C --
         
v->capacity *= 2;
uint64_t *p = realloc(v->p, v->capacity * sizeof(uint64_t));
... //omitted code
v->p = p;
    \end{minted}
    \vspace{-7pt}
\begin{minted}[frame=lines,framesep=1.1mm,baselinestretch=0.47, fontsize=\scriptsize,
    breaklines, breakanywhere, linenos,numbersep=2pt,
    highlightlines={5}]{rust}
         -- Manually Debugged Translation --
    
v.capacity *= 2;
if let Some(ref mut inner_vec) = v.p {
    inner_vec.resize(v.capacity, 0);
} 
\end{minted}
\end{minipage}\hfill

%% file: Listings/memresize.tex
\begin{minipage}{.48\linewidth}
    % --- C Source ---
    \begin{minted}[frame=lines,framesep=1mm,baselinestretch=0.47, fontsize=\scriptsize,
        breaklines, breakanywhere, linenos,numbersep=2pt,
        highlightlines={2}]{c}
           -- buffer Source Code in C --        
char *arr = realloc(t->arr, t->capacity * (sizeof(char)));
    \end{minted}
    \vspace{-8pt}
    % --- Rustine Translation ---
    \begin{minted}[frame=lines,framesep=1mm,baselinestretch=0.47, fontsize=\scriptsize,
        breaklines, breakanywhere, linenos,numbersep=2pt,
        highlightlines={4}]{rust}
            -- Rustine Translation --
match &mut t.arr {
    Some(arr) => {
        arr.resize(t.capacity, '\0');
    },
    None => {
        t.arr = Some(vec!['\0'; t.capacity]);
    }
}
    \end{minted}
\end{minipage}\hfill
\begin{minipage}{.48\linewidth}
    % --- Rustine Debug ---
    \begin{minted}[frame=lines,framesep=1mm,baselinestretch=0.47, fontsize=\scriptsize,
        breaklines, breakanywhere, linenos,numbersep=2pt,
        highlightlines={6}]{rust}
              -- Rustine Debugging --

match &mut t.arr {
    Some(arr) => {
        let current_capacity = arr.capacity();
        if t.capacity > current_capacity {
        arr.reserve(t.capacity - current_capacity);
        }
    },
    None => {
        t.arr = Some(Vec::with_capacity(t.capacity));
    }
}
    \end{minted}
\end{minipage}

%% file: Listings/global.tex
\begin{minipage}{.47\linewidth}
\begin{minted}[frame=lines,framesep=1.1mm,baselinestretch=0.47, fontsize=\scriptsize,
    breaklines, breakanywhere, linenos,numbersep=2pt,
    highlightlines={7-14}]{rust}
         -- Manually Debugged Translation --

lazy_static ! {
 pub static ref TI_INDICATORS
    : Mutex<Vec<TiIndicatorInfo>> = {
     let mut indicators = Vec::new ();
     indicators.push(
       TiIndicatorInfo {
        name: Some("abs".to_string()),
        full_name: Some("Vector Absolute Value".to_string()),
        start: Some(Box::new(|slice: &[f64]|
            { ti_abs_start(Some(slice))})
            as Box<dyn Fn(&[f64]) -> i32>),
        indicator: ti_abs,
        ... //omitted code
      }

\end{minted}
\end{minipage}\hfill
\begin{minipage}{.46\linewidth}
\begin{minted}[frame=lines,framesep=1.1mm,baselinestretch=0.47, fontsize=\scriptsize,
        breaklines, breakanywhere, linenos,numbersep=2pt,
        highlightlines={3}]{c}
         -- tulipindicator Source Code in C --
struct ti_indicator_info ti_indicators[] = {
    { "abs", "Vector Absolute Value", ti_abs_start, ti_abs, 0, TI_TYPE_SIMPLE, 1, 0, 1, { "real", 0 }, { "", 0 }, { "abs", 0 }, 0, 0, 0 },
 ... //omitted code
    \end{minted}
    \vspace{-8pt}
\begin{minted}[frame=lines,framesep=1mm,baselinestretch=0.47, fontsize=\scriptsize,
    breaklines, breakanywhere, linenos,numbersep=2pt,framesep=1mm,
    highlightlines={7}]{rust}
              -- Rustine Translation --
lazy_static ! {
 pub static ref TI_INDICATORS
  : Mutex<Vec<TiIndicatorInfo>> = {
   let mut indicators = Vec::new ();
   let empty_indicator = unsafe{ zeroed::<TiIndicatorInfo>() };
   indicators.push(empty_indicator);
   Mutex::new (indicators)
};}
\end{minted}
\end{minipage}\hfill

%% file: Listings/lock.tex
\begin{minipage}{.47\linewidth}
\begin{minted}[frame=lines,framesep=1mm,baselinestretch=0.47, fontsize=\scriptsize,
    breaklines, breakanywhere, linenos,numbersep=2pt,
    highlightlines={6}]{rust}
              -- Rustine Translation --
pub fn process_input(input: i32) {
    let mut robot = ROBOT.lock().unwrap();
    let mut check_x = robot.as_ref().unwrap().x;
    let mut check_y = robot.as_ref().unwrap().y;
    // Lock is still held here!
    // Following function calls will read the global variable
    // But they will face a deadlock because lock is not released

    ... //omitted code
}

\end{minted}
\end{minipage}\hfill
\begin{minipage}{.46\linewidth}
\begin{minted}[frame=lines,framesep=1.1mm,baselinestretch=0.47, fontsize=\scriptsize,
        breaklines, breakanywhere, linenos,numbersep=2pt,
        highlightlines={5}]{c}
         -- robotfindskitten Source Code in C --
void process_input(int input) {
    int check_x = robot.x;
    int check_y = robot.y;
    ... //Following function calls will read robot
    \end{minted}
    \vspace{-8pt}
\begin{minted}[frame=lines,framesep=1.1mm,baselinestretch=0.47, fontsize=\scriptsize,
    breaklines, breakanywhere, linenos,numbersep=2pt,framesep=1mm,
    highlightlines={3,7}]{rust}
         -- Manually Debugged Translation --
pub fn process_input(input: i32) {
    let mut robot = ROBOT.lock().unwrap();
    let mut check_x = robot.as_ref().unwrap().x;
    let mut check_y = robot.as_ref().unwrap().y;
    // explicitly drop the lock
    drop(robot);
    ... //omitted code
\end{minted}
\end{minipage}\hfill

%% file: Listings/binn-debug.tex
\begin{minipage}{.48\linewidth}

    % --- C Source ---
    \begin{minted}[
        frame=lines,
        framesep=1mm,
        baselinestretch=0.47,
        fontsize=\scriptsize,
        breaklines,
        breakanywhere,
        linenos,
        numbersep=2pt,
        highlightlines={5,6,7,8}
    ]{c}
            -- binn Source Code in C --
static void copy_be32(u32 *pdest, u32 *psource) {
  unsigned char *source = (unsigned char *) psource;
  unsigned char *dest = (unsigned char *) pdest;
  dest[0] = source[3]; dest[1] = source[2];
  dest[2] = source[1]; dest[3] = source[0];
    \end{minted}

    \vspace{-8pt}

    % ----- Rustine Debug -----
    \begin{minted}[
        frame=lines,
        framesep=1mm,
        baselinestretch=0.47,
        fontsize=\scriptsize,
        breaklines,
        breakanywhere,
        linenos,
        numbersep=2pt,
        highlightlines={2,5}
    ]{rust}
              -- Rustine Debugging --
  let source = psource.to_ne_bytes();
  let mut dest_bytes = [0u8; 4];
  for i in 0..4 {dest_bytes[i] = source[3 - i];}
  *pdest = u32::from_ne_bytes(dest_bytes);
    \end{minted}

\end{minipage}
\hspace{0.01\linewidth}
\begin{minipage}{.48\linewidth}
    % --- Rustine Translation ---
    \begin{minted}[
        frame=lines,
        framesep=1mm,
        baselinestretch=0.47,
        fontsize=\scriptsize,
        breaklines,
        breakanywhere,
        linenos,
        numbersep=2pt,
        highlightlines={4,6-11}
    ]{rust}
             -- Rustine Translation --

pub fn copy_be32(pdest: &mut u32, psource: &u32) {
  let source = psource.to_be_bytes();
  let dest = &mut *pdest;
  *dest = u32::from_be_bytes([
     source[3],
     source[2],
     source[1],
     source[0]
  ]);
}
    \end{minted}
\end{minipage}

%% file: Sections/Evaluation-RQ4.tex
\subsection{RQ4. Performance and Cost}
\label{subsec:rq-performance}

\begin{figure}[t]
    \centering
    %\vspace{-10pt}
    \includegraphics[width=0.94\linewidth]{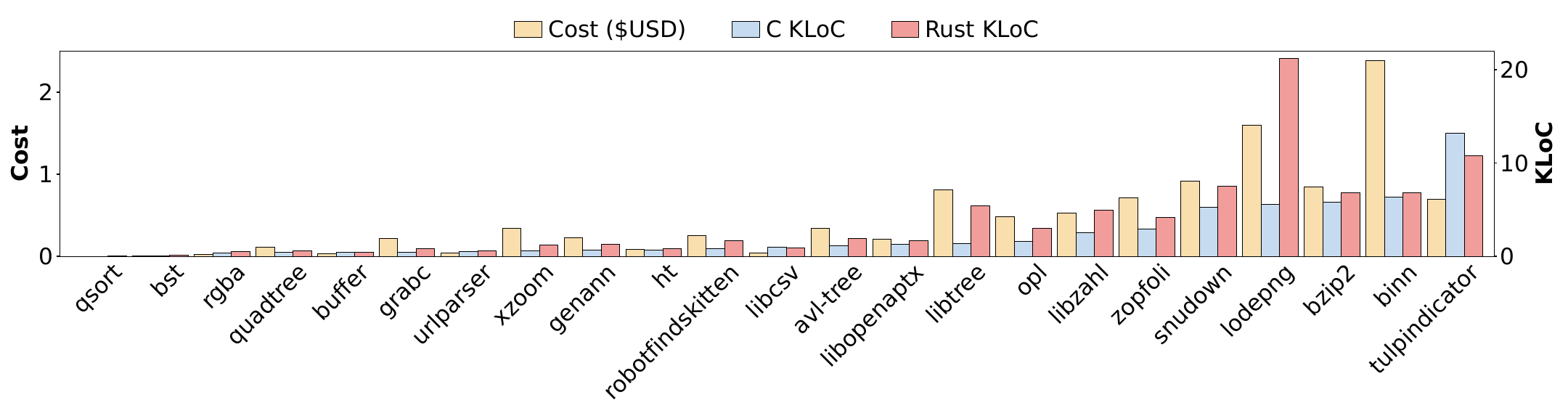}
    \vspace{-10pt}
    \caption{Subject projects' size (Rust and C), and corresponding translation costs}
    %\vspace{-10pt}
    \label{fig:cost}
\end{figure}

The cost of translation using \approach ranged from $\$0.0026$ to $\$2.39$ (average=$\$0.48$), \textbf{demonstrating cost-efficiency}. The cost of automated debugging is negligible compared to actual translation, as the LLM attempts fixing 
%focuses only on fixing 
a few suspicious functions. For example, in the \emph{libtree} project, translation and debugging cost $\$0.82$ and $\$0.049$ ($6\%$ of the total cost), respectively.
%accounting for roughly $6\%$ of the total translation cost. 
\textbf{There is a significant correlation between the translation cost and project size} in C ($\rho=0.90$ with strong statistical significance, p-value$=4.96e-9<0.05$) and Rust ($\rho=0.95$ with strong statistical significance, p-value$=1.64e-12<0.05$). Besides the size, the complex logic and specific program properties, e.g., heavy use of global variables or dependencies, can also contribute to the cost of translation.

Given that the cost depends on the models, and most of the alternative approaches used GPT-4o in their experiments for translations, we also estimated the price based on the total number of input and output tokens in our experiments. 
%Figure~\ref{fig:cost} demonstrates the estimated cost of using GPT-4o in our experiments. 
In contrast to Syzygy~\cite{shetty2024syzygy}, which reported around $\$800$ for translating the single project \emph{zopfli}, the estimated cost for \approach would be $\$10.92$ (actual cost of $\$0.71$). The highest translation cost using GPT-4o would be $\$37$ (actual cost of $\$2.39$), showing \approach potential to scale to large real-world repositories without incurring high costs.

The translation process dominates the overall runtime, ranging from $1$ minute to $20$ hours (average = $3.92$ hours), with other components (\emph{Pre-processing and Refactoring} and \emph{Analysis and Decomposition}) taking less than a few seconds, \textbf{demonstrating scalability of \approach}. 
%i.e., \emph{Pre-processing and Refactoring} and \emph{Analysis and Decomposition} take less than a few seconds. The translation time ranges from $1$ minute to $20$ hours (average = $3.87$ hours), \textbf{demonstrating scalability of \approach}. 
The Spearman Rank Order Correlation test shows a \emph{strong correlation} between the translation time and project size in C ($\rho=0.88$ with strong statistical significance, p-value$=2.41e-8<0.05$), corroborating that translation of bigger projects takes more time. 

%There is no correlation between the translation cost and the size of projects in C and Rust, with 
%{\color{blue}{$\rho = 0.88$}}, {\color{red}{$p = 2.41\times10^{-8}$}} 
%and 
%{\color{blue}{$\rho = 0.94$}}, {\color{red}{$p = 2.21\times10^{-11}$}}.

%Moreover, our low cost is not solely due to the lower API price. Even when recalculated using GPT-4o’s unit price, the project with the highest translation cost would require only \$37, further reinforcing the effectiveness of our approach in maximizing prompt utilization, providing informative ICL examples, and minimizing redundant iterations.

%\tianran{Should I put the figure here to visualize the cost for each project?}
%Regarding the cost of post-translation debugging, since the debugging process is relatively lightweight and focuses only on suspicious functions, its cost is almost negligible compared to translation. For example, in the \emph{libtree} project, translation costs \$0.82, while debugging costs only about \$0.049, accounting for roughly 6\% of the total translation cost. Given that the translation itself incurs only minimal cost, this further reinforces the cost-effectiveness of the overall \approach framework.

%\subsubsection{Refactoring and Translation Performance}

%The relatively long duration is mainly due to continuous API calls, which are sensitive to network conditions, but this can be mitigated by running the translation locally. Moreover, the current translation is executed sequentially, and performance can be further improved by enabling parallel translation of independent functions.

%% file: Sections/RelatedWork.tex
\vspace{-5pt}
\section{Related Work}
\label{sec:related-work}

Automated code translation techniques can be categorized as rule-based techniques or learning-based approaches. 
For <C,Rust> translation pairs, transpilers such as \ctorust~\cite{ctorust}, Corrode~\cite{corrode}, \crown~\cite{zhang2023ownership}, and \laertes~\cite{emre2021translating} rely on Clang-based parsing and systematic rewrites. These approaches prioritize preserving semantics but often yield non-idiomatic Rust due to the overuse of unsafe constructs.
%and the neglect of ownership abstractions. 
%Translation frameworks for other programming language pairs, such as <Java,C\#>~\cite{java2csharp,sharpen} and <C,Go>~\cite{c2go}, share the same limitation of sacrificing idiomaticity for syntactical correctness or maintaining semantic equivalence.

The emergence of neural networks and LLMs has greatly advanced automated code translation. Neural machine translation methods~\cite{nguyen2013lexical,nguyen2014migrating,nguyen2015divide,chen2018tree} have been shown to effectively automate translation between different programming language pairs, such as <Java,C\#>, <Python,JavaScript>, <Java,Python>, and <C++,Java>, although poorly scaled to real-world repositories. 
Pan et al.~\cite{pan2024lost} demonstrated the possible use of LLMs in automating code translation, while highlighting the challenges in real-world code translation. Vert~\cite{yang2024vert} utilized LLMs for verified translation of small program slices from real-world C projects to Rust, and AlphaTrans~\cite{ibrahimzada2025alphatrans} proposed a scalable technique for  repository-level code translation from Java to Python. Subsequent work showed that ideas of compositional or skeleton-guided translation are effective for repository-level C to Rust translation~\cite{nitin2025c2saferrust,shetty2024syzygy,sim2025large,wang2025evoc2rust,yuan2025project,cai2025rustmap,zhang2025scalable,wang2025effireasontrans,hong2025forcrat,bai2025rustassure,zhou2025llm,wang2025program,ibrahimzada2025matchfixagent,luo2025integrating,li2025adversarial,guan2025repotransagent,eniser2024towards}.

Prior work on translating real-world C projects to Rust suffers from (1) the generation of non-idiomatic Rust code, (2) a lack of scalability due to high cost, (3) limited debugging support, and insufficient translation validation.
%of assertions, and coverage of the tests, failing to determine the bias towards success given low quality test suites. 
%Several related work, while report the LoC for the entire projects and claim translation of entire repository, only end up translating a subset of code. 
Assessment of safety in most related work remains at measuring the number of raw pointers.
%, without proper analysis of different quality aspects. 
LLM-based techniques, specifically agentic translations, rarely report translation costs, making it impossible to assess the applicability of the developed techniques. \approach, in contrast, automatically translates C projects to fully compilable, safe, and idiomatic Rust at a reasonable cost and with a high level of semantic equivalence. Due to the high quality of translations and debugging support, the human developer quickly fixed semantic mismatches. 
%In terms of idiomaticity and readability, \approach surpasses existing techniques given a large-scale experimental setting. 

There have also been several attempts to construct C to Rust translation benchmarks~\cite{khatry2025crust,ou2024repository,xue2025new,xue2025classeval,chang2025code}. We were not able to use these benchmarks because of (1) quality issues in the artifacts, e.g., not being able to compile or build, or lack of representativeness of challenges in C to Rust translation, e.g., excessive use of raw pointers and pointer arithmetic; or (2) not being able to run LLM-based alternative approaches on them due to technical issues or high cost of evaluation. 

%% file: Sections/Conclusion.tex
\vspace{-3pt}
\section{Threats to the Validity}
\label{sec:threats}

\textbf{External Validity.}
To ensure generalizability of the tool design and applicability in practice, we evaluated \approach on $23$ projects that were translated by recent automated C to Rust translation techniques. The size of these projects ranged from $27$ to \num{13200} LoCs, and their implementations contained a diverse set of C-specific features that could challenge translation to Rust, such as excessive raw pointer declarations and dereferences, global mutable variables, micros, and pointer arithmetic. Concerning the generalizability of claims, we compared \approach with several transpilation or LLM-based techniques, with the results consistently showing the superiority of \approach. %We have also translated two widely used C projects, mimalloc~\cite{mimalloc} and littlefs~\cite{littlefs}, with no compilation error and to pure compilable idiomatic Rust to show usability of \approach in practice.

\vspace{3pt}
\noindent \textbf{Internal Validity.} One threat to the internal validity of code translation results could be inflated functional equivalence success. To account for such bias caused by limited, low coverage tests, we manually augmented the existing test suite of subjects with diverse unit tests, including high-quality assertions.
%improving the function and line coverage by \hl{$...\%$} and \hl{$...\%$}, respectively. 
We evaluated the quality of translations using existing and new metrics, providing all the details for transparency. Our artifacts are publicly available~\cite{website}. 
%\reyhan{please update the highlighted numbers}

\vspace{3pt}
\noindent \textbf{Construct Validity.} We obtained the subjects from the artifacts of prior techniques to ensure \approach translates the same version of a code translated by other approaches. Through careful manual investigation of alternative approaches artifacts, we identified several bugs in calculating the metrics or targeted suppression of Clippy rules. We fixed those issues and used widely-used tools for calculating the metrics on their translation, as explained in \S \ref{subsec:eval-setup}

\vspace{-3pt}
\section{Conclusion}
\label{sec:conclusion}

This paper proposes \approach for scalable and efficient translation of C repositories into safe, idiomatic Rust. 
%\approach is a neuro-symbolic technique, i.e., it combines sound program analysis techniques such as pointer analysis, dependency analysis, and refactoring, as well as AI concepts such as adaptive in-context learning and RAG, to produce high-quality Rust translations. An extensive evaluation of \approach on $23$ C projects, ranging from $27$ to \num{13200} LoCs, demonstrates the effectiveness and efficiency of \approach compared to six alternative approaches in repository-level C to Rust translation. 
\approach is an important and essential first step to advance the research in repository-level code translation: it generates a fully compilable idiomatic Rust project as the baseline, helping future research to focus on resolving complex issues concerning C to Rust translation, such as concurrency. As the next step, we aim to ensure semantic equivalence in translation in the presence of concurrent implementation in C projects. This requires technical advancements in the translation as well as validation of concurrent behavior under two different concurrency systems. 
%Another important problem in code translation is library mismatch and migration, which we plan to explore as the next step to advance C to Rust translation. 